\documentclass[reqno]{amsart}

\usepackage{amsaddr}
\usepackage{amsmath}
\usepackage{amssymb}
\usepackage{amsthm}
\usepackage{bm}
\usepackage{color}
\usepackage{eucal}
\usepackage{fullpage}
\usepackage{graphicx}
\usepackage{hhline}
\usepackage{lineno}
\usepackage{makecell}
\usepackage[sc]{mathpazo}
\usepackage{mathrsfs}
\usepackage{mathtools}
\usepackage[numbers,sort&compress,square]{natbib}
\usepackage{setspace}
\usepackage{stmaryrd}
\usepackage{subfig}
\usepackage[all,cmtip]{xy}

\newcommand*\patchAmsMathEnvironmentForLineno[1]{%
	\expandafter\let\csname old#1\expandafter\endcsname\csname #1\endcsname
	\expandafter\let\csname oldend#1\expandafter\endcsname\csname end#1\endcsname
	\renewenvironment{#1}%
	{\linenomath\csname old#1\endcsname}%
	{\csname oldend#1\endcsname\endlinenomath}}%
\newcommand*\patchBothAmsMathEnvironmentsForLineno[1]{%
	\patchAmsMathEnvironmentForLineno{#1}%
	\patchAmsMathEnvironmentForLineno{#1*}}%
\AtBeginDocument{%
	\patchBothAmsMathEnvironmentsForLineno{equation}%
	\patchBothAmsMathEnvironmentsForLineno{align}%
	\patchBothAmsMathEnvironmentsForLineno{flalign}%
	\patchBothAmsMathEnvironmentsForLineno{alignat}%
	\patchBothAmsMathEnvironmentsForLineno{gather}%
	\patchBothAmsMathEnvironmentsForLineno{multline}%
}

\begin{document}

\title{Evolutionary dynamics with game transitions}
\author{\small Qi Su$^{1,2}$, Alex McAvoy$^{3}$, Long Wang$^{1}$, and Martin A. Nowak$^{2,3,4}$}
\address{\small $^{1}$Center for Systems and Control, College of Engineering, Peking University, Beijing 100871, China \\
	$^{2}$Program for Evolutionary Dynamics, Harvard University, Cambridge, MA 02138, USA \\
	$^{3}$Department of Organismic and Evolutionary Biology, Harvard University, Cambridge, MA 02138, USA \\
	$^{4}$Department of Mathematics, Harvard University, Cambridge, MA 02138, USA}

\begin{abstract}
The environment has a strong influence on a population's evolutionary dynamics. Driven by both intrinsic and external factors, the environment is subject to continual change in nature. To capture an ever-changing environment, we consider a model of evolutionary dynamics with game transitions, where individuals' behaviors together with the games they play in one time step influence the games to be played next time step. Within this model, we study the evolution of cooperation in structured populations and find a simple rule: weak selection favors cooperation over defection if the ratio of the benefit provided by an altruistic behavior, $b$, to the corresponding cost, $c$, exceeds $k-k'$, where $k$ is the average number of neighbors of an individual and $k'$ captures the effects of the game transitions. Even if cooperation cannot be favored in each individual game, allowing for a transition to a relatively valuable game after mutual cooperation and to a less valuable game after defection can result in a favorable outcome for cooperation. In particular, small variations in different games being played can promote cooperation markedly. Our results suggest that simple game transitions can serve as a mechanism for supporting prosocial behaviors in highly-connected populations.
\end{abstract}

\maketitle

\section{Introduction}
The prosocial act of bearing a cost to provide another individual with a benefit, which is often referred to as ``cooperation'' \cite{sigmund:PUP:2010}, reduces the survival advantage of the donor and fosters that of the recipient. Understanding how such a trait can be maintained in a competitive world has long been a focal issue in evolutionary biology and ecology \cite{2006-Nowak-p1560-1563}. The spatial distribution of a population makes an individual more likely to interact with neighbors than with those who are more distant.
Population structures can affect the evolution of cooperation \cite{1964-Hamilton-p1-16,hamilton:JTB:1964b,1992-Nowak-p826-829,2006-Ohtsuki-p502-505,2007-Taylor-p469-469,2017-Allen-p227-230,2019-Qi-p20190041-20190041}.
In ``viscous'' populations, one's offspring often stay close to their places of birth.
Relatives thus interact more often than two random individuals.
Compared with the well-mixed setting, population ``viscosity'' is known to promote cooperation \cite{mitteldorf:JTB:2000} (although, when the population density is fixed, local competition can cancel the cooperation-promoting effect of viscosity \cite{wilson:EE:1992,taylor:EE:1992}).
Past decades have seen an intensive investigation of the evolution of cooperation in graph-structured populations \cite{2006-Ohtsuki-p502-505,2007-Taylor-p469-469,2017-Allen-p227-230,2019-Qi-p20190041-20190041}.
One of the best-known findings is that weak selection favors cooperation if the ratio of the benefit provided by an altruistic act, $b$, to the cost of expressing such an altruistic trait, $c$, exceeds the average number of neighbors, $k$, i.e. $b/c>k$ \cite{2006-Ohtsuki-p502-505,2014-Rand-p17093-17098}.
This simple rule strongly supports the proposition that population structure is one of the major mechanisms responsible for the evolution of cooperation \cite{2006-Nowak-p1560-1563}.

On the other hand, many realistic systems are highly-connected, with each individual having many neighbors on average.
For example, in a contact network consisting of students from a French high school, each student has $36$ neighbors on average,
meaning $k=36$ \cite{2015-Mastrandrea-p136497-136497}.
In such cases, the threshold for establishing cooperation, based on the rule ``$b/c>k$,'' is quite high: the benefit from an altruistic act must be at least $36$ times greater than its cost.
Somewhat large mean degrees have also been observed in collegiate Facebook networks, with well-known examples ranging from $39$ neighbors to well over $100$ \citep{traud:SIAMR:2011,traud:PA:2012,nr}. Such networks can (and do) involve the expression of social behaviors much more complex than those captured by the simple model of cooperation described previously. However, even for such a simple model, it is not understood if and when the threshold for the evolution of cooperation can be reduced to something less than the mean number of neighbors. Here, we consider a natural way in which this threshold can be relaxed using ``game transitions.''

In evolutionary game theory, an individual's reproductive success is determined by games played within the population. Many prior studies have relied on an assumption that the environment in which individuals evolve is time-invariant, meaning the individuals play a single, fixed game.
However, this assumption is not always realistic and can represent an oversimplification of reality \cite{2018-Hilbe-p246-249}, as many experimental studies have shown that the environment individuals face changes over time (and often) \cite{2014-Levin-p10838-10845,2013-Franzenburg-p781-781,2013-McFall-Ngai-p3229-3236,2008-Acar-p471-471}. As a simple example, overgrazing typically leads to the degradation of the common pasture land, leaving herders with fewer resources to utilize in subsequent seasons. By constraining the number of livestock within a reasonable range, herders can achieve a more sustainable use of pasture land \cite{2007-Rankin-p643-651}. In this kind of population, individuals' actions influence the state of environment, which in turn impacts the actions taken by its members. Apart from endogenous factors like individuals' actions, exogenous factors like seasonal climate fluctuations and soil conditions can also modify the environment experienced by the individuals. Examples are not limited to human-related activities but also appear in various microbial systems including bacteria and viruses \cite{2013-McFall-Ngai-p3229-3236,2008-Acar-p471-471}.

In this study, we use graphs to model a population's spatial structure, where nodes represent individuals and edges describe their interactions. We propose a model of evolutionary dynamics with game transitions: individuals sharing an edge interact (``play a game'') in each time step, and their strategic actions together with the game played determine the game to be played in the next time step. We find that game transitions can lower the threshold for establishing cooperation by $k'$, which means that the condition for cooperation to evolve is $b/c>k-k'$, where $k'$ captures the effects of the game transitions. Even if cooperation is disfavored in each individual game, transitions between the games can be favorable for the evolution of cooperation. In fact, just slight differences between games can dramatically lower the barrier for the success of cooperators. Our results suggest that game transitions can play a critical role in the evolution of prosocial behaviors.

\section{Model}
\begin{figure}[!h]
\centering
\includegraphics[width=0.8\linewidth]{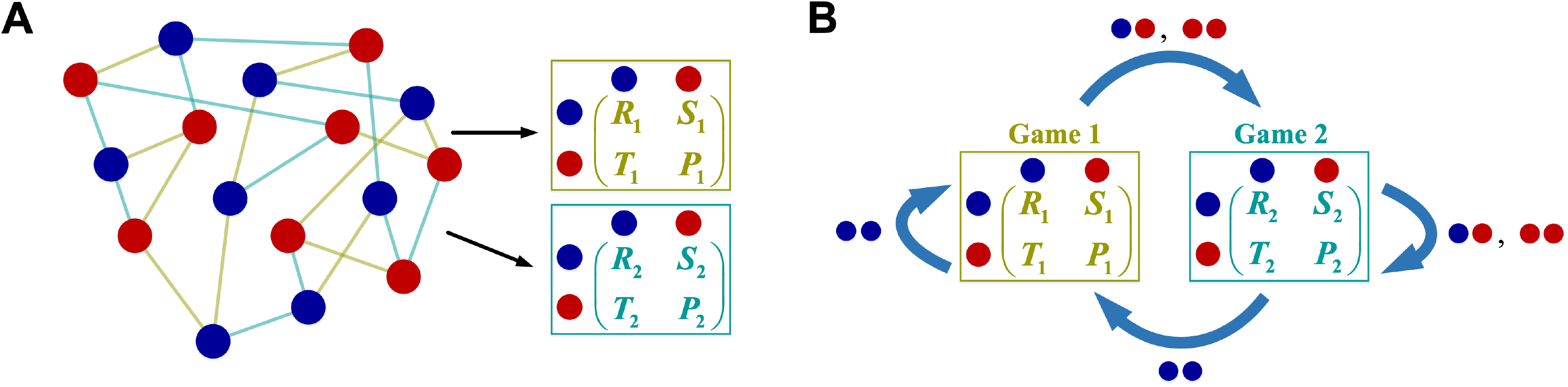}
\caption{\label{fig:1} \textbf{Game transitions on graphs}.
	Each player occupies a node on the graph and has a strategic behavior, blue (``cooperate'') or red (``defect''), used in interactions with neighbors (\textbf{A}).
	In every time step, each player plays a game with every neighbor and accumulates its payoffs from all interactions.
	Games in different interactions can be different, highlighted by the color of edges and relevant payoff matrices.
	At the end of each time step, a random player is selected to be replaced and all games update.
	Players' behaviors and the game they played in one time step determine the game to be played in the next time step (\textbf{B}). For example, if both players choose to take ``red'' behaviors in game 1, i.e. mutual defection, they will play game 2 in the subsequent time step.
}
\end{figure}

We study a population of $N$ players consisting of cooperators, $C$, and defectors, $D$. The population structure is described by a graph. Each player occupies a node on the graph. Edges between nodes describe the events related to interactions and biological reproduction (or behavior imitation). In each time step, each player interacts separately with every neighbor, and the games played in different interactions can be distinct (Fig.~\ref{fig:1}A). When playing game $i$, mutual cooperation brings each player a ``reward,'' $R_{i}$, whereas mutual defection leads to an outcome of ``punishment,'' $P_{i}$; unilateral cooperation leads to a ``sucker's payoff,'' $S_{i}$, for the cooperator and a ``temptation,'' $T_{i}$, for the defector. We assume that each game is a prisoner's dilemma, which is defined by the payoff ranking $T_{i}>R_{i}>P_{i}>S_{i}$. 
Each player derives an accumulated payoff, $\pi$, from all interactions, and this payoff is translated into reproductive fitness, $f=1-\delta +\delta\pi$, where $\delta\geqslant 0$ represents the intensity of selection \cite{2004-Nowak-p646-650}. We are particularly concerned with the effects of weak selection \cite{2010-Wu-p46106-46106,2013-Wu-p1003381-1003381}, meaning $0<\delta\ll 1$.

At the end of each time step, one player is selected for death uniformly at random from the population. The neighbors of this player then compete for the empty site, with each neighbor sending an offspring to this location with probability proportional to fitness. Following this ``death-birth'' update step, the games played in the population also update based on the previous games played and the actions taken in those games (Fig.~\ref{fig:1}B). For the player occupying the empty site, the games it will play are determined by the interactions of the prior occupant.

The game transition can be deterministic or stochastic (probabilistic). If the game to be played is independent of the previous game, the game transition is ``state-independent'' \cite{2018-Hilbe-p246-249}. When the game that will be played depends entirely on the previous game, the game transition is ``behavior-independent.'' The simplest case is when the games in all interactions are identical initially and remain constant throughout the evolutionary process, which is the setup of most prior studies \cite{2006-Ohtsuki-p502-505}.

\section{Results}
In the absence of mutation, a finite population will eventually reach a monomorphic state in which all players have the same strategy, either all-cooperation or all-defection. We study the competition between cooperation and defection by comparing the fixation probability of a single cooperator, $\rho_{C}$, to that of a single defector, $\rho_{D}$. Concretely, $\rho_{C}$ is the probability that a cooperator starting in a random location generates a lineage that takes over the entire population. Analogously, $\rho_{D}$ is the probability that a defector in a random position turns a population of cooperators into defectors. Selection favors cooperators relative to defectors if $\rho_{C}>\rho_{D}$ \cite{2004-Nowak-p646-650}.

\subsection{Game transitions between two states}
We begin with the case of deterministic game transitions between two states. Each state corresponds to a donation game (see SI Appendix, sections 3 and 4 for a comprehensive investigation of two-state games). In game $1$, a cooperator bears a cost of $c$ to bring its opponent a benefit of $b_1$, and a defector does nothing.
Analogously, in game $2$, a cooperator pays a cost of $c$ to bring its opponent a benefit of $b_2$.
That is, $R_i=b_i-c$, $S_i=-c$, $T_i = b_i$, and $P_i = 0$ in game $i$. Both $b_1$ and $b_2$ are larger than $c$.
The preferred choice for each player is defection, but $R_{i}>P_{i}$ in each game, resulting in the dilemma of cooperation.
We say that game $i$ is ``more valuable'' than game $j$ if $b_{i}>b_{j}$.
We take $b_1>b_2$ and explore a natural transition structure in which only mutual cooperation leads to the most valuable game.

\begin{figure}
\centering
\includegraphics[width=0.8\linewidth]{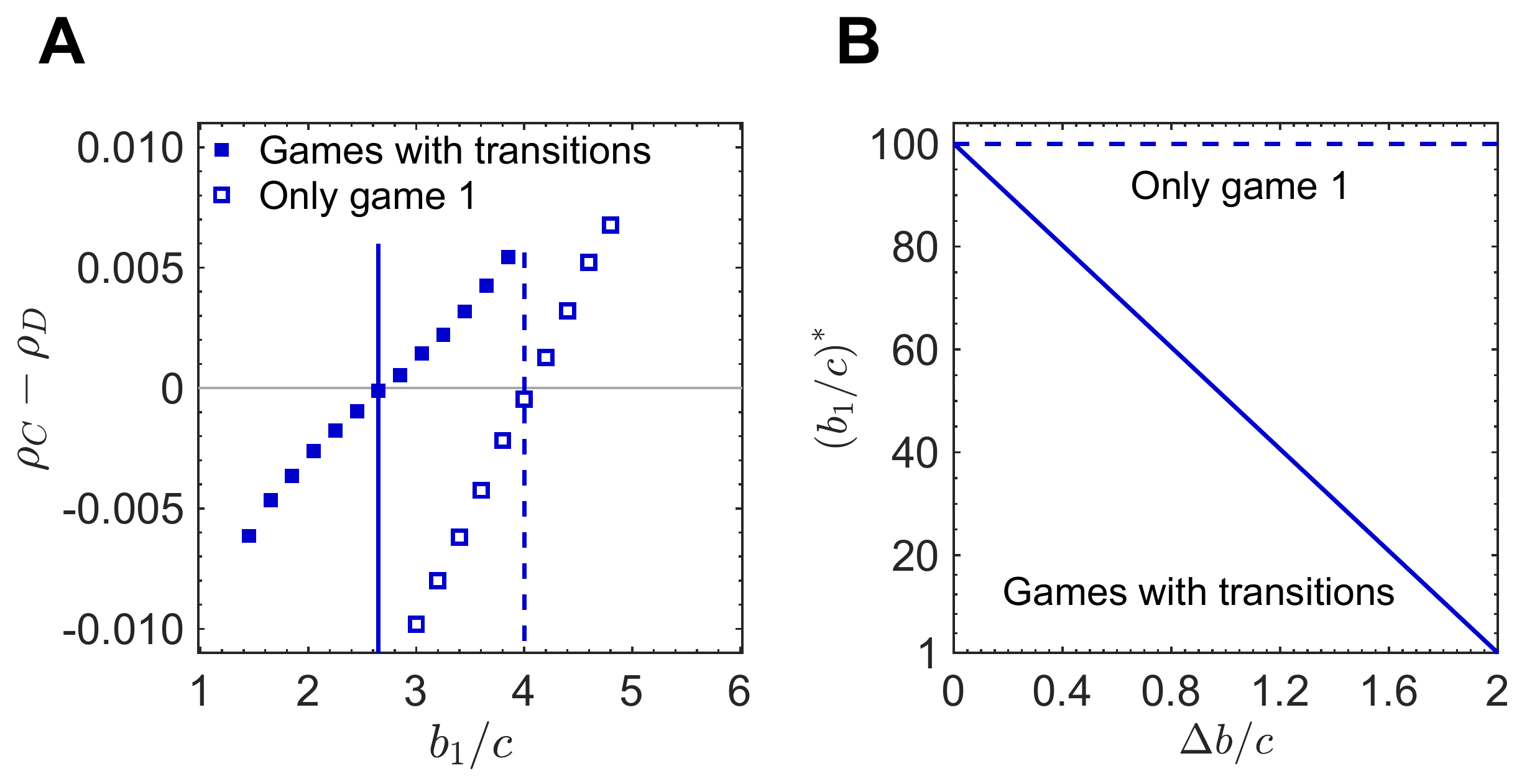}
\caption{\label{fig:2} \textbf{Game transitions can promote cooperation under death-birth updating.}
	We study a transition structure between two donation games. A cooperator pays a cost $c$ to bring its opponent a benefit $b_1$ in game 1 or $b_2$ in game 2;
	defectors pay nothing and provide no benefits.
	$b_1$ is larger than $b_2$. Mutual cooperation leads to game 1 and all other action profiles lead to game 2.
	Compared with only playing game 1, game transitions reduce the critical benefit-to-cost ratio, $\left(b_1/c\right)^{*}$, for the evolution of cooperation (see cross points of dots and the horizontal line in \textbf{A}).
	Dots show simulation data and vertical lines represent analytical results.
	Parameter values: $N=500$, $k=4$, $\delta =0.01$, $c = 1$,  and $b_2 = b_1-0.9$.
	In each simulation, all players play game 1 initially.
	Each simulation runs until the population reaches fixation (all-$C$ or all-$D$), and each point is averaged over $10^6$ independent runs.
	A small difference between $b_1$ and $b_2$ ($\Delta b = b_1-b_2$) remarkably reduces the critical benefit-to-cost ratio $\left(b_1/c\right)^{*}$ (\textbf{B}).
	We take $k=100$ in $\textbf{B}$.
}
\end{figure}

If every player has $k$ neighbors (i.e. the graph is ``$k$-regular''), we find that
\begin{equation}\label{eq:ruleDB}
	\rho_{C}>\rho_{D} \Longleftrightarrow \frac{b_{1}}{c} > k - \xi\frac{\Delta b}{c} ,
\end{equation}
where $\Delta b=b_{1}-b_{2}$ and $\xi =\left(k-1\right) /2$. Note that $\xi$ is positive and independent of payoff values such as $b_1$, $b_2$, and $c$. We obtain this condition under weak selection, based on the assumption that the population size $N$ is much larger than $k$. When $b_1=b_2$, the two games are the same, which leads to the well-known rule of $b_1/c>k$ for cooperation to evolve on regular graphs \cite{2006-Ohtsuki-p502-505}. The existence of the term $\xi\Delta b/c$ indicates that transitions between different games can reduce the barrier for the success of cooperation. Even when both games oppose cooperation individually, i.e. $b_1/c<k$ and $b_2/c<k$, transitions between them can promote cooperation (Fig.~\ref{fig:2}A). Our analytical results agree well with numerical simulations.

The beneficial effects of game transitions on cooperation become more prominent on graphs of large degree, $k$.
We find that a slight difference between games $1$ and $2$, $\Delta b$, can remarkably lower the barrier for cooperation to evolve.
For example, when $k = 100$ and $c = 1$, the critical benefit-to-cost ratio, $\left(b_1/c\right)^{*}$, decreases from $100$ to $50.5$ for $\Delta b=1.0$ (Fig. \ref{fig:2}B).
Therefore, game transitions can significantly promote cooperation in realistic and highly-connected societies \cite{1998-Watts-p440-442}. We find that similar results hold under the closely-related ``imitation'' updating (SI Appendix, Fig. S1 and section 3).

\begin{figure}
\centering
\includegraphics[width=0.8\linewidth]{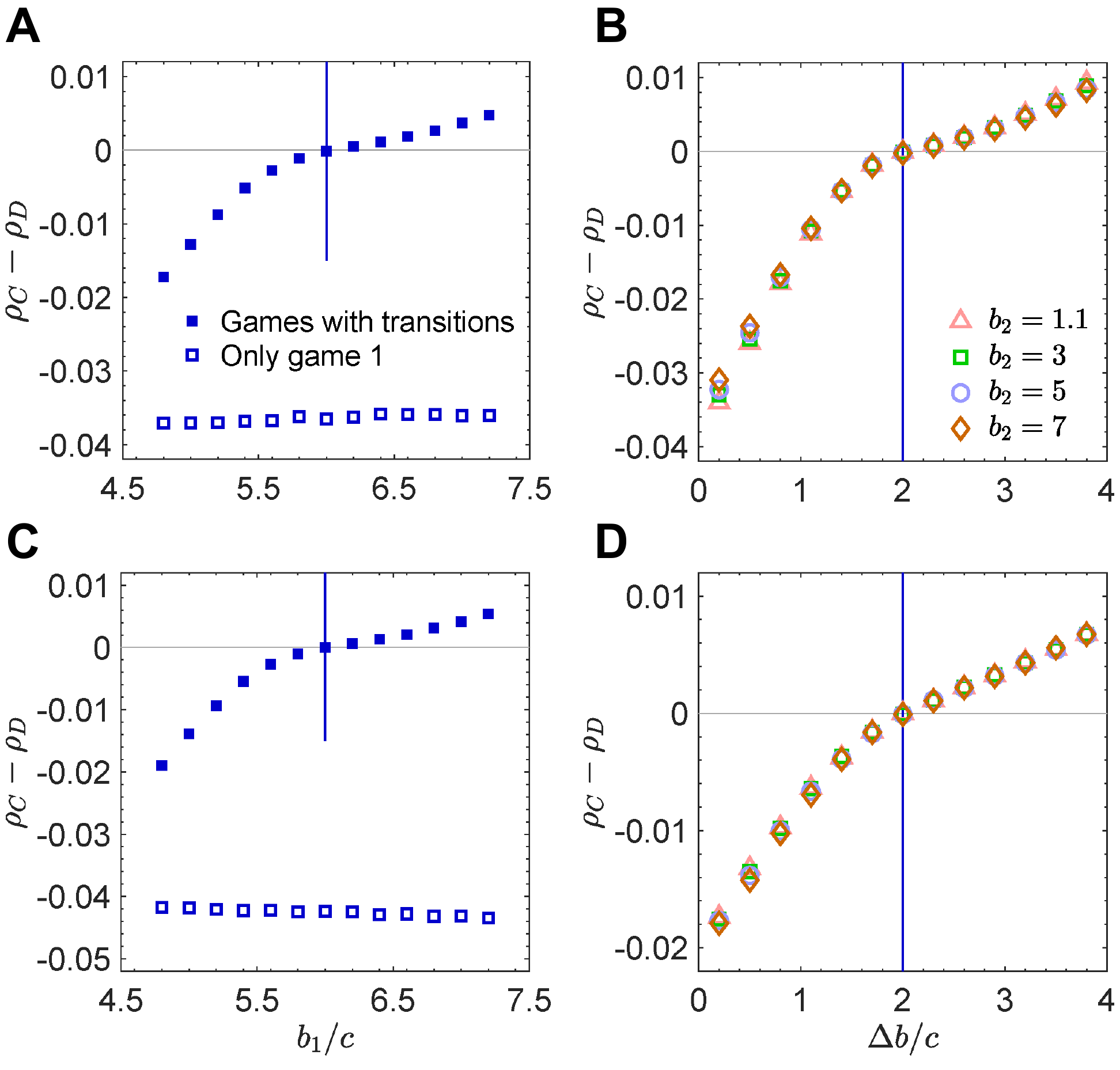}
\caption{\label{fig:3} \textbf{Game transitions can favor cooperation under birth-death (A, B) and pairwise-comparison updating (C, D).}
	When individuals play only game 1, cooperation is disfavored over defection for any benefit-to-cost ratio, $b_1/c$ (\textbf{A}, \textbf{C}).
	When mutual cooperation leads to game 1 and other action profiles lead to game 2, cooperation can evolve.
	With game transitions, the difference between the two games, $\Delta b = b_1-b_2$, rather than the individual value of $b_1$ and $b_2$, determines the success of cooperation (\textbf{B}, \textbf{D}).
	Parameter values: $N=500$, $k=4$, $\delta=0.01$, $c = 1$, and $b_2=4$ (\textbf{A}, \textbf{C}).
	In each simulation all players play game 1 initially.
	Each simulation runs until the population reaches fixation, and each point is averaged over $10^6$ independent runs.
}
\end{figure}

Next, we consider ``birth-death'' \cite{1958-Moran-p60-60} and ``pairwise-comparison'' \cite{1998-Szabo-p69-73,2007-Traulsen-p522-529} updating. Under birth-death updating, in each time step a random player is selected for reproduction with probability proportional to fitness. The offspring replaces a random neighbor. Under pairwise-comparison updating, a player is first selected uniformly-at-random to update his or her strategy. When player $i$ is chosen for a strategy updating, it randomly chooses a neighbor $j$ and compares payoffs. If $\pi_{i}$ and $\pi_{j}$ are the payoffs to $i$ and $j$, respectively, player $i$ adopts $j$'s strategy with probability $1/\left[1+\exp\left(-\delta\left(\pi_{j}-\pi_{i}\right)\right)\right]$ and retains its old strategy otherwise. When mutual cooperation leads to game 1 and other action profiles lead to game 2, under both birth-death and pairwise-comparison updating we have the rule
\begin{equation}
	\rho_{C}>\rho_{D} \Longleftrightarrow \xi\frac{\Delta b}{c} > 1 , \label{eq:ruleBDPC}
\end{equation}
where $\xi=1/2$ (SI Appendix, sections 3 and 4). When the two games are the same, $\Delta b=0$, and cooperators are never favored over defectors (Fig. \ref{fig:3}AC). Game transitions provide an opportunity for cooperation to thrive as long as $b_1-b_2>c/\xi$, which opens an avenue for the evolution of cooperation under birth-death and pairwise-comparison updating.
One can attribute this result to the fact that under this transition structure, mutual cooperation results in $b_{1}-c$ but when two players use different actions, the cooperator gets $-c$ and the defector gets $b_{2}$. If $b_{1}-b_{2}>c/\xi$, then it must be true that $b_{1}-c>b_{2}$, which means that the players are effectively in a coordination game with a preferred outcome of mutual cooperation.

More intriguingly, Eq.~\ref{eq:ruleBDPC} shows that the success of cooperators depends on the difference between benefits provided by an altruistic behavior in game $1$ and game $2$, and it is independent of the exact value in each game (Fig. \ref{fig:3}BD). Thus, in a dense population where individuals have many neighbors, even if the benefits provided by an altruistic behavior are low in both game 1 and game 2, transitions between them can still support the evolution of cooperation. We stress that the difference between the two games required to favor cooperation is surprisingly small. For example, $b_1-b_2>2c$ warrants the success of cooperation over defection on graphs of any degree.

We further examine random graphs \cite{1960-Erdoes-p17-61} and scale-free networks \cite{1999-Barabasi-p509-512}, where players differ in the number of their neighbors (SI Appendix, Fig. S2). We find that game transitions can provide more advantages for the evolution of cooperation than their static counterparts under death-birth and imitation updating, and they also give a way for cooperation to evolve under birth-death and pairwise-comparison updating. In addition, we study evolutionary processes with mutation and/or behavior exploration (SI Appendix, Fig. S3). The results demonstrate the robustness of the effects of game transitions on the evolution of cooperation.

\subsection{Stochastic, state-independent transitions}
\begin{figure}
\centering
\includegraphics[width=0.8\linewidth]{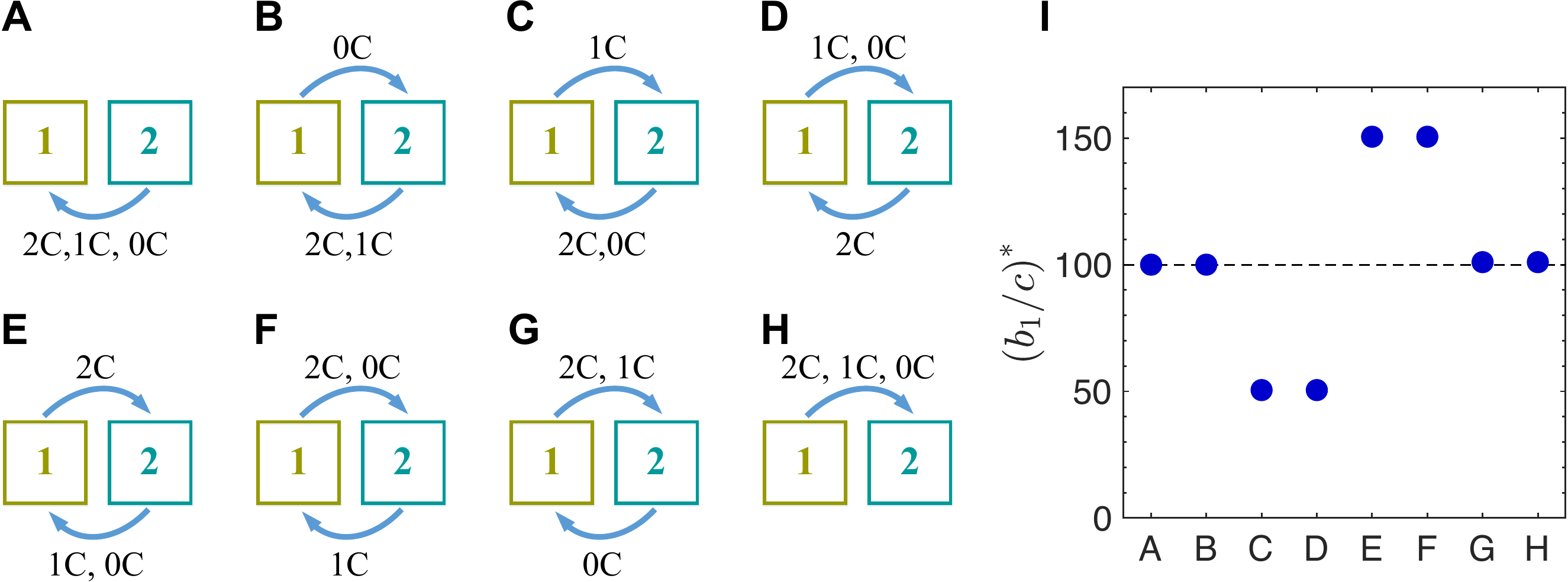}
\caption{\label{fig:4} \textbf{Critical ratio for the evolution of cooperation as a function of the game transition pattern}.
	Game transitions are state-independent, which means that the game to be played in the next time step depends on only the number of cooperators.
	For $\Delta b_{12}=1$ and $k=100$, we calculate the threshold $\left(b_1/c\right)^{*}$ for all eight deterministic transitions between two states ($\textbf{A}-\textbf{H}$).
	Game transitions promote cooperation only when mutual cooperation always allows for a more valuable game $1$ and unilateral defection leads to a less valuable game $2$ ($\textbf{C}, \textbf{D}$).
	The transition after mutual cooperation or unilateral defection is critical to the evolutionary outcome.
	For example, modifying the transition responding to mutual cooperation ($\textbf{B}$, $\textbf{F}$) or unilateral defection ($\textbf{B}$, $\textbf{D}$) changes $\left(b_1/c\right)^{*}$ significantly.
	However, the transitions responding to mutual defection have negligible effects. Critical ratios, $\left(b_{1}/c\right)^{\ast}$: $100$ ($\textbf{A}$, $\textbf{B}$), $50.5$ ($\textbf{C}$, $\textbf{D}$), $150.5$ ($\textbf{E}$, $\textbf{F}$), and $101$ ($\textbf{G}$, $\textbf{H}$).
}
\end{figure}

For more general state-independent transitions between two games, let $p$ and $q$ represent the probabilities of transitioning to game 2 (the less-valuable game) after mutual cooperation and after unilateral cooperation/defection, respectively. Under death-birth updating, the condition for cooperation to be favored over defection follows the format of Eq.~\ref{eq:ruleDB} with
\begin{equation} \label{alleightcase}
	\xi = \frac{k-1}{2}q-\frac{k+1}{2}p .
\end{equation}
The example in Fig.~\ref{fig:2} corresponds to $p=0$ and $q=1$. We explore all eight deterministic game transitions in Fig.~\ref{fig:4}. We see that game transitions promote cooperation only when mutual cooperation leads to a more profitable game 1 and unilateral defection leads to a less profitable game 2 (Fig.~\ref{fig:4}CD). However, when mutual cooperation leads to a detrimental state 1 and unilateral defection leads to a beneficial state 2, it is more difficult for cooperation to evolve (Fig.~\ref{fig:4}EF). In particular, whether or not the game transitions affect the evolution of cooperation depends strongly on the transition after mutual cooperation and the transition after unilateral cooperation/defection.
For example, in Fig.~\ref{fig:4}BF, changing the transition following mutual cooperation influences $\left(b_1/c\right)^{*}$ considerably. Transitions following mutual defection play a less prominent role (Fig.~\ref{fig:4}CD).

\subsection{Game transitions among $n$ states}
We turn now to the general setup of game transitions among $n$ states (i.e. games $1$ through $n$). If two players play game $i$ in the current time step and among them there are $s\in\left\{0,1,2\right\}$ cooperators, they will play game $j$ in the next time step with probability $p_{ij}^{\left(s\right)}$. $s$ is $2$ for mutual cooperation, $1$ for unilateral cooperation/defection, and $0$ for mutual defection.
In the prior example of the state transitioning to game 1 after mutual cooperation and to game 2 otherwise, we have $n=2$ and $p_{21}^{\left(2\right)}=1$.
This setup can recover deterministic or probabilistic transitions, state-dependent or -independent, behavior-dependent or -independent, and the traditional models involving only a single game \cite{2006-Ohtsuki-p502-505,2014-Rand-p17093-17098}, as specific cases. We assume that all games are donation games (see SI Appendix, section 3 for any two-player, two-strategy game). In game $i$, a cooperator pays a cost of $c$ to bring its opponent a benefit of $b_i$. Game $1$ is the most valuable, meaning $b_1\geqslant b_i$ for every $i$.

Under death-birth updating, we find that
\begin{equation}
	\rho_{C}>\rho_{D} \Longleftrightarrow \frac{b_{1}}{c} > k - \sum_{i=2}^n \xi_{i}\frac{\Delta b_{1i}}{c},  \label{eq:generalruleDB}
\end{equation}
where, for every $i$, $\Delta b_{1i}=b_1-b_i$, and $\xi_i$ depends on the game transition pattern (i.e. $p_{ij}^{(s)}$) but is independent of the benefit in each game, $b_{i}$, and cost, $c$ (see Appendix~A for the calculation of $\xi_i$). The term $\sum_{i=2}^n \xi_{i}\Delta b_{1i}/c$ captures how game transitions influence this threshold. The effects of game transitions on cooperation actually arise from two sources: the game transition pattern and the variation in different games.
$\xi_i$ captures the former and $\Delta b_{1i}/c$ the latter.

Importantly, these two components are independent, which makes it easier to understand the role of each. Let $k'$ denote $\sum_{i=2}^n \xi_{i}\Delta b_{1i}/c$ and let $b$ denote $b_1$.
We can interpret Eq.~\ref{eq:generalruleDB} intuitively: weak selection favors cooperation if the ratio of the benefit from an altruistic behavior, $b$, to its cost, $c$, exceeds the average \textit{effective} number of neighbors, $k-k'$.
Analogously, under birth-death or pairwise-comparison updating, we find that
\begin{equation}
	\rho_{C}>\rho_{D} \Longleftrightarrow \sum_{i=2}^n \xi_{i} \frac{\Delta b_{1i}}{c} > 1 . \label{eq:generalrulePC}
\end{equation}
We refer the reader to Appendix~A for the calculation of $\xi_i$.

Our study above assumes that in each time step, games played by any two players are likely to update (``global'' transitions). We also consider the case that games in only a fraction of interactions have chance to update. When games to be updated are randomly selected from the whole population, such a game transition can be transformed to the global transition with a modified transition matrix (see SI Appendix, section 3). Therefore, Eqs.~\ref{eq:generalruleDB}--\ref{eq:generalrulePC} still predict the evolutionary outcome.

We also study the case in which the games to be updated are spatially correlated, with only those nearby an individual who competes to reproduce being affected (``local'' transitions). Under death-birth and pairwise-comparison updating, global and local transitions lead to decidedly different models. We show that, however, the simple rules for cooperation to evolve (Eqs.~\ref{eq:generalruleDB} and \ref{eq:generalrulePC}) still hold provided $\xi_i$ is modified (SI Appendix, section 1 and Figs. S4-S6). We give a brief overview of local game transitions in Appendix~B.

\section{Pure versus stochastic strategies}
So far, in every time step each player is either a cooperator or a defector. But the model we propose here has a much broader scope than just two pure strategies. For example, we also investigate the competition between stochastic strategies under game transitions. Let $s_p$ denote a stochastic strategy with which, in each time step, a player chooses cooperation with probability $p$ and defects otherwise. $s_1$ thus corresponds to a pure cooperator and $s_0$ to a pure defector.

We find that the condition for $s_p$ being favored by selection over $s_q$ still follows the format of Eq.~\ref{eq:generalruleDB} under death-birth updating and Eq.~\ref{eq:generalrulePC} under birth-death or pairwise-comparison updating, provided that $\xi_i$ is modified (SI Appendix, section 3). When mutual cooperation leads to a more valuable game and other action profiles lead to a less valuable game, under death-birth updating game transitions lower the threshold for a cooperative strategy (i.e. $s_p$ with a large $p$) being favored relative to a less cooperative strategy. We also find that game transitions can favor the evolution of a cooperative stochastic strategy under birth-death and pairwise-comparison updating.

\section{Discussion}
We consider evolutionary dynamics with game transitions, coupling individuals' actions with the environment.
Individuals' behaviors modify the environment, which in turn affects the viability of future actions in that environment. We find a simple rule for the success of cooperators in an environment that can switch between an arbitrary number of states, namely $b/c>k-k'$, where $k'$ exactly captures how game transitions affect the evolution of cooperation. When all environmental states are identical, we recover the rule $b/c>k$ \cite{2006-Ohtsuki-p502-505}.

In a two-action game governed by a single payoff matrix with entries $R$, $S$, $T$, and $P$, the so-called ``sigma rule'' of Tarnita et al. \cite{2009-Tarnita-p570-581} says that there exists $\sigma$ for which cooperators are favored over defectors if and only if $\sigma R+S>T+\sigma P$. The coefficient $\sigma$, which is independent of the payoffs, captures how the spatial model and its associated update rule affect evolutionary dynamics. For an infinite random regular graph under death-birth updating, $\sigma =\left(k+1\right) /\left(k-1\right)$. When all interactions are governed by a donation game with a donation cost $c$ and benefit $b_1$, substituting $R=b_1-c$, $S=-c$, $T=b_1$ and $P=0$ into the sigma rule gives the condition of cooperation being favored over defection. Intriguingly, Eq.~\ref{eq:ruleDB} can be phrased in the form of a sigma rule, with $R=b_1-c+\left(b_1-b_2\right)\left(k-1\right) /\left(k+1\right)$, $S=-c$, $T=b_1$, and $P=0$. With game transitions, evolution proceeds ``as if'' all interactions are governed by an effective game with $R=b_1-c+\left(b_1-b_2\right)\left(k-1\right) /\left(k+1\right)$, $S=-c$, $T=b_1$, and $P=0$. Compared with the donation game, mutual cooperation brings each player an extra benefit of $\left(b_1-b_2\right)\left(k-1\right) /\left(k+1\right)$ in this effective game. That is, the game transitions create a situation in which two cooperators play a synergistic game and obtain synergistic benefits (see more discussions in SI Appendix, section 3E).

This intuition also holds for birth-death and pairwise-comparison updating. For a prisoner's dilemma in a constant environment, weak selection disfavors cooperation in any homogeneous structured population \cite{2006-Ohtsuki-p502-505,2014-Allen-p113-151}. With game transitions, the synergistic benefit to each cooperator upon their mutual cooperation induces a transformation of the payoff structure. In particular, the synergistic benefit can transform the nature of the interaction from a prisoner's dilemma to a coordination game with a preferred outcome of mutual cooperation.

The fact that game transitions allow cooperation to evolve is related to the idea of partner-fidelity feedback in evolutionary biology \cite{1991-Bull-p63-74,2004-Sachs-p135-160}. Partner-fidelity feedback describes that one's cooperation increases its partner's fitness, which ultimately feeds back as a fitness increase to the cooperator. Unlike reactive strategies like Tit-for-Tat, this feedback is an automatic process and does not require the partner's conditional response. In the classic example of grass-endophyte mutualism \cite{schardl:PRPB:1997,cheplick:OUP:2009}, by producing secondary compounds to protect the grass host, endophytes obtain more nutritional provisioning from the host. By providing nutrients to the endophytes, the grass host is more resistant to herbivores due to the increased delivery of secondary compounds. Similarly, in our study mutual cooperation could generate a synergistic benefit, which in turn promotes the evolution of cooperation.

When mutual cooperation allows for a more profitable game and other actions profiles lead to a less profitable game, a slight difference between games considerably reduces the threshold for the evolution of cooperation. The reason is that although the variation in games might be orders of magnitude smaller than the threshold for establishing cooperation, transitions among such games generate a synergistic benefit upon mutual cooperation that is of the same order of magnitude as the cost of a cooperative act. Since the synergistic benefit partly makes up for the loss from a cooperative act, a slight difference between games makes cooperation less costly. This finding is of significance to understanding large-scale cooperation in many highly-connected social networks. In these networks, an individual can have hundreds of neighbors \cite{1998-Watts-p440-442,2000-Amaral-p11149-11152} and cooperators thus face the risk of being exploited by many neighboring defectors. If the environment remains constant, cooperation must be profitable enough to make up for exploitation by defection \cite{2006-Ohtsuki-p502-505}. Game transitions can act to reduce the threshold for maintaining cooperation considerably.

We also find that game transitions can stabilize cooperation even when mutation or random strategy exploration is allowed. In a constant environment, when a mutant defector arises within a cluster of cooperators, it dilutes the spatial assortment of cooperators and thus hinders the evolution of cooperation \cite{2012-Allen-p97-105}. When the environment changes as a result of individuals' behaviors, although the defecting mutant indeed exploits its neighboring cooperators temporarily, the environment in which this happens deteriorates rapidly. As a result, the temptation to defect is weakened. In a constant environment, selection also favors the establishment of spatial assortment while mutation destroys it continuously. The population state finally reaches a ``mutation-selection stationary (MSS)'' distribution. But when the environment is subject to transitions, the interaction environment would also be a part of this distribution. In this case, the joint distribution over individuals' states and games could be described as a ``game-mutation-selection stationary (GMSS)'' distribution.

Recent years have seen a growing interest in exploring evolutionary dynamics in a changing and/or heterogeneous environments \cite{2019-Peter-p20140663-20140663,2013-Assaf-p238101-238101,2015-McAvoy-p1004349-1004349,2016-Weitz-p7518-7525,2016-Gokhale-p28-42,2006-Hauert-p2565-2571,2014-Stewart-p17558-17563,2019-Tilman-p-,2019-Hauert-p347-360,2019-Su-p1006947-1006947}. Our model is somewhat different. First, our study accounts for both exogenous factors and individuals' behaviors in the change of the environment, modeling general environmental feedback. In addition, the environment that two players face is independent of that of another pair of players. Individuals' strategic behaviors directly influence the environment in which they evolve, which enables an individual to reciprocate with the opponent in a single interaction through environmental feedback. Therefore, even if cooperators are disfavored in each individual environment, cooperators can still be favored over defectors through environmental reciprocity. Such an effect has never been observed in prior studies where all individuals interact in a homogeneous environment \cite{2019-Peter-p20140663-20140663,2016-Weitz-p7518-7525}. In those studies, although the environments the individuals face can change with time, at any specific stage the environment is identical for all individuals. When defection is a dominant strategy in each individual environment, defection also dominates cooperation in the context of an ever-changing environment \cite{2019-Peter-p20140663-20140663,2013-Assaf-p238101-238101,2016-Weitz-p7518-7525}. In a recent work, Hilbe et al. \cite{2018-Hilbe-p246-249} found that individuals can rely on repeated interactions and continuous strategies to achieve environmental reciprocity. Compared with their model, in our setup individuals play a one-shot game with a pure, unconditional strategy. Our model shows that without relying on direct reciprocity and any strategic complexity, game transitions can still promote the evolution of cooperation.

\setcounter{section}{0}
\setcounter{equation}{0}
\renewcommand{\thesection}{Appendix~\Alph{section}}
\renewcommand{\theequation}{A.\arabic{equation}}

\section{Calculation of $\xi_i$}
Let $p_{ij}^{\left(s\right)}$ ($i,j\in \{1,2,\cdots,n\}$ and $s\in\left\{0,1,2\right\}$) be the probability that the state transitions from game $i$ to game $j$ after $s$ players cooperate.
Let $\mathbf{P}^{\left(s\right)}$ denote a game transition matrix, where $p_{ij}^{\left(s\right)}$ is the element in the $i$th row and $j$th column.
We present the formula of $\xi_i$ for a class of game transition patterns here and show the calculation of $\xi$ for general transitions in SI Appendix, section 3.

For every $s\in\left\{0,1,2\right\}$, suppose that the Markov chain with state space $\left\{1,2,\dots ,n\right\}$ and transition matrix $\mathbf{P}^{\left(s\right)}$ has only one recurrence class (and the states therein are aperiodic). Let $\textbf{u}^{\left(s\right)}=\left(u_{1}^{\left(s\right)},\cdots,u_{n}^{\left(s\right)}\right)$ denote the stationary distribution of this chain, i.e. the solution to $\mathbf{u}^{\left(s\right)}=\mathbf{u}^{\left(s\right)}\mathbf{P}^{\left(s\right)}$ with $\sum_{j=1}^n u_{i}^{\left(s\right)}=1$. We have (SI Appendix, section 3)
\begin{equation}
	\xi_i = \frac{\left(k-1\right)}{2}u_{i}^{\left(1\right)}-\frac{\left(k+1\right)}{2}u_i^{\left(2\right)}
\end{equation}
for death-birth updating and
\begin{equation}
	\xi_i = \frac{u_i^{\left(1\right)}}{2}-\frac{u_i^{\left(2\right)}}{2}
\end{equation}
for birth-death or pairwise-comparison updating. In particular, for game transitions between two states, we have
\begin{equation}
	\xi_2 = \frac{\left(k-1\right) p_{12}^{\left(1\right)}}{2\left(p_{12}^{\left(1\right)}+p_{21}^{\left(1\right)}\right)}-\frac{\left(k+1\right) p_{12}^{\left(2\right)}}{2\left(p_{12}^{\left(2\right)}+p_{21}^{\left(2\right)}\right)}
\end{equation}
for death-birth updating and
\begin{equation}
	\xi_2 = \frac{p_{12}^{\left(1\right)}}{2\left(p_{12}^{\left(1\right)}+p_{21}^{\left(1\right)}\right)}-\frac{p_{12}^{\left(2\right)}}{2\left(p_{12}^{\left(2\right)}+p_{21}^{\left(2\right)}\right)},
\end{equation}
for birth-death or pairwise-comparison updating. For other game transitions, the evolutionary dynamics (and thus $\xi_{i}$) may be sensitive to the initial condition, i.e. the initial fractions of various games. We illustrate an example calculation of $\xi_i$ in SI Appendix, section 4.

\section{Global versus local game transitions}
\begin{figure}
\centering
\includegraphics[width=0.8\linewidth]{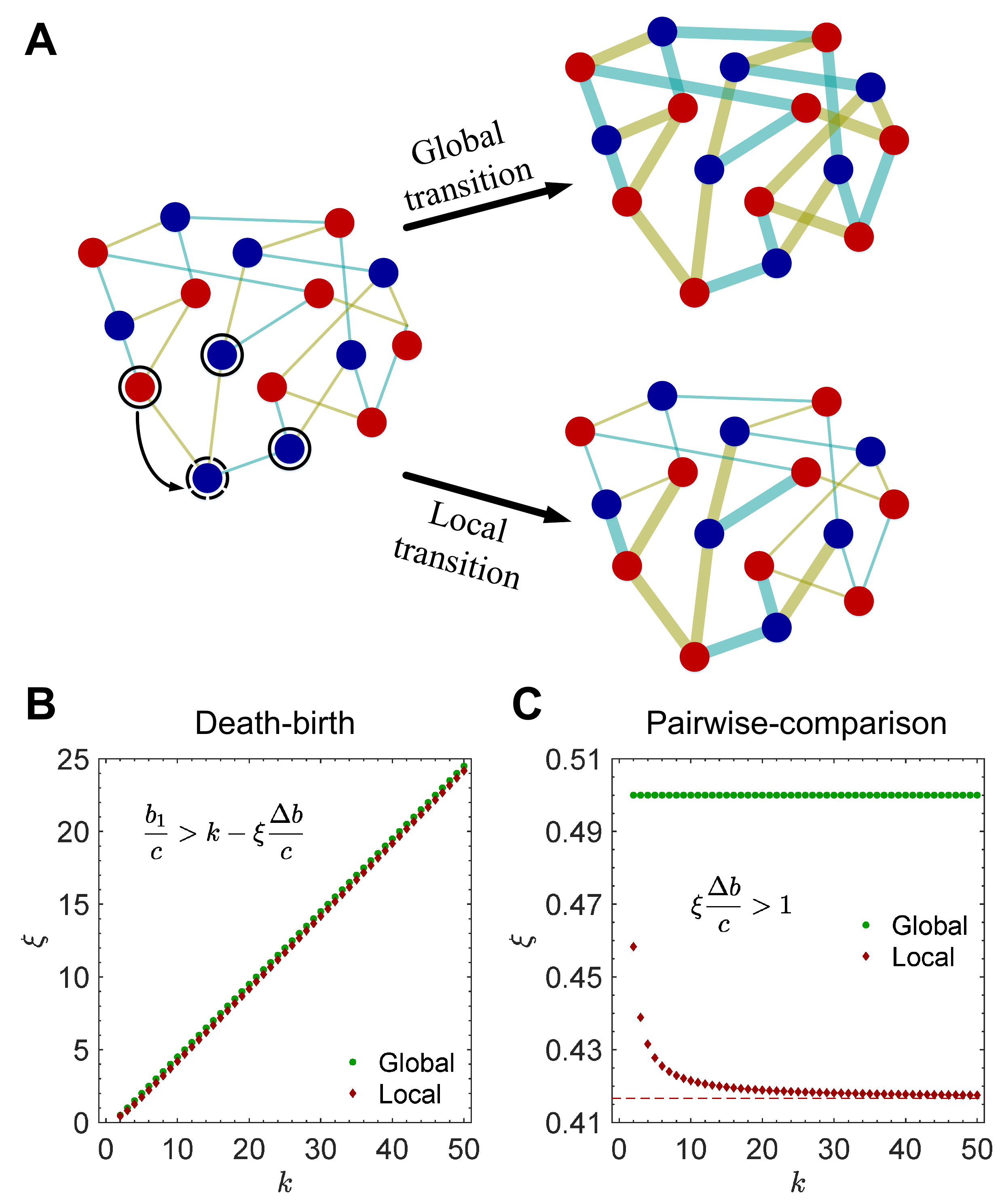}
\caption{\label{fig:5} \textbf{Global and local game transitions}.
	Depicted in $\textbf{A}$ is an example of game transitions in one time step under death-birth updating.
	A random player (dashed circle) is chosen for death;
	subsequently this individual's neighbors (solid circles) compete to reproduce and send an offspring into the vacancy with a probability proportional to fitness. With global game transitions, games in all interactions update in each time step.
	With local game transitions, only the games involved with players who compete to reproduce update (depicted by bold edges).
	We examine both global and local transitions under death-birth (\textbf{B}) and pairwise-comparison updating (\textbf{C}).
	When a game has an opportunity to update, it transitions to a more valuable game 1 after mutual cooperation and to a less valuable game 2 after defection.
	Game transitions, regardless of whether they are global or local, can promote cooperation markedly, although in this case global transitions result in a more relaxed condition for the evolution of cooperation than do local transitions.
}
\end{figure}

Our study above assumes that game transitions are an automatic (and exogenous) responses to interactions. Thus, in each time step, the games played by any two players are likely to update (``global'' transitions). But when the game transitions are subject to individuals' willingness to play a game, players could present different tendencies to modify the environments in which they evolve.
For example, under death-birth updating, if player $i$ is selected for death, then only $i$'s nearest neighbors compete to reproduce and replace $i$ with an offspring.
Compared with those not involved in competition around the vacant site, individuals close to the individual to be replaced have stronger incentives to change the environment they face since this environment affects their success in filling the vacancy.
In other words, games induced by the nearest neighbors of the deceased drive the evolution of a system. Therefore, one could impose transitions only on these games, leading to ``local'' transitions (see Fig.~\ref{fig:5}A).

Birth-death updating requires competition at the population level, so global and local transitions are identical in this case.
For death-birth and pairwise-comparison updating, however, global and local transitions lead to different models.
We show that the simple rules for cooperation to evolve (Eqs.~\ref{eq:generalruleDB} and \ref{eq:generalrulePC}) still hold provided $\xi_i$ is modified (SI Appendix, sections 1 and 4).
Specifically, we consider the following transition pattern: when a game has an opportunity to update, it transitions to a more valuable game 1 after mutual cooperation and to a less valuable game 2 after defection.
Under death-birth updating we have $\rho_{C}>\rho_{D}$ if and only if $b_{1}/c>k-\xi\Delta b/c$, where $\xi =\left(6k^{4}-10k^{3}+3k^{2}+6k+2\right) /\left(12k^{3}\right)$ (compared to $\xi = (k-1)/2$ for global transitions). For pairwise-comparison updating, $\rho_{C}>\rho_{D}$ if and only if $\xi\Delta b/c>1$, where $\xi =\left(10k^{2}-4k+1\right) /\left(24k^{2}-12k\right)$ (compared to $\xi=1/2$ for global transitions).

According to the nature of the critical threshold ($b_{1}/c>k-\xi\Delta b/c$ for death-birth updating and $\xi\Delta b/c>1$ for pairwise-comparison updating), global transitions act as a more effective promoter of cooperation than local transitions do (Fig. \ref{fig:5}BC). But for both kinds of game transitions, many messages are qualitatively the same: game transitions can promote cooperation (Figs. \ref{fig:2}--\ref{fig:3}, SI Appendix, Fig. S4); game transitions can amplify the beneficial effects of game variations on cooperation (Fig. \ref{fig:2}, Fig. \ref{fig:3}, SI Appendix, Fig. S5); and game transitions responding to mutual cooperation or unilateral cooperation/defection strongly affect cooperation. We include a more detailed discussion of global versus local game transitions in SI Appendix, section 4.
	
\section*{Acknowledgments}
We thank the referees for many helpful comments and for suggesting the connection to partner-fidelity feedback. The authors gratefully acknowledge support from the National Natural Science Foundation of China (grants 61751301, 61533001), the China Scholar Council (grant 201706010277), the Bill \& Melinda Gates Foundation (grant OPP1148627), the Army Research Laboratory (grant W911NF-18-2-0265), and the Office of Naval Research (grant N00014-16-1-2914).

\newpage

\setcounter{equation}{0}
\setcounter{figure}{0}
\setcounter{section}{0}
\setcounter{table}{0}
\renewcommand{\thesection}{SI.\arabic{section}}
\renewcommand{\thesubsection}{SI.\arabic{section}.\arabic{subsection}}
\renewcommand{\theequation}{SI.\arabic{equation}}
\renewcommand{\thefigure}{SI.\arabic{figure}}
\renewcommand{\thetable}{SI.\arabic{table}}

\begin{center}
\textbf{Supporting Information}
\end{center}

The supplementary information is structured as follows.

In Section 1, we study evolutionary dynamics with local game transitions.
We derive an analytical condition for one strategy to be favored over the other.
A further analysis gives the mathematical formula for the critical benefit-to-cost ratio for cooperation to evolve.

In Section 2, we study how the initial condition (initial fractions of various games played in the population) affects the evolutionary outcomes.
We provide an effective approach to evaluate whether or not the evolutionary dynamics is sensitive to the initial condition.

In Section 3, we study evolutionary dynamics with global game transitions.
We derive an analytical condition of one strategy to be favored over the other, as well as the critical benefit-to-cost ratio for cooperation to evolve.
We prove that our rules also hold when players use stochastic strategies (cooperate or defect with a probability rather than unconditionally).

In Section 4, we study four representative examples, including state-independent game transitions (the game to be played is independent of games played in the past), strategy-independent game transitions (the game to be played is independent of players' actions in the past), game transitions between two states (including the example presented in the main text), and probabilistic game transitions between three states (transitions between different games with a probability).
We show that how probabilistic game transitions affect the favorable effects of game transitions on cooperation may depend on the variations in different games.

\section{Evolutionary dynamics with local game transitions}\label{section_local}
We consider game transitions among $n$ states, described by games $1,2,\dots ,n$.
The payoff structure of game $i$ is
\begin{equation}
\bordermatrix{
& A & B \cr
A & R_i & S_i \cr
B & T_i & P_i \cr
},
\end{equation}
where each value corresponds to a payoff derived by a player with a strategy in the column against a player with a strategy in the row.
The game transition pattern is described by three matrices, i.e.
\begin{equation} \label{game_transition}
\textbf{P}^{(2)}
=
\left[
\begin{array}{ccc}
p_{11}^{(2)} & \cdots & p_{1n}^{(2)} \\
\vdots & \ddots & \vdots \\
p_{n1}^{(2)} & \cdots & p_{nn}^{(2)}
\end{array}
\right],
\quad
\textbf{P}^{(1)}
=
\left[
\begin{array}{ccc}
p_{11}^{(1)} & \cdots & p_{1n}^{(1)}\\
\vdots & \ddots & \vdots \\
p_{n1}^{(1)} & \cdots & p_{nn}^{(1)}
\end{array}
\right],
\quad
\textbf{P}^{(0)}
=
\left[
\begin{array}{ccc}
p_{11}^{(0)} & \cdots & p_{1n}^{(0)} \\
\vdots & \ddots & \vdots \\
p_{n1}^{(0)} & \cdots & p_{nn}^{(0)}
\end{array}
\right],
\end{equation}
where $p_{ij}^{(s)}$ represents the probability that players play game $j$ in the next time step conditioned on that they play game $i$ in the current time step and there are $s$ $A$-players, where $i,j\in \left\{1,2,\dots, n\right\}$ and $s\in \left\{0,1,2\right\}$.

On graphs or social networks, each player occupies a node.
If two players (or nodes occupied by players) are connected by an edge or a social tie, they play a one-shot game in each time step.
The main idea of the theoretical analysis is to couple the game played by two connected players and their strategy profiles into edges.
Let $E_{XY}^{(i)}$ denote an edge in which the two connected players take strategies $X$ and $Y$ ($X,Y\in\left\{A,B\right\}$), respectively, in game $i\in\left\{1,2,\dots ,n\right\}$.
For example, in edge $E_{AA}^{(1)}$, both of the players take $A$ strategy and they play game $1$.
We then introduce the following variables to describe this evolving system: \\
$p_A$: the frequency of $A$-players; \\
$p_B$: the frequency of $B$-players; \\
$p_{XY}^{(i)}$: the frequency of edge $E_{XY}^{(i)}$; \\
$q_{X|Y}^{(i)}$: the probability to find an edge $E_{XY}^{(i)}$ given that one node of this edge is occupied by a $Y-$player; \\
$p_{XY}$: the frequency of edges that connect an $X$-player and a $Y$-player; \\
$q_{X|Y}$: the conditional probability to find an $X-$player given that the adjacent node is occupied by a $Y-$player.

Then we have the identities
\begin{subequations}
\begin{equation}\label{notation_3a}
p_A+p_B = 1 ; \\
\end{equation}
\begin{equation}
\sum_{i=1}^n q_{X|Y}^{(i)} = q_{X|Y} ; \\
\end{equation}
\begin{equation}
p_{XY}^{(i)} = q_{X|Y}^{(i)} p_Y ; \\
\end{equation}
\begin{equation}
p_{XY}^{(i)} = p_{YX}^{(i)} ; \\
\end{equation}
\begin{equation}
q_{A|X}+q_{B|X} = 1 ; \\
\end{equation}
\begin{equation}\label{notaitonf}
p_{XY} = q_{X|Y} p_Y .
\end{equation}
\end{subequations}

Note that players' strategies and the game they play coevolve throughout the evolutionary process.
From the perspective of network dynamics, we need to consider the change in the frequency of nodes occupied by $A$-players and the frequency of edge $E_{XY}^{(i)}$.
Based on above identities, we can use $p_A$ and $q_{X|Y}^{(i)}$ to describe the whole system.
In the following, we study a random regular graph, where each node is linked to other $k$ nodes.

\subsection{Interactions}
In each time step, each player interacts separately with every neighbor, and the games played in different interactions can be distinct.
Each player derives an accumulated payoff, $\pi$, from all interactions, and this payoff is translated into reproductive fitness, $f=1-\delta +\delta\pi$, where $\delta\geqslant 0$ represents the intensity of selection \cite{si2004-Nowak-p646-650}.
$\delta$ scales the contribution of games played to one's fitness/reproductive rates.
The assumption of $\delta\ll 1$, termed weak selection, describes that the game plays only a very small role or it represents only one of many factors influencing the overall reproductive rate.
Besides, this assumption allows to derive analytical results and has been widely used in evolutionary biology \cite{si2004-Ewens,si2004-Nowak-p646-650}.
We focus on the effects of weak selection \cite{si2010-Wu-p46106-46106,si2013-Wu-p1003381-1003381}.

\subsection{Death-birth updating}
Under death-birth updating, in each time step, a random player is selected to die; all neighbors then compete to reproduce and send an offspring to the vacant site (with the probability proportional to fitness) \cite{si2006-Ohtsuki-p502-505}.
We can also interpret this updat rule in a social setting:
a random player $i$ decides to update his or her strategy;
subsequently, he or she adopts a neighbor's strategy with a probability proportional to the neighbor's fitness.
Local transitions account for the fact that only the nearest neighbors compete for the vacancy.
When the environment change is subject to human's willingness, these neighbors, compared with other players not involved in the competition, are more incentivized to modify the environment (games) in which they evolve.
Therefore, under local game transitions, only games played by the nearest neighbors of the dead can update.
We first investigate the change in the frequency of $A-$players.

\subsubsection{Change in $p_A$---updating a B-player} \label{subsubsection_pA_B}
A $B$-player is chosen to die with probability $p_B$.
Let $k_{A|B}^{(i)}$ denote the number of neighbors who adopt strategy $A$ and play game $i$ with the focal (dead) player.
Analogously, $k_{B|B}^{(i)}$ denotes the number of neighbors who adopt strategy $B$ and play game $i$ with the focal player.
Therefore, $\sum_{i=1}^n \left(k_{A|B}^{(i)}+k_{B|B}^{(i)}\right)=k$.
The probability for such a neighborhood configuration is
\begin{equation}
\mathcal{B}\left(k_{A|B}^{(i)},k_{B|B}^{(i)}|i=1,\dots,n\right)=
\frac{k!}{\Pi_{i=1}^n\left(k_{A|B}^{(i)}!k_{B|B}^{(i)}!\right)}
\Pi_{i=1}^n\left[\left(q_{A|B}^{(i)}\right)^{k_{A|B}^{(i)}}\left(q_{B|B}^{(i)}\right)^{k_{B|B}^{(i)}}\right].
\end{equation}
The fitness of a neighbor who adopts strategy $A$ and plays game $i$ with the focal player is
\begin{equation}
f_{A|B}^{(i)}=1-\delta+\delta\left[(k-1)\sum_{j=1}^nq_{A|A}^{(j)}R_j+(k-1)\sum_{j=1}^nq_{B|A}^{(j)}S_j+S_i\right].
\end{equation}
The fitness of a neighbor who adopts strategy $B$ and plays game $i$ with the focal player is
\begin{equation}
f_{B|B}^{(i)}=1-\delta+\delta\left[(k-1)\sum_{j=1}^nq_{A|B}^{(j)}T_j+(k-1)\sum_{j=1}^nq_{B|B}^{(j)}P_j+P_i\right].
\end{equation}
The probability that one of neighboring $A$-players replaces the vacancy under such a neighborhood configuration is given by
\begin{equation}
\mathbb{P}\left(A\rightarrow B\right)
=\frac{\sum_{i=1}^nk_{A|B}^{(i)}f_{A|B}^{(i)}}{\sum_{i=1}^n\left(k_{A|B}^{(i)}f_{A|B}^{(i)}+k_{B|B}^{(i)}f_{B|B}^{(i)}\right)}.
\end{equation}
The probability that one of neighboring $B-$players replaces the vacancy under such a neighborhood configuration is given by
\begin{equation}
\mathbb{P}\left(B\rightarrow B\right)
=\frac{\sum_{i=1}^nk_{B|B}^{(i)}f_{B|B}^{(i)}}{\sum_{i=1}^n\left(k_{A|B}^{(i)}f_{A|B}^{(i)}+k_{B|B}^{(i)}f_{B|B}^{(i)}\right)}.
\end{equation}
Therefore, $p_A$ increases by $1/N$ with probability
\begin{equation}
\mathbb{P}\left(\Delta p_A=\frac{1}{N}\right)=p_B\sum_{\sum_{i=1}^n \left(k_{A|B}^{(i)}+k_{B|B}^{(i)}\right)=k}
\mathcal{B}\left(k_{A|B}^{(i)},k_{B|B}^{(i)}|i=1,\dots,n\right)
\mathbb{P}\left(A\rightarrow B\right) .
\end{equation}

\subsubsection{Change in $p_A$---updating a $A$-player} \label{subsubsection_pA_A}
An $A$-player is chosen to die with probability $p_A$.
Let $k_{A|A}^{(i)}$ denote the number of neighbors who adopt strategy $A$ and play game $i$ with the focal player.
Analogously, $k_{B|A}^{(i)}$ denotes the number of neighbors who adopt strategy $B$ and play game $i$ with the focal player.
Therefore, $\sum_{i=1}^n \left(k_{A|A}^{(i)}+k_{B|A}^{(i)}\right)=k$.
The probability for such a neighborhood configuration is
\begin{equation}
\mathcal{A}\left(k_{A|A}^{(i)},k_{B|A}^{(i)}|i=1,\dots,n\right)=
\frac{k!}{\Pi_{i=1}^n\left(k_{A|A}^{(i)}!k_{B|A}^{(i)}!\right)}
\Pi_{i=1}^n\left[\left(q_{A|A}^{(i)}\right)^{k_{A|A}^{(i)}}\left(q_{B|A}^{(i)}\right)^{k_{B|A}^{(i)}}\right].
\end{equation}
The fitness of a neighbor who adopts strategy $A$ and plays game $i$ with the focal player is
\begin{equation}
f_{A|A}^{(i)}=1-\delta+\delta\left[(k-1)\sum_{j=1}^nq_{A|A}^{(j)}R_j+(k-1)\sum_{j=1}^nq_{B|A}^{(j)}S_j+R_i\right].
\end{equation}
The fitness of a neighbor who adopts strategy $B$ and plays game $i$ with the focal player is
\begin{equation}
f_{B|A}^{(i)}=1-\delta+\delta\left[(k-1)\sum_{j=1}^nq_{A|B}^{(j)}T_j+(k-1)\sum_{j=1}^nq_{B|B}^{(j)}P_j+T_i\right].
\end{equation}
The probability that one of neighboring $A$-players replaces the vacancy under such a neighborhood configuration is given by
\begin{equation}
\mathbb{P}\left(A\rightarrow A\right)
=\frac{\sum_{i=1}^nk_{A|A}^{(i)}f_{A|A}^{(i)}}{\sum_{i=1}^n\left(k_{A|A}^{(i)}f_{A|A}^{(i)}+k_{B|A}^{(i)}f_{B|A}^{(i)}\right)}.
\end{equation}
The probability that one of neighboring $B-$players replaces the vacancy under such a neighborhood configuration is given by
\begin{equation}
\mathbb{P}\left(B\rightarrow A\right)
=\frac{\sum_{i=1}^nk_{B|A}^{(i)}f_{B|A}^{(i)}}{\sum_{i=1}^n\left(k_{A|A}^{(i)}f_{A|A}^{(i)}+k_{B|A}^{(i)}f_{B|A}^{(i)}\right)}.
\end{equation}
Therefore, $p_A$ decreases by $1/N$ with probability
\begin{equation}
\mathbb{P}\left(\Delta p_A=-\frac{1}{N}\right)=p_A\sum_{\sum_{i=1}^n \left(k_{A|A}^{(i)}+k_{B|A}^{(i)}\right)=k}
\mathcal{A}\left(k_{A|A}^{(i)},k_{B|A}^{(i)}|i=1,\dots,n\right)
\mathbb{P}\left(B\rightarrow A\right) .
\end{equation}

\subsubsection{Change in $p_A$}
Let us now suppose that one strategy replacement event takes place in one unit of time.
The time derivative of $p_A$ is given by
\begin{equation} \label{pA_derivative_DB}
\begin{split}
\dot{p}_A
&= \frac{1}{N}\mathbb{P}\left(\Delta p_A=\frac{1}{N}\right)+\left(-\frac{1}{N}\right)\mathbb{P}\left(\Delta p_A=-\frac{1}{N}\right) \\
&= \delta\frac{1}{kN} \sum_{i=1}^{n}\left( I_{R_i}R_i+I_{S_i}S_i+I_{T_i}T_i+I_{P_i}P_i \right)+O\left(\delta^{2}\right) ,
\end{split}
\end{equation}
where
\begin{subequations}
\begin{equation}
I_{R_i} = p_{AA}^{(i)}(k-1)\left[(k-1)q_{B|A}(q_{A|A}+q_{B|B})+q_{B|A}\right] ; \label{I_ai} \\
\end{equation}
\begin{equation}
I_{S_i} = p_{AB}^{(i)}(k-1)\left[(k-1)q_{B|A}(q_{A|A}+q_{B|B})+q_{B|B}\right] ; \label{I_bi}\\
\end{equation}
\begin{equation}
I_{T_i} = -p_{AB}^{(i)}(k-1)\left[(k-1)q_{A|B}(q_{A|A}+q_{B|B})+q_{A|A}\right] ; \label{I_ci}\\
\end{equation}
\begin{equation}
I_{P_i} = -p_{BB}^{(i)}(k-1)\left[(k-1)q_{A|B}(q_{A|A}+q_{B|B})+q_{A|B}\right] . \label{I_di}
\end{equation}
\end{subequations}

\subsubsection{Change in $p_{AA}^{(i)}$}
We proceed with the change in the frequency of each type of edge.
Note that when a random player $l$ is chosen to die, the edges between (i) $l$ and its nearest neighbors and (ii) $l$'s nearest neighbors and next-nearest neighbors have chance to update (see the description of local game transitions).
We stress that the change in $p_{AA}^{(i)}$ is different from that in $p_A$.
$p_A$ does not change when neighboring $A$-players replace the focal $A$-player (the dead $A$-player) or neighboring $B$-players replace the focal $B$-player (the dead $B$-player).
However, in the same case $p_{AA}^{(i)}$ likely changes since games in these edges may switch, which changes the edge type.

We first consider the case that a random $B$-player is chosen to die.
We take the same neighborhood configuration as we do in Section \ref{subsubsection_pA_B}, i.e. $k_{A|B}^{(j)},k_{B|B}^{(j)}$ for $j=1,\dots,n$.
The change in $p_{AA}^{(i)}$ results from two parts: the switching of edges connecting the focal $B$-player and its nearest neighbors, and the switching of edges connecting the nearest neighbors and the next-nearest neighbors.
Under the given neighborhood configuration, the change in $p_{AA}^{(i)}$ based on the former part is
\begin{equation} \label{pAAi_AtakeB_inside}
\mathbb{P}\left(\Delta p_{AA}^{(i)}=\frac{2\left[\sum_{j=1}^n p_{ji}^{(1)}k_{A|B}^{(j)}\right]}{kN}\right)
=\mathcal{B}\left(k_{A|B}^{(j)},k_{B|B}^{(j)}|j=1,\dots,n\right)\mathbb{P}\left(A\rightarrow B\right).
\end{equation}
Eq. \ref{pAAi_AtakeB_inside} describes the edge switching of $E_{BA}^{(j)}\rightarrow E_{AA}^{(i)}$, which occurs when
(i) a neighboring $A$-player reproduces and its offspring replaces the dead $B$-player, i.e., $BA \rightarrow AA$; (ii) neighboring $A$-players who plays game $j$ with the dead in the current time step then plays game $i$ in the next time step, $(j)\rightarrow (i)$.

The change in $p_{AA}^{(i)}$ due to edges between the nearest and the next-nearest neighbors is
\begin{equation} \label{pAAi_AtakeB_outside}
\begin{split}
&\mathbb{P}\left(\Delta p_{AA}^{(i)}=\frac{2(k-1)\left[\sum_{j=1}^n p_{ji}^{(2)}q_{A|A}^{(j)}-\sum_{j=1}^np_{ij}^{(2)}q_{A|A}^{(i)}\right]\sum_{j=1}^nk_{A|B}^{(j)}}{kN}\right)  \\ &=\mathcal{B}\left(k_{A|B}^{(j)},k_{B|B}^{(j)}|j=1,\dots,n\right)\left[\mathbb{P}\left(A\rightarrow B\right)+\mathbb{P}\left(B\rightarrow B\right)\right].
\end{split}
\end{equation}
Eq. \ref{pAAi_AtakeB_outside} indicates that regardless of which neighbor replaces the focal $B$-player, the change in $p_{AA}^{(i)}$ due to the edges between the nearest and next-nearest neighbors remains the same.

Next, we consider the case in which a random $A$-player is chosen to die.
We take the same neighborhood configuration as we do in Section \ref{subsubsection_pA_A}, i.e., $k_{A|A}^{(j)},k_{B|A}^{(j)}$ for $j=1,\dots,n$.
The change in $p_{AA}^{(i)}$ due to edges between the focal $A$-player and its nearest neighbors is
\begin{equation} \label{pAAi_AtakeA_inside}
\mathbb{P}\left(\Delta p_{AA}^{(i)}=\frac{2\left[\sum_{j=1}^n p_{ji}^{(2)}k_{A|A}^{(j)}-\sum_{j=1}^n p_{ij}^{(2)}k_{A|A}^{(i)}\right]}{kN}\right)
=\mathcal{A}\left(k_{A|A}^{(j)},k_{B|A}^{(j)}|j=1,\dots,n\right)\mathbb{P}\left(A\rightarrow A\right)
\end{equation}
and
\begin{equation} \label{pAAi_BtakeA_inside}
\mathbb{P}\left(\Delta p_{AA}^{(i)}=\frac{-2k_{A|A}^{(i)}}{kN}\right)= \mathcal{A}\left(k_{A|A}^{(j)},k_{B|A}^{(j)}|j=1,\dots,n\right)\mathbb{P}\left(B\rightarrow A\right).
\end{equation}
Eq. \ref{pAAi_AtakeA_inside} (resp. Eq. \ref{pAAi_BtakeA_inside}) captures the case in which a neighboring $A$-player (resp. $B$-player) successfully occupies the vacant site.

The change in $p_{AA}^{(i)}$ due to edges between the nearest and next-nearest neighbors is
\begin{equation} \label{pAAi_AtakeA_outside}
\begin{split}
&\mathbb{P}\left(\Delta p_{AA}^{(i)}=\frac{2(k-1)\left[\sum_{j=1}^n p_{ji}^{(2)}q_{A|A}^{(j)}-\sum_{j=1}^np_{ij}^{(2)}q_{A|A}^{(i)}\right]\sum_{j=1}^nk_{A|A}^{(j)}}{kN}\right) \\ &=\mathcal{A}\left(k_{A|A}^{(j)},k_{B|A}^{(j)}|j=1,\dots,n\right)\left[\mathbb{P}\left(A\rightarrow A\right)+\mathbb{P}\left(B\rightarrow A\right)\right] .
\end{split}
\end{equation}

From Eqs. \ref{pAAi_AtakeB_inside}-\ref{pAAi_AtakeA_outside}, the time derivative of $p_{AA}^{(i)}$ is given by
\begin{equation} \label{pAAi_derivative}
\begin{split}
\dot{p}_{AA}^{(i)} &=
\sum_{\sum_{j=1}^n \left(k_{A|B}^{(j)}+k_{B|B}^{(j)}\right)=k}p_B\mathbb{P}\left(\Delta p_{AA}^{(i)}=
\frac{2\sum_{j=1}^n p_{ji}^{(1)}k_{A|B}^{(j)}}{kN}\right)\frac{2\sum_{j=1}^n p_{ji}^{(1)}k_{A|B}^{(j)}}{kN}  \\
&+\sum_{\sum_{j=1}^n \left(k_{A|B}^{(j)}+k_{B|B}^{(j)}\right)=k}p_B\mathbb{P}\left(\Delta p_{AA}^{(i)}=\frac{2(k-1)\left[\sum_{j=1}^n p_{ji}^{(2)}q_{A|A}^{(j)}-\sum_{j=1}^np_{ij}^{(2)}q_{A|A}^{(i)}\right]\sum_{j=1}^nk_{A|B}^{(j)}}{kN}\right)  \\
&\hspace{3.5cm} \frac{2(k-1)\left[\sum_{j=1}^n p_{ji}^{(2)}q_{A|A}^{(j)}-\sum_{j=1}^np_{ij}^{(2)}q_{A|A}^{(i)}\right]\sum_{j=1}^nk_{A|B}^{(j)}}{kN}  \\
&+\sum_{\sum_{j=1}^n \left(k_{A|A}^{(j)}+k_{B|A}^{(j)}\right)=k}p_A\mathbb{P}\left(\Delta p_{AA}^{(i)}=\frac{2\left[\sum_{j=1}^n p_{ji}^{(2)}k_{A|A}^{(j)}-\sum_{j=1}^n p_{ij}^{(2)}k_{A|A}^{(i)}\right]}{kN}\right)  \\
&\hspace{3.5cm} \frac{2\left[\sum_{j=1}^n p_{ji}^{(2)}k_{A|A}^{(j)}-\sum_{j=1}^n p_{ij}^{(2)}k_{A|A}^{(i)}\right]}{kN} \\
&+\sum_{\sum_{j=1}^n \left(k_{A|A}^{(j)}+k_{B|A}^{(j)}\right)=k}p_A\mathbb{P}\left(\Delta p_{AA}^{(i)}=\frac{-2k_{A|A}^{(i)}}{kN}\right)\frac{-2k_{A|A}^{(i)}}{kN} \\
&+\sum_{\sum_{j=1}^n \left(k_{A|A}^{(j)}+k_{B|A}^{(j)}\right)=k}p_A\mathbb{P}\left(\Delta p_{AA}^{(i)}=\frac{2(k-1)\left[\sum_{j=1}^n p_{ji}^{(2)}q_{A|A}^{(j)}-\sum_{j=1}^np_{ij}^{(2)}q_{A|A}^{(i)}\right]\sum_{j=1}^nk_{A|A}^{(j)}}{kN}\right)  \\
&\hspace{3.5cm} \frac{2(k-1)\left[\sum_{j=1}^n p_{ji}^{(2)}q_{A|A}^{(j)}-\sum_{j=1}^np_{ij}^{(2)}q_{A|A}^{(i)}\right]\sum_{j=1}^nk_{A|A}^{(j)}}{kN} \\
&=\frac{2}{kN}\sum_{j=1}^n\Big\{\left[k^2-(k-1)q_{B|A}\right]p_{ji}^{(2)}-k^2\delta_{ji}\Big\}p_{AA}^{(j)}  \\
& +\frac{2}{kN}\sum_{j=1}^n \left[(k-1)q_{A|B}+1\right]p_{ji}^{(1)}p_{AB}^{(j)}+O\left(\delta\right) ,
\end{split}
\end{equation}
where $\delta_{ij}=1$ if $i=j$ and $\delta_{ij}=0$ otherwise.

\subsubsection{Change in $p_{AB}^{(i)}$}
When a $B$-player is selected to die and its neighbourhood configuration is the same as that in Section \ref{subsubsection_pA_B}, the change in $p_{AB}^{(i)}$ due to edges between the nearest and next-nearest neighbors is
\begin{equation} \label{pABi_AtakeB_inside}
\mathbb{P}\left(\Delta p_{AB}^{(i)}=\frac{-k_{A|B}^{(i)}+\sum_{j=1}^n p_{ji}^{(0)}k_{B|B}^{(j)}}{kN}\right)
=\mathcal{B}\left(k_{A|B}^{(j)},k_{B|B}^{(j)}|j=1,\dots,n\right)\mathbb{P}\left(A\rightarrow B\right)
\end{equation}
and
\begin{equation} \label{pABi_BtakeB_inside}
\mathbb{P}\left(\Delta p_{AB}^{(i)}=\frac{\sum_{j=1}^n p_{ji}^{(1)}k_{A|B}^{(j)}-\sum_{j=1}^n p_{ij}^{(1)}k_{A|B}^{(i)}}{kN}\right)
=\mathcal{B}\left(k_{A|B}^{(j)},k_{B|B}^{(j)}|j=1,\dots,n\right)\mathbb{P}\left(B\rightarrow B\right).
\end{equation}
Eq. \ref{pABi_AtakeB_inside} (resp. Eq. \ref{pABi_BtakeB_inside}) captures the case when a neighboring $A$-player (resp. $B$-player) successfully occupies the vacant site.

The change in $p_{AB}^{(i)}$ due to edges between the nearest and next-nearest neighbors is
\begin{equation} \label{pABi_AtakeB_outside}
\begin{split}
&\mathbb{P}\Bigg(\Delta p_{AB}^{(i)}=\frac{
(k-1)\left[\sum_{j=1}^n p_{ji}^{(1)}q_{B|A}^{(j)}-\sum_{j=1}^np_{ij}^{(1)}q_{B|A}^{(i)}\right]\sum_{j=1}^nk_{A|B}^{(j)}}{kN}  \\
&\hspace{2.5cm} +\frac{(k-1)\left[\sum_{j=1}^n p_{ji}^{(1)}q_{A|B}^{(j)}-\sum_{j=1}^np_{ij}^{(1)}q_{A|B}^{(i)}\right]\sum_{j=1}^nk_{B|B}^{(j)}}{kN}
\Bigg)  \\
&=\mathcal{B}\left(k_{A|B}^{(j)},k_{B|B}^{(j)}|j=1,\dots,n\right)\left[\mathbb{P}\left(A\rightarrow B\right)+\mathbb{P}\left(B\rightarrow B\right)\right].
\end{split}
\end{equation}

When an $A$-player is selected to die and its neighbourhood configuration is the same as that in Section \ref{subsubsection_pA_A}, the change in $p_{AB}^{(i)}$ due to edges between the nearest and next nearest neighbors is
\begin{equation} \label{pABi_AtakeA_inside}
\mathbb{P}\left(\Delta p_{AB}^{(i)}=\frac{\sum_{j=1}^n p_{ji}^{(1)}k_{B|A}^{(j)}-\sum_{j=1}^n p_{ij}^{(1)}k_{B|A}^{(i)}}{kN}\right)
=\mathcal{A}\left(k_{A|A}^{(j)},k_{B|A}^{(j)}|j=1,\dots,n\right)\mathbb{P}\left(A\rightarrow A\right)
\end{equation}
and
\begin{equation} \label{pABi_BtakeA_inside}
\mathbb{P}\left(\Delta p_{AB}^{(i)}=\frac{-k_{B|A}^{(i)}+\sum_{j=1}^n p_{ji}^{(2)}k_{A|A}^{(j)}}{kN}\right)
=\mathcal{A}\left(k_{A|A}^{(j)},k_{B|A}^{(j)}|j=1,\dots,n\right)\mathbb{P}\left(B\rightarrow A\right).
\end{equation}
Eq.~\ref{pABi_AtakeA_inside} (resp. Eq.~\ref{pABi_BtakeA_inside}) captures the case when a neighboring $A$-player (resp. $B$-player) successfully occupies the vacant site.

The change in $p_{AB}^{(i)}$ due to edges between the nearest and next-nearest neighbors is
\begin{equation} \label{pABi_AtakeA_outside}
\begin{split}
&\mathbb{P}\Bigg(\Delta p_{AB}^{(i)}=\frac{
(k-1)\left[\sum_{j=1}^n p_{ji}^{(1)}q_{B|A}^{(j)}-\sum_{j=1}^np_{ij}^{(1)}q_{B|A}^{(i)}\right]\sum_{j=1}^nk_{A|A}^{(j)}}{kN}  \\
&\hspace{2.5cm} +\frac{(k-1)\left[\sum_{j=1}^n p_{ji}^{(1)}q_{A|B}^{(j)}-\sum_{j=1}^np_{ij}^{(1)}q_{A|B}^{(i)}\right]\sum_{j=1}^nk_{B|A}^{(j)}}{kN}
\Bigg)  \\
&=\mathcal{A}\left(k_{A|A}^{(j)},k_{B|A}^{(j)}|j=1,\dots,n\right)\left[\mathbb{P}\left(A\rightarrow A\right)+\mathbb{P}\left(B\rightarrow A\right)\right].
\end{split}
\end{equation}

Analogously, we have
\begin{equation}  \label{pABi_derivative}
\begin{split}
\dot{p}_{AB}^{(i)}=
& \frac{1}{kN}\sum_{j=1}^n (k-1)q_{B|A}p_{ji}^{(2)}p_{AA}^{(j)}   \\
& +\frac{1}{kN}\sum_{j=1}^n \left[(k-1)(q_{A|A}+q_{B|B}+2k)p_{ji}^{(1)}-2k^2\delta_{ji}\right]p_{AB}^{(j)}  \\
& +\frac{1}{kN}\sum_{j=1}^n (k-1)q_{A|B}p_{ji}^{(0)}p_{BB}^{(j)}+O(\delta).
\end{split}
\end{equation}

\subsubsection{Change in $p_{BB}^{(i)}$}
When a $B$-player is selected to die and its neighbourhood configuration is the same as that in Section \ref{subsubsection_pA_B}, the change in $p_{AB}^{(i)}$ due to edges between the nearest and next-nearest neighbors is
\begin{equation} \label{pBBi_AtakeB_inside}
\mathbb{P}\left(\Delta p_{BB}^{(i)}=\frac{-2k_{B|B}^{(i)}}{kN}\right)
=\mathcal{B}\left(k_{A|B}^{(j)},k_{B|B}^{(j)}|j=1,\dots,n\right)\mathbb{P}\left(A\rightarrow B\right)
\end{equation}
and
\begin{equation} \label{pBBi_BtakeB_inside}
\mathbb{P}\left(\Delta p_{BB}^{(i)}=\frac{\sum_{j=1}^n p_{ji}^{(0)}k_{B|B}^{(j)}-\sum_{j=1}^n p_{ij}^{(0)}k_{B|B}^{(i)}}{kN}\right)
=\mathcal{B}\left(k_{A|B}^{(j)},k_{B|B}^{(j)}|j=1,\dots,n\right)\mathbb{P}\left(B\rightarrow B\right).
\end{equation}
Eq.~\ref{pBBi_AtakeB_inside} (resp. Eq.~\ref{pBBi_BtakeB_inside}) captures the case when a neighboring $A$-player (resp. $B$-player) successfully occupies the vacant site.

The change in $p_{BB}^{(i)}$ due to edges between the nearest and next-nearest neighbors is
\begin{equation} \label{pBBi_AtakeB_outside}
\begin{split}
&\mathbb{P}\left(\Delta p_{BB}^{(i)}=\frac{2(k-1)\left[\sum_{j=1}^n p_{ji}^{(0)}q_{B|B}^{(j)}-\sum_{j=1}^np_{ij}^{(0)}q_{B|B}^{(i)}\right]\sum_{j=1}^nk_{B|B}^{(j)}}{kN}\right) \ \\ &=\mathcal{B}\left(k_{A|B}^{(j)},k_{B|B}^{(j)}|j=1,\dots,n\right)\left[\mathbb{P}\left(A\rightarrow B\right)+\mathbb{P}\left(B\rightarrow B\right)\right].
\end{split}
\end{equation}

When an $A$-player is selected to die and its neighbourhood configuration is the same as that in Section \ref{subsubsection_pA_A}, the change in $p_{BB}^{(i)}$ due to edges between the nearest and next nearest neighbors is
\begin{equation} \label{pBBi_AtakeA_inside}
\mathbb{P}\left(\Delta p_{BB}^{(i)}=\frac{2\sum_{j=1}^n p_{ji}^{(1)}k_{B|A}^{(j)}}{kN}\right)
=\mathcal{A}\left(k_{A|A}^{(j)},k_{B|A}^{(j)}|j=1,\dots,n\right)\mathbb{P}\left(B\rightarrow A\right).
\end{equation}
The change in $p_{BB}^{(i)}$ due to edges between the nearest and next nearest neighbors is
\begin{equation} \label{pBBi_AtakeA_outside}
\begin{split}
&\mathbb{P}\left(\Delta p_{AA}^{(i)}=\frac{2(k-1)\left[\sum_{j=1}^n p_{ji}^{(0)}q_{B|B}^{(j)}-\sum_{j=1}^np_{ij}^{(0)}q_{B|B}^{(i)}\right]\sum_{j=1}^nk_{B|A}^{(j)}}{kN}\right) \ \\ &=\mathcal{A}\left(k_{A|A}^{(j)},k_{B|A}^{(j)}|j=1,\dots,n\right)\left[\mathbb{P}\left(A\rightarrow A\right)+\mathbb{P}\left(B\rightarrow A\right)\right].
\end{split}\end{equation}

The derivative of $p_{BB}^{(i)}$ is
\begin{equation} \label{pBBi_derivative}
\dot{p}_{BB}^{(i)}=\frac{2}{kN}\sum_{j=1}^n \left[(k-1)q_{B|A}+1\right]p_{ji}^{(1)}p_{AB}^{(j)}+\frac{2}{kN}\sum_{j=1}^n \Big\{\left[k^2-(k-1)q_{A|B}\right]p_{ji}^{(0)}-k^2\delta_{ji}\Big\}p_{BB}^{(j)}+O\left(\delta\right) .
\end{equation}

\subsubsection{Different time scales}
From Eq.~\ref{pAAi_derivative}, we have
\begin{equation} \label{qAA_derivative_former}
\dot{p}_{AA}=
\sum_{i=1}^n \dot{p}_{AA}^{(i)}
=\frac{2p_A}{kN(1-p_A)}(q_{A|A}-1)\left[(k-1)q_{A|A}-(k-2)p_A-1\right]+O\left(\delta\right)
\end{equation}
and
\begin{equation} \label{qAA_derivative}
\dot{q}_{A|A}=\frac{d}{dt}\left(\frac{p_{AA}}{p_A}\right) =\frac{2}{kN(1-p_A)}(q_{A|A}-1)\left[(k-1)q_{A|A}-(k-2)p_A-1\right]+O\left(\delta\right) .
\end{equation}
When the intensity of selection is weak ($\delta\ll 1$), $q_{A|A}$ reaches its equilibrium much faster than $p_A$ (see Eqs. \ref{pA_derivative_DB},\ref{qAA_derivative}).
Thus, the dynamical system converges quickly onto the slow manifold with $\dot{q}_{A|A}=0$, so we have
\begin{equation}\label{q_AA}
q_{A|A}=\frac{k-2}{k-1}p_A+\frac{1}{k-1}.
\end{equation}
From Eqs.~\ref{notation_3a}-\ref{notaitonf} and \ref{q_AA}, we find that for all $X,Y \in \left\{A,B\right\}$, $p_{XY}$ and $q_{X|Y}$ are a function of $p_A$.

We define a function $\textbf{A}\left(\textbf{R}^{(s)}\right)$ mapping a set of $(n-1)\times(n-1)$ matrix $\textbf{R}^{(s)}$ to a $3(n-1)\times 3(n-1)$ matrix, given by
\begin{equation} \label{A_DB}
\textbf{A}\left(\textbf{R}^{(s)}\right)=
\left[
\begin{array}{ccc}
2(k^2-\alpha)\textbf{R}^{(2)} & 2(\beta+1)\textbf{R}^{(1)} & \textbf{0} \\
\alpha\textbf{R}^{(2)} & (2k^2-k)\textbf{R}^{(1)} & \beta\textbf{R}^{(0)} \\
\textbf{0} & 2(\alpha+1)\textbf{R}^{(1)} & 2(k^2-\beta)\textbf{R}^{(0)}
\end{array}
\right],
\end{equation}
where $\alpha=(k-2)(1-p_A)$ and $\beta=(k-2)p_A$.
Then we use $\textbf{P}^{(s)}$ in Eq.~\ref{game_transition} to define two $(n-1)\times (n-1)$ matrices as follow
\begin{equation} \label{Ps}
\bar{\textbf{P}}^{(s)}
=
\left[
\begin{matrix}
p_{11}^{(s)}-p_{n1}^{(s)} & \cdots & p_{(n-1)1}^{(s)}-p_{n1}^{(s)} \\
\vdots & \ddots & \vdots \\
p_{1(n-1)}^{(s)}-p_{n(n-1)}^{(s)} & \cdots & p_{(n-1)(n-1)}^{(s)}-p_{n(n-1)}^{(s)}
\end{matrix}
\right], \quad
\tilde{\textbf{P}}^{(s)}
=
\left[
\begin{matrix}
p_{n1}^{(s)} & \cdots & 0 \\
\vdots & \ddots & \vdots \\
0 & \cdots & p_{n(n-1)}^{(s)}
\end{matrix}
\right],
\end{equation}
where $s\in\left\{0,1,2\right\}$.
Let $\textbf{b}$ denote a column vector with $3(n-1)$ entries: the first $n-1$ entries are $p_{AA}$; the next $n-1$ entries are
$p_{AB}$; the last $n-1$ entries are $p_{BB}$.
Let $\textbf{v}$ denote a column vector $\left(p_{AA}^{(1)},\dots,p_{AA}^{(n-1)},p_{AB}^{(1)},\dots,p_{AB}^{(n-1)},p_{BB}^{(1)},\dots,p_{BB}^{(n-1)}\right)^{T}$.
Combining Eq.~\ref{q_AA} and $p_{XY}^{(n)}=p_{XY}-\sum_{i=1}^{n-1}p_{XY}^{(i)}$, we can reduce the system of Eqs.~\ref{pAAi_derivative},\ref{pABi_derivative},\ref{pBBi_derivative} to
\begin{equation} \label{system_matrix}
\dot{\textbf{v}}=\frac{1}{kN}\left[\textbf{A}\left(\bar{\textbf{P}}^{(s)}\right)-2k^2\textbf{I}\right]\textbf{v}+\frac{1}{kN}\textbf{A}\left(\tilde{\textbf{P}}^{(s)}\right)\textbf{b}
\equiv \bar{\textbf{A}}\textbf{v}+\bar{\textbf{b}} .
\end{equation}

For a linear system described by Eq.~\ref{system_matrix}, its equilibrium points can be obtained by solving the equation $\left[\textbf{A}\left(\bar{\textbf{P}}^{(s)}\right)-2k^2\textbf{I}\right]\textbf{v}+\textbf{A}\left(\tilde{\textbf{P}}^{(s)}\right)\textbf{b}=\textbf{0}$.
If for $0<p_A<1$, all eigenvalues of $\bar{\textbf{A}}$ are negative numbers or complex numbers with negative real parts, the system is asymptotically stable and has a single equilibrium point given by \cite{si2001-Khalil}
\begin{equation} \label{root_v_DB}
\textbf{v} = -\left[\textbf{A}\left(\bar{\textbf{P}}^{(s)}\right)-2k^2\textbf{I}\right]^{-1}\textbf{A}\left(\tilde{\textbf{P}}^{(s)}\right)\textbf{b} .
\end{equation}
Regardless of the initial state of $\left(p_{AA}^{(1)},\dots,p_{AA}^{(n-1)},p_{AB}^{(1)},\dots,p_{AB}^{(n-1)},p_{BB}^{(1)},\dots,p_{BB}^{(n-1)}\right)^{T}$, the system ultimately approaches to the equilibrium point.
In other words, the initial fractions of various games do not affect the evolutionary outcome.
We state that none of $\bar{\textbf{A}}$'s eigenvalues can be positive, since this leads to a few terms in $\textbf{v}$ increasing above $1$ or decreasing below $0$ \cite{si2001-Khalil}, which is unrealistic in the current system.
But $\bar{\textbf{A}}$ may have zero eigenvalues.
In such cases, the system described by Eq.~\ref{system_matrix} has more than one equilibrium point.
The initial state of $\left(p_{AA}^{(1)},\dots,p_{AA}^{(n-1)},p_{AB}^{(1)},\dots,p_{AB}^{(n-1)},p_{BB}^{(1)},\dots,p_{BB}^{(n-1)}\right)^{T}$ determines the equilibrium point that the system approaches.
That is, the initial fractions of various games influence the evolutionary outcome.
In Section 2, we provide an approach to efficiently evaluate the dependence of the evolutionary outcome to the initial fractions of various games.

\subsubsection{Diffusion approximation}
For given game transition matrices and the initial fractions of various games, by solving Eq.~\ref{system_matrix}, we obtain $p_{AA}^{(i)}$, $p_{AB}^{(i)}$, and $p_{BB}^{(i)}$ as functions of $p_A$.
Substituting $p_{AA}^{(i)}$, $p_{AB}^{(i)}$, and $p_{BB}^{(i)}$ into Eqs.~\ref{I_ai}-\ref{I_di} and combining with Eq.~\ref{q_AA}, we have
\begin{subequations}
\begin{equation}
I_{R_i} = (k-2)(k+1)(1-p_A)v_i \quad (\equiv I_{R_i}(p_A)) ; \label{Iai}\\
\end{equation}
\begin{equation}
I_{S_i} = \left[-(k-2)(k+1)p_A+k^2-k-1\right]v_{n+i-1} \quad (\equiv I_{S_i}(p_A)) ; \label{Ibi}\\
\end{equation}
\begin{equation}
I_{T_i} = -\left[(k-2)(k+1)p_A+1\right]v_{n+i-1} \quad (\equiv I_{T_i}(p_A)) ; \label{Ici}\\
\end{equation}
\begin{equation}
I_{P_i} = -(k-2)(k+1)p_Av_{2n+i-2} \quad (\equiv I_{P_i}(p_A)) \label{Idi}
\end{equation}
\end{subequations}
for $1\leqslant i\leqslant n-1$. We obtain $I_{R_n}$, $I_{S_n}$, $I_{T_n}$ and $I_{P_n}$ by separately replacing $v_i$ in Eq.~\ref{Iai} with $\left(p_{AA}-\sum_{i=1}^{n-1}v_i\right)$, $v_{n+i-1}$ in Eqs.~\ref{Ibi} and \ref{Ici} with $\left(p_{AB}-\sum_{i=1}^{n-1}v_{n+i-1}\right)$, and $v_{2n+i-2}$ in Eq.~\ref{Idi} with $\left(p_{BB}-\sum_{i=1}^{n-1}v_{2n+i-2}\right)$.

We consider a one-dimensional diffusion process of the random variable $p_A$.
Within a short time interval $\Delta t$, we have
\begin{subequations}
\begin{equation}
\begin{split}
\mathbb{E}\left[\Delta p_A\right]
&= \frac{1}{N}\mathbb{P}\left(\Delta p_A=\frac{1}{N}\right)+\left(-\frac{1}{N}\right)\mathbb{P}\left(\Delta p_A=-\frac{1}{N}\right) \\\
&= \delta\frac{1}{kN} \sum_{i=1}^{n}\left( I_{R_i}R_i+I_{S_i}S_i+I_{T_i}T_i+I_{P_i}P_i \right)\Delta t \equiv \bar{E}(p_A)\Delta t ;
\end{split}\end{equation}
\begin{equation}
\begin{split}
\text{Var}\left[\Delta p_A\right] &= \left(\frac{1}{N}\right)^2\mathbb{P}\left(\Delta p_A=\frac{1}{N}\right)+\left(-\frac{1}{N}\right)^2\mathbb{P}\left(\Delta p_A=-\frac{1}{N}\right) \\\
&=\frac{2(k-2)}{N^2(k-1)}p_A(1-p_A)\Delta t \equiv \bar{V}(p_A)\Delta t.
\end{split}
\end{equation}
\end{subequations}

The fixation probability $\phi_A(x)$ of $A$-players with initial frequency $p_A(t=0)=x$, satisfies the following differential equation [see Eq. (5.2.186) in Ref~\cite{si2004-Gardiner} and detailed derivation]:
\begin{equation} \label{diffusion_Eq.}
0=\bar{E}(x)\frac{d\phi_A(x)}{dx}+\frac{\bar{V}(x)}{2}\frac{d^2\phi_A(x)}{dx^2}.
\end{equation}

The solution to Eq.~\ref{diffusion_Eq.} is [see Eq (5.2.189) in Ref~\cite{si2004-Gardiner}]
\begin{equation}
\phi_A(x)=\frac{\int_{0}^{x}G(y)dy}{\int_{0}^1G(y)dy},
\end{equation}
where
\begin{equation} \label{Gy}
\begin{split}
G(y)&=\exp\left(-\int \frac{2\bar{E}(y)}{\bar{V}(y)}dy\right) \ \\
&= \exp\left(-\int \delta\frac{N(k-1)}{k(k-2)}\sum_{i=1}^n\left(\frac{I_{R_i}(y)}{y(1-y)}R_i+\frac{I_{S_i}(y)}{y(1-y)}S_i+
\frac{I_{T_i}(y)}{y(1-y)}T_i+\frac{I_{P_i}(y)}{y(1-y)}P_i\right)dy\right) \ \\
&= 1-\delta\frac{N(k-1)}{k(k-2)}\sum_{i=1}^n\int\left(R_i\frac{I_{R_i}(y)}{y(1-y)}+S_i\frac{I_{S_i}(y)}{y(1-y)}
+T_i\frac{I_{T_i}(y)}{y(1-y)}+P_i\frac{I_{P_i}(y)}{y(1-y)}\right)dy+O\left(\delta^{2}\right) .
\end{split}
\end{equation}
In Eq.~\ref{Gy}, the third equality holds when $\delta$ is sufficiently small.

\subsubsection{Fixation probability}
In a population of $B$-players, when a fraction $x$ of $B$-players mutates to $A$-players, the fixation probability of these $A$-players is
\begin{equation}
\begin{split}
\phi_A(x)=x+
&\delta\frac{N(k-1)}{k(k-2)}\sum_{i=1}^n\Bigg\{ \ \\
&x\int_{0}^{1}\left[\int\left(R_i\frac{I_{R_i}(y)}{y(1-y)}+S_i\frac{I_{S_i}(y)}{y(1-y)}
+T_i\frac{I_{T_i}(y)}{y(1-y)}+P_i\frac{I_{P_i}(y)}{y(1-y)}\right)dy\right]dy \ \\
&-\int_{0}^{x}\left[\int\left(R_i\frac{I_{R_i}(y)}{y(1-y)}+S_i\frac{I_{S_i}(y)}{y(1-y)}
+T_i\frac{I_{T_i}(y)}{y(1-y)}+P_i\frac{I_{P_i}(y)}{y(1-y)}\right)dy\right]dy \Bigg\}+O\left(\delta^{2}\right) .
\end{split}
\end{equation}
The fixation probability of a fraction $x$ of $B$-players is
\begin{equation}
\phi_B(x)=1-\phi_A(1-x).
\end{equation}
Then the ratio of fixation probabilities is
\begin{equation}
\begin{split}
\frac{\phi_A(x)}{\phi_B(x)}=1+
&\delta\frac{N(k-1)}{k(k-2)x}\sum_{i=1}^n\Big\{ \ \\
&\int_{1-x}^{1}\left[\int\left(R_i\frac{I_{R_i}(y)}{y(1-y)}+S_i\frac{I_{S_i}(y)}{y(1-y)}
+T_i\frac{I_{T_i}(y)}{y(1-y)}+P_i\frac{I_{P_i}(y)}{y(1-y)}\right)dy\right]dy \ \\
&-\int_{0}^{x}\left[\int\left(R_i\frac{I_{R_i}(y)}{y(1-y)}+S_i\frac{I_{S_i}(y)}{y(1-y)}
+T_i\frac{I_{T_i}(y)}{y(1-y)}+P_i\frac{I_{P_i}(y)}{y(1-y)}\right)dy\right]dy \Big\}+O\left(\delta^{2}\right) .
\end{split}
\end{equation}
For sufficiently small $x$, we have
\begin{equation}
\begin{split}
\frac{\phi_A(x)}{\phi_B(x)}&=1+\delta\frac{N(k-1)}{k(k-2)x}\sum_{i=1}^n\Bigg\{ \ \\
& \quad\quad x\left[\int\left(R_i\frac{I_{R_i}(y)}{y(1-y)}+S_i\frac{I_{S_i}(y)}{y(1-y)}
+T_i\frac{I_{T_i}(y)}{y(1-y)}+P_i\frac{I_{P_i}(y)}{y(1-y)}\right)dy\right]_{y=1} \ \\
& \quad\quad  -x\left[\int\left(R_i\frac{I_{R_i}(y)}{y(1-y)}+S_i\frac{I_{S_i}(y)}{y(1-y)}
+T_i\frac{I_{T_i}(y)}{y(1-y)}+P_i\frac{I_{P_i}(y)}{y(1-y)}\right)dy\right]_{y=0} \Bigg\}+O\left(\delta^{2}\right) \ \\
&=1+\delta\frac{N(k-1)}{k(k-2)}\sum_{i=1}^n\Big[R_i\int_{0}^{1}\frac{I_{R_i}(y)}{y(1-y)}dy+S_i\int_{0}^{1}\frac{I_{S_i}(y)}{y(1-y)}dy \ \\
&\hspace{4cm}+T_i\int_{0}^{1}\frac{I_{T_i}(y)}{y(1-y)}dy+P_i\int_{0}^{1}\frac{I_{P_i}(y)}{y(1-y)}dy\Big]+O\left(\delta^{2}\right) .
\end{split}
\end{equation}

Overall, for a sufficiently large population and $x=1/N$, the condition of $A$-players being favored over $B$-players ($\rho_A>\rho_B$) is
\begin{equation} \label{general_rule_DB}
\sum_{i=1}^n\left[R_i\int_{0}^{1}\frac{I_{R_i}(y)}{y(1-y)}dy+S_i\int_{0}^{1}\frac{I_{S_i}(y)}{y(1-y)}dy
+T_i\int_{0}^{1}\frac{I_{T_i}(y)}{y(1-y)}dy+P_i\int_{0}^{1}\frac{I_{P_i}(y)}{y(1-y)}dy\right]>0.
\end{equation}
Eq.~\ref{general_rule_DB} holds for not only death-birth updating, but also for imitation (see Section 1\ref{section_imitation} for details) and pairwise-comparison (see Section 1\ref{section_pairwise_comparison} for details) updating.
Note that for different updating rules, $I_{R_i}$, $I_{S_i}$, $I_{T_i}$, $I_{P_i}$ differ.

\subsubsection{The rule $b/c>k-k'$}
We now turn to donation games.
The payoff structure for game $i$ is
\begin{equation}\label{donation_games}
\bordermatrix{
& A & B \cr
A & b_i-c & -c \cr
B & b_i & 0 \cr
}.
\end{equation}
Substituting payoff structures into Eq.~\ref{general_rule_DB} and using Eqs.~\ref{I_ai}-\ref{I_di}, we obtain the condition for $\rho_A>\rho_B$ under death-birth updating, given by
\begin{equation} \label{db_equality}
\sum_{i=1}^n \alpha_ib_i+\alpha_c c >0,
\end{equation}
where
\begin{equation} \label{rule_donation_DB}
\alpha_i = \int_{0}^{1}\frac{I_{R_i}(y)+I_{T_i}(y)}{y(1-y)}dy, \quad\quad \alpha_c = -\frac{k^2(k-2)}{k-1}.
\end{equation}
Furthermore, we have
\begin{equation}
\sum_{i=1}^n \alpha_i=\frac{k(k-2)}{k-1}\left(=-\frac{\alpha_c}{k}\right).
\end{equation}
Using $\alpha_1=-\alpha_c/k-\sum_{i=2}^n\alpha_i$ and denoting $(b_1-b_i)/c$ by $\Delta b_{1i}/c$, we can rewrite Eq.~\ref{db_equality} as
\begin{equation} \label{DB_bckxi}
\frac{b_1}{c}>k-\sum_{i=2}^n \xi_i \frac{\Delta b_{1i}}{c} ,
\end{equation}
where
\begin{equation} \label{xi_DB}
\xi_i = -\frac{k-1}{k(k-2)}\int_{0}^{1}\frac{I_{R_i}(y)+I_{T_i}(y)}{y(1-y)}dy.
\end{equation}
Inserting Eqs.~\ref{Iai} and \ref{Ici} into Eq.~\ref{xi_DB}, we get the formula of $\xi_i$ for death-birth updating.
Letting $k'= \sum_{i=2}^n \xi_i \Delta b_{1i}/c$ and $b = b_1$, we obtain the rule $b/c>k-k'$.

\subsection{Imitation updating} \label{section_imitation}
In each time step, a random player $i$ is selected to evaluate its strategy.
This player retains its own strategy or imitates a neighbor's strategy with probability proportional to fitness.
Analyzing the evolutionary process as we do under death-birth updating, we have
\begin{equation} \label{pA_derivative}
\dot{p}_A = \delta\frac{k}{(k+1)^2N} \sum_{i=1}^{n}\left(I_{R_i}R_i+I_{S_i}S_i+I_{T_i}T_i+I_{P_i}P_i \right)+O\left(\delta^{2}\right) ,
\end{equation}
where
\begin{subequations}
\begin{equation}
I_{R_i} = p_{AA}^{(i)}(k-1)q_{B|A}\left[(k-1)(q_{A|A}+q_{B|B})+3\right] ; \\
\end{equation}
\begin{equation}
I_{S_i} = p_{AB}^{(i)}\left\{(k-1)q_{B|A}\left[(k-1)(q_{A|A}+q_{B|B})+2\right]+(k-1)q_{B|B}+2\right\} ; \\
\end{equation}
\begin{equation}
I_{T_i} = -p_{AB}^{(i)}\left\{(k-1)q_{A|B}\left[(k-1)(q_{A|A}+q_{B|B})+2\right]+(k-1)q_{A|A}+2\right\} ; \\
\end{equation}
\begin{equation}
I_{P_i} = -p_{BB}^{(i)}(k-1)q_{A|B}\left[(k-1)(q_{A|A}+q_{B|B})+3\right] .
\end{equation}
\end{subequations}
We redefine the function $\textbf{A}\left(\textbf{R}^{(s)}\right)$ to be
\begin{equation}
\textbf{A}\left(\textbf{R}^{(s)}\right)=
\left[
\begin{array}{ccc}
2(k^2+k-\alpha)\textbf{R}^{(2)} & 2(\beta+1)\textbf{R}^{(1)} & \textbf{0} \\
\alpha\textbf{R}^{(2)} & (2k^2+k)\textbf{R}^{(1)} & \beta\textbf{R}^{(0)} \\
\textbf{0} & 2(\alpha+1)\textbf{R}^{(1)} & 2(k^2+k-\beta)\textbf{R}^{(0)}
\end{array}
\right].
\end{equation}
Then the system under imitation updating can be reduced to
\begin{equation} \label{system_matrix_IM}
\dot{\textbf{v}}=\frac{1}{kN}\left[\textbf{A}\left(\bar{\textbf{P}}^{(s)}\right)-2k(k+1)\textbf{I}\right]\textbf{v}+\frac{1}{kN}\textbf{A}\left(\tilde{\textbf{P}}^{(s)}\right)\textbf{b}.
\end{equation}
All other variables such as $\alpha,\beta,\bar{\textbf{P}}^{(s)},\tilde{\textbf{P}}^{(s)},\textbf{b},\textbf{v}$ follow those defined for death-birth updating.

For donation games described by Eq.~\ref{donation_games}, we have the condition for $\rho_A>\rho_B$,
\begin{equation} \label{IM_bckxi}
\frac{b_1}{c}>k+2-\sum_{i=2}^n \xi_i\frac{\Delta b_{1i}}{c},
\end{equation}
where
\begin{equation} \label{xi_IM}
\xi_i= -\frac{(k-1)}{k(k-2)}\int_{0}^{1}\frac{I_{R_i}(y)+I_{T_i}(y)}{y(1-y)}dy.
\end{equation}
Solving Eq.~\ref{system_matrix_IM} and inserting $I_{R_i} ,I_{T_i}$ in Eq.~\ref{xi_IM}, we get the expression for $\xi_i$.

\subsection{Pairwise-comparison updating} \label{section_pairwise_comparison}
In each generation, a random player $i$ is selected to evaluate its strategy.
This player randomly selects a neighbor $j$ and compares payoffs.
Player $i$ then adopts $j$'s strategy with probability
\begin{equation}
\frac{1}{1+e^{-\delta(\pi_j-\pi_i)}} ,
\end{equation}
where $\pi_i$ and $\pi_j$ denote the payoffs of $i$ and $j$, respectively. Otherwise, player $i$ retains its strategy.

Analogously, we have
\begin{equation}
\dot{p}_A
=\delta\frac{1}{2N}\sum_{i=1}^{n}\left(I_{R_i}R_i+I_{S_i}S_i+I_{T_i}T_i+I_{P_i}P_i \right)+O\left(\delta^{2}\right) ,
\end{equation}
where
\begin{subequations}
\begin{equation}
I_{R_i} = p_{AA}^{(i)}(k-1)q_{B|A} ; \\
\end{equation}
\begin{equation}
I_{S_i} = p_{AB}^{(i)}\left[(k-1)q_{B|A}+1\right] ; \\
\end{equation}
\begin{equation}
I_{T_i} = -p_{AB}^{(i)}\left[(k-1)q_{A|B}+1\right] ; \\
\end{equation}
\begin{equation}
I_{P_i} = -p_{BB}^{(i)}(k-1)p_{A|B} .
\end{equation}
\end{subequations}
We redefine the function $\textbf{A}\left(\textbf{R}^{(s)}\right)$ to be
\begin{equation}
\textbf{A}\left(\textbf{R}^{(s)}\right) =
\left[
\begin{array}{ccc}
2(4k-2-\alpha)\textbf{R}^{(2)} & 2(\beta+1)\textbf{R}^{(1)} & \textbf{0} \\
\alpha\textbf{R}^{(2)} & (7k-4)\textbf{R}^{(1)} & \beta\textbf{R}^{(0)} \\
\textbf{0} & 2(\alpha+1)\textbf{R}^{(1)} & 2(4k-2-\beta)\textbf{R}^{(0)}
\end{array}
\right].
\end{equation}
Then the system under pairwise-comparison updating can be reduced to
\begin{equation} \label{system_matrix_PC}
\dot{\textbf{v}}=\frac{1}{kN}\left[\textbf{A}\left(\bar{\textbf{P}}^{(s)}\right)-(8k-4)\textbf{I}\right]\textbf{v}+\frac{1}{kN}\textbf{A}\left(\tilde{\textbf{P}}^{(s)}\right)\textbf{b}.
\end{equation}
All other variables such as $\alpha,\beta,\bar{\textbf{P}}^{(s)},\tilde{\textbf{P}}^{(s)},\textbf{b},\textbf{v}$ follow those defined for death-birth updating.

For donation games described by Eq.~\ref{donation_games}, we have the condition for $\rho_A>\rho_B$,
\begin{equation} \label{PC_bckxi}
\sum_{i=2}^n \xi_i \frac{\Delta b_{1i}}{c}>1,
\end{equation}
where
\begin{equation} \label{xi_PC}
\xi_i = -\frac{(k-1)}{k(k-2)}\int_{0}^{1}\frac{I_{R_i}(y)+I_{T_i}(y)}{y(1-y)}dy .
\end{equation}
By solving Eq.~\ref{system_matrix_PC} and inserting $I_{R_i} ,I_{T_i}$ in Eq.~\ref{xi_PC}, we get the expression for $\xi_i$.

\section{Approach to evaluate the sensitivity of evolutionary dynamics to the initial condition} \label{section_sensitivity}
Here, we consider the initial condition, which refers to the initial fractions of various games played in the population.
By calculating the eigenvalues of matrix $\left[\textbf{A}\left(\bar{\textbf{P}}^{(s)}\right)-2k^2\textbf{I}\right]/(kN)$ in Eq.~\ref{system_matrix} and evaluating the sign of all eigenvalues, we can tell whether or not under a given game transition pattern the evolutionary outcome is sensitive to the initial condition under death-birth updating for local game transitions.
Analogously, we can study the matrix $\left[\textbf{A}\left(\bar{\textbf{P}}^{(s)}\right)-2k(k+1)\textbf{I}\right]/(kN)$ in Eq.~\ref{system_matrix_IM} under imitation updating and the matrix $\left[\textbf{A}\left(\bar{\textbf{P}}^{(s)}\right)-(8k-4)\textbf{I}\right]/(kN)$ in Eq.~\ref{system_matrix_PC} under pairwise-comparison updating.

In this section, we provide an alternative approach to determine the dependence of the evolutionary outcome on the initial conditions. Based on the game transition matrix $\textbf{P}^{(2)}$, $\textbf{P}^{(1)}$, and $\textbf{P}^{(0)}$ in Eq.~\ref{game_transition}, we define a Markov chain with a state space $\textbf{E}=\left\{1,2,\dots,3n\right\}$.
The probability transition matrix for this Markov chain is given by
\begin{equation} \label{M_proposition}
\textbf{M}=
\left[
\begin{array}{ccc}
\textbf{P}^{(2)}/2 & \textbf{P}^{(2)}/2 & \textbf{0} \\
\textbf{P}^{(1)}/3 & \textbf{P}^{(1)}/3 & \textbf{P}^{(1)}/3 \\
\textbf{0} & \textbf{P}^{(0)}/2 & \textbf{P}^{(0)}/2
\end{array}
\right].
\end{equation}
The entry in the $i$th row and the $j$th column of $\textbf{M}$ is the transition probability from state $i$ to state $j$.
If the defined random process has only one closed communicating class, the evolutionary outcome is independent of the initial condition, regardless of the update rule. However, if it has more than one such class, the evolutionary outcome is sensitive to the initial condition.

For a random process defined by a state space $\textbf{E}$ and a probability transition matrix $\textbf{M}$, we can examine its communicating class structure as follows: letting $\bar{\textbf{M}}=\sum_{i=1}^{3n}\textbf{M}^i$, the above random process has only one closed communicating class if and only if in $\bar{\textbf{M}}$ there exists at least some $i$ ($1\leqslant i\leqslant 3n$) such that all entries in the $i$th column are positive. The random process has more than one closed communicating classes if and only if in $\bar{\textbf{M}}$ for every $i$ ($1\leqslant i\leqslant 3n$) there exists at least one entry of $0$ in the $i$th column.
~\\

The sign of each entry in $\bar{\textbf{M}}$, taking $\bar{M}_{ij}$ (the entry in the $i$th row and the $j$th column in $\bar{\textbf{M}}$) for example, actually indicates the transition possibility (not probability) from state $i$ to state $j$ within $3n$-step transitions (less than or equal to $3n$ steps).
$\bar{M}_{ij}>0$ means that the system can transition from state $i$ to $j$ in at most $3n$ steps.
For $\bar{M}_{ij}=0$, the transition is unlikely to happen within $3n$ steps, which indicates that the system entering into state $i$ can never transition to state $j$.
If there exists some $j$ ($1\leqslant j\leqslant 3n$) such that all entries in the $j$th column of $\bar{\textbf{M}}$ are positive, any state can transition to state $j$ and thus state $j$ lies in a closed communicating class.
In such a situation, if there is another closed communicating class, any state lying in the second class is unlikely to transition to state $j$, which leads to a contradiction.
Therefore, a column of positive entries suggests a single closed communicating class.
Similarly, if there is only one closed communicating class, any state can transition to one state of the closed communicating class.
Thus, there must exist a column of positive entries.
Analogously, we can prove that the absence of a column of positive entries implies the existence of more than one closed communicating class.
We provide examples with two states for a better understanding of this approach.
The game transition matrices are
\begin{equation} \label{pattern_state2}
\textbf{P}^{(2)}
=
\left[
\begin{matrix}
1 & 0 \\
0 & 1
\end{matrix}
\right],
\quad
\textbf{P}^{(1)}
=
\left[
\begin{matrix}
1 & 0 \\
0 & 1
\end{matrix}
\right],
\quad
\textbf{P}^{(0)}
=
\left[
\begin{matrix}
1 & 0 \\
0 & 1
\end{matrix}
\right].
\end{equation}
We have
\begin{equation}
\textbf{M}=
\left[
\begin{array}{cccccc}
1/2 & 0 & 1/2 & 0 & 0 & 0   \\
0 & 1/2 & 0 & 1/2 & 0 & 0  \\
1/3 & 0 & 1/3 & 0 & 1/3 & 0   \\
0 & 1/3 & 0 & 1/3 & 0 & 1/3   \\
0 & 0 & 1/2 & 0 & 1/2 & 0 \\
0 & 0 & 0 & 1/2 & 0 & 1/2
\end{array}
\right],
\end{equation}
which gives
\begin{equation}
\bar{\textbf{M}}=
\left[
\begin{array}{cccccc}
631/290 & 0       & 129/49 & 0      & 56/47   & 0   \\
0       & 631/290 & 0      & 129/49 & 0       & 56/47  \\
86/49   & 0       & 122/49 & 0      & 86/49   & 0   \\
0       & 86/49   & 0      & 122/49 & 0       & 86/49   \\
56/47   & 0       & 129/49 & 0      & 631/290 & 0 \\
0       & 56/47   & 0      & 122/49 & 0       & 631/290
\end{array}
\right].
\end{equation}
There exist entries of $0$ in every column of $\bar{\textbf{M}}$.
Thus, there is more than one closed communicating class, and the initial fractions of various games affect the evolutionary outcomes.
As a consistency check, we calculate the eigenvalues of $\left[\textbf{A}\left(\bar{\textbf{P}}^{(s)}\right)-2k^2\textbf{I}\right]/(kN)$ in Eq.~\ref{system_matrix}, which are given by $\lambda_1=0$, $\lambda_2=-(k-2)/(Nk)$, and $\lambda_3=-(2k-2)/(Nk)$.
The eigenvalue $\lambda_1=0$ confirms the sensitivity of the evolutionary outcome to the initial conditions.

Actually, Eq.~\ref{pattern_state2} intuitively shows that the game remains fixed throughout the evolutionary process.
According to prior studies about edge-dependent games, the evolution proceeds ``as if'' all interactions are governed by an ``effective'' game \cite{si2015-McAvoy-p1004349-1004349,si2019-Su-p1006947-1006947}.
This ``effective'' game corresponds to the averaged of games played in all interactions, which suggests that its payoff structure depends on the fractions of various games.
Thus the evolutionary outcome is sensitive to the initial condition, in line with the analysis based on the above approach.
In Section~4\ref{section_example2}, we illustrate how to calculate $\xi_i$ for such a system.

Then, we present an example with game transition matrices
\begin{equation} \label{fig2_matrix}
\textbf{P}^{(2)}
=
\left[
\begin{matrix}
1 & 0 \\
1 & 0
\end{matrix}
\right],
\quad
\textbf{P}^{(1)}
=
\left[
\begin{matrix}
0 & 1 \\
0 & 1
\end{matrix}
\right],
\quad
\textbf{P}^{(0)}
=
\left[
\begin{matrix}
0 & 1 \\
0 & 1
\end{matrix}
\right].
\end{equation}
Actually, this case corresponds to the game transition pattern used in Fig. 1 in the main text: both taking strategy $A$ leads to game 1 and other strategy profiles lead to game 2.
We have
\begin{equation}
\textbf{M}=
\left[
\begin{array}{cccccc}
1/2 & 0 & 1/2 & 0 & 0 & 0   \\
1/2 & 0 & 1/2 & 0 & 0 & 0  \\
0 & 1/3 & 0 & 1/3 & 0 & 1/3   \\
0 & 1/3 & 0 & 1/3 & 0 & 1/3   \\
0 & 0 & 0 & 1/2 & 0 & 1/2 \\
0 & 0 & 0 & 1/2 & 0 & 1/2
\end{array}
\right],
\end{equation}
which gives
\begin{equation}
\bar{\textbf{M}}=
\left[
\begin{array}{cccccc}
49/34  &    36/49  &      49/34  &    56/47  &       0   &        56/47   \\
49/34  &    36/49  &      49/34  &    56/47  &       0   &        56/47   \\
36/49  &      50/49  &      36/49  &      86/49  &        0   &        86/49   \\
36/49  &      50/49  &      36/49  &      86/49  &        0   &        86/49   \\
37/81  &      36/49  &      37/81  &     631/290  &      0   &       631/290   \\
37/81  &      36/49  &      37/81  &     631/290  &      0   &       631/290
\end{array}
\right].
\end{equation}
Except for entries in the fifth column, all other entries in $\bar{\textbf{M}}$ are positive.
There is only one closed communicating class.
The evolutionary outcome therefore is insensitive to the initial condition.
As a consistency check, we calculate the eigenvalues of $\left[\textbf{A}\left(\bar{\textbf{P}}^{(s)}\right)-2k^2\textbf{I}\right]/(kN)$ in Eq.~\ref{system_matrix}, which are given by $\lambda_1=-2k/N$, $\lambda_2=-2k/N$, and $\lambda_3=-2k/N$.
The system has a unique equilibrium point and the evolutionary outcome is independent of the initial condition.
In Section 4\ref{section_example2},
we illustrate how to calculate $\xi_i$ for this system.
~\\

We now briefly explain why the closed communicating class of this random process can predict the sensitivity of the evolutionary outcome to the initial condition.
The main idea is whether or not the initially-assigned game between two connected players constrains the game they play in the long-term evolutionary process.
For example, the transition pattern we illustrate in Eq.~\ref{pattern_state2} describes that if two individuals initially play game 1, regardless of their strategic actions, they will play game 1 throughout the process.
Obviously, the initial game decides the game they play later.
Therefore, the initial condition affects the evolutionary outcome.

We rename the states in $\textbf{E}$ as $\left\{E_{AA}^{(1)},\dots,E_{AA}^{(n)},E_{AB}^{(1)},\dots,E_{AB}^{(n)},E_{BB}^{(1)},\dots,E_{BB}^{(n)}\right\}$, where the $i_{th}$ entry corresponds to the original state $i$.
In the following, we show that $\textbf{M}$ actually captures the state transition of an edge throughout the process.
As defined in Section 1,
the state of an edge is given by $E_{XY}^{(i)}$, where $X,Y\in \left\{A,B\right\}$ and $i\in \left\{1,2,\dots,n\right\}$.
$X$ and $Y$ are strategies of the two connected players and $i$ is the game they play.
The transition of an edge state can arise from two parts: the change in players' strategies and the change in the game they play.
The illustrated below is the transition of an edge in one time step:
\begin{equation}
\begin{matrix}
E_{AA}^{(i)} & \xrightleftharpoons[p_{ji}^{(1)}]{p_{ij}^{(2)}} & E_{AB}^{(j)} & \xrightleftharpoons[p_{lj}^{(0)}]{p_{jl}^{(1)}} & E_{BB}^{(l)}\\
%   & & & & \\
p_{ii_1}^{(2)}\downharpoonleft\upharpoonright p_{i_1i}^{(2)} &  & p_{jj_1}^{(1)}\downharpoonleft\upharpoonright p_{j_1j}^{(1)} & & p_{ll_1}^{(0)}\downharpoonleft\upharpoonright p_{l_1l}^{(0)}\\
& & & & \\
E_{AA}^{(i_1)} &  & E_{AB}^{(j_1)} & & E_{BB}^{(l_1)}
\end{matrix}
\end{equation}
Let $AA$ denote both players taking $A$-strategies, $AB$ denote one player taking $A-$strategy and the other taking $B$-strategy, and $BB$ denote both players taking $B-$strategies.
Note that in the current model, in each generation, only a player has the opportunity to update its strategy.
Thus, the strategy transition follows
(i) $AA$ can remain in $AA$ or transition to $AB$ but can not transition to $BB$;
(ii) $AB$ can remain in $AB$ or transition to $AA$ or $BB$;
(iii) $BB$ can remain in $BB$ or transition to $AB$ but can not transition to $AA$.
That is, for any $i$ and $l$, $E_{AA}^{(i)}$ and $E_{BB}^{(l)}$ are unlikely to transition to each other, corresponding to the two null matrices in $\textbf{M}$ (see $\textbf{0}$ in Eq.~\ref{M_proposition}).
The transition of the game is governed by $\textbf{P}^{(2)}$, $\textbf{P}^{(1)}$, and $\textbf{P}^{(0)}$.
For example, $\textbf{P}^{(2)}$ determines whether or not an edge of $E_{AA}^{(i)}$ can transition to $E_{AA}^{(i_1)}$ or $E_{AB}^{(j)}$ in a time step ($i_1,j\in\{1,2,\dots,n\}$), corresponding to those terms including $\textbf{P}^{(2)}$ in $\textbf{M}$.
Note that the realistic evolutionary process is much more complicated and it is impossible to obtain the exact transition probability of an edge from one state to another, but the matrix $\textbf{M}$ can describe the possibility that an edge of $E_{X_1Y_1}^{(i)}$ transitions to that of $E_{X_2Y_2}^{(j)}$ in a time step.
The zero entries in $\textbf{M}$ indicate the transition can never happen and the nonzero entries suggest the transition is can happen.

When the random process has only closed communicating class, let $\textbf{c}$ denote the set of all states lying in this class.
Eqs.~\ref{pA_derivative_DB},\ref{pAAi_derivative},\ref{pABi_derivative},\ref{pBBi_derivative} show that for a sufficiently small $\delta$,
the fractions of various edges change much faster than the fractions of the two strategies.
That is, although there is a frequent transition between $A$-players and $B$-players, namely an $A$-player transitioning to a $B$-player or a $B$-player transitioning to an $A$-player, the fraction of $A$-players varies at a relatively low rate.
In the evolutionary process, an edge transitions among various states as the strategies adopted by the connected individuals and the game they play change.
The state transition possibility of an edge is described by $\textbf{M}$.
Eventually, the state of this edge enters into the closed communicating class $\textbf{c}$ and can never escape from $\textbf{c}$, regardless of its initial state.
Thus, if the random process defined above has only one closed communicating class, the evolutionary outcome is independent of the initial condition.

When the random process has $m$ ($>1$) closed communicating classes, we denote them by $\textbf{c}_1, \textbf{c}_2, \dots, \textbf{c}_m$.
If it is possible for an edge of $E_{X_1Y_1}^{(i)}$ to transition to $E_{X_2Y_2}^{(j)}$ within one update step, we denote it by $E_{X_1Y_1}^{(i)}\rightarrow E_{X_2Y_2}^{(j)}$.
We apply two propositions below:\\
(i) for $E_{X_1Y_1}^{(i_1)}\in \textbf{c}_{j_1}$ and $E_{X_2Y_2}^{(i_2)}\in \textbf{c}_{j_2}$, $i_1\ne i_2$ for any $j_1\ne j_2$ regardless of $X_1Y_1$ and $X_2Y_2$; \\
(ii) in every closed communicating class $\textbf{c}_j$, there exists some $i$ satisfying $E_{AA}^{(i)}\in \textbf{c}_j$, $E_{AB}^{(i)}\in\textbf{c}_j$ and $E_{BB}^{(i)}\in \textbf{c}_j$. For different closed communicating classes, $i$ can be different.

About proposition (i), for $E_{AA}^{(i)}\in \textbf{c}_j$, since $E_{AA}^{(i)}$ is a recurrent state, there exists $l$ making either $E_{AA}^{(l)}\rightarrow E_{AA}^{(i)}$ or $E_{AB}^{(l)}\rightarrow E_{AA}^{(i)}$.
Note that $l$ can be $i$.
$E_{BB}^{(l)}\rightarrow E_{AA}^{(i)}$ is impossible since in each generation just one player has the change to update its strategy.
$E_{AA}^{(l)}\rightarrow E_{AA}^{(i)}$ means $E_{AA}^{(l)}\rightarrow E_{AB}^{(i)}$ due to the strategy transition, leading to $E_{AB}^{(i)} \in \textbf{c}_j$.
$E_{AB}^{(l)}\rightarrow E_{AA}^{(i)}$ means $E_{AB}^{(l)}\rightarrow E_{AB}^{(i)}$ due to the game transition, leading to $E_{AB}^{(i)} \in \textbf{c}_j$.
Thus, $E_{AA}^{(i)}\in \textbf{c}_j$ always gives $E_{AB}^{(i)} \in \textbf{c}_j$.
Similarly, $E_{BB}^{(i)}\in \textbf{c}_j$ gives $E_{AB}^{(i)} \in \textbf{c}_j$.
That is, $E_{X_1Y_1}^{(i_1)}\in \textbf{c}_{j_1}$ means $E_{AB}^{(i_1)}\in \textbf{c}_{j_1}$ and $E_{X_2Y_2}^{(i_2)}\in \textbf{c}_{j_2}$ means $E_{AB}^{(i_2)}\in \textbf{c}_{j_2}$.
For $j_1\ne j_2$, $i_1\ne i_2$.
Otherwise, an edge state lies in two different closed communicating classes, which leads to a contradiction.

Based on the proof of proposition (i), every closed communicating class includes at lease one state with the form of $E_{AB}^{(i)}$.
Note that $i$ in different closed communicating classes is different.
Let $E_{AB}^{(i)}\in \textbf{c}_j$.
Due to the game transition, there must exist $l$ with $E_{AB}^{(i)}\rightarrow E_{AB}^{(l)}$ and $E_{AB}^{(l)}\in \textbf{c}_j$.
In addition, due to the strategy transition, we have $E_{AB}^{(i)}\rightarrow E_{AA}^{(l)}$ and $E_{AB}^{(i)}\rightarrow E_{BB}^{(l)}$, which gives $E_{AA}^{(l)}\in \textbf{c}_j$ and $E_{BB}^{(l)}\in \textbf{c}_j$.
Overall, $\textbf{c}_j$ includes $E_{AA}^{(l)}$, $E_{AB}^{(l)}$ and $E_{BB}^{(l)}$.

Proposition (i) stresses that when an edge transitions into a state that lies in a closed communicating class like $\textbf{c}_{j_1}$, the games to be played by the two connected players are limited by $\textbf{c}_{j_1}$.
If the edge transitions into a state in another closed communicating class like $\textbf{c}_{j_2}$, the two connected players can only play games limited by $\textbf{c}_{j_2}$.
In particular, the games limited by $\textbf{c}_{j_1}$ and those by $\textbf{c}_{j_2}$ are completely different.
The initial condition affects which closed communicating class an edge will transition into.
A representative example is that for $E_{AA}^{(i_1)}\in \textbf{c}_{j_1}$, $E_{AB}^{(i_1)}\in \textbf{c}_{j_1}$, $E_{BB}^{(i_1)}\in \textbf{c}_{j_1}$ and
$E_{AA}^{(i_2)}\in \textbf{c}_{j_2}$, $E_{AB}^{(i_2)}\in \textbf{c}_{j_2}$, $E_{BB}^{(i_2)}\in \textbf{c}_{j_2}$, when all players play game $i_1$ initially, the games to be played are limited by $\textbf{c}_{j_1}$ throughout the process.
However, if initially all players play game $i_2$, $\textbf{c}_{j_2}$ constrains the games to be played throughout the process.
~\\

We examine the above approach with $10^8$ numerical examples.
In every example, we generate three $4\times 4$ random matrices in which all entries are nonnegative.
We normalize these matrices to make the sum of entries in each row $1$.
The three matrices are assigned to $\textbf{P}^{(2)}$, $\textbf{P}^{(1)}$, and $\textbf{P}^{(0)}$.
On the one hand, based on $\textbf{P}^{(2)}$, $\textbf{P}^{(1)}$, and $\textbf{P}^{(0)}$, we calculate the eigenvalues of $\left[\textbf{A}\left(\bar{\textbf{P}}^{(s)}\right)-2k^2\textbf{I}\right]/(kN)$ in Eq.~\ref{system_matrix} and record whether or not there are zero eigenvalues.
On the other, based on $\textbf{P}^{(2)}$, $\textbf{P}^{(1)}$, and $\textbf{P}^{(0)}$, we calculate $\bar{\textbf{M}}$ and record whether or not there is some $i$ such that all entries in the $i$th column of $\bar{\textbf{M}}$ are positive.
In all examples, whenever there is zero entry in each column of $\bar{\textbf{M}}$, there are zero eigenvalues.
As long as there exists some $i$ such that all entries in the $i$th column of $\bar{\textbf{M}}$ are positive, there is no zero eigenvalue.
Thus, our approach predicts well the sensitivity of the evolutionary outcomes to the initial condition.
Furthermore, for $\textbf{P}^{(2)}$, $\textbf{P}^{(1)}$, and $\textbf{P}^{(0)}$ under which the evolutionary outcome is sensitive to the initial condition,
a slight perturbation or noise to the game transition pattern (to $\textbf{P}^{(2)}$, $\textbf{P}^{(1)}$, and $\textbf{P}^{(0)}$) can turn this system into one insensitive to the initial condition.
A simple way to achieve this is adding to each entry in $\textbf{P}^{(2)}$, $\textbf{P}^{(1)}$, and $\textbf{P}^{(0)}$ an arbitrary small number $\delta_1$, $\delta_2$, and $\delta_3$, respectively, where $\delta_1$, $\delta_2$, and $\delta_3$ are not necessary identical.

\section{Evolutionary dynamics with global game transitions} \label{section_global}

\subsection{Global game transitions}
In this section, we study evolutionary dynamics with global game transitions.
In each time step, games in all interactions have chances to update.
We proceed with the mathematical analysis as we do in Section \ref{section_local}.
We take the same variables and notations (Eqs.~\ref{game_transition}-\ref{notaitonf}).
The change in $p_A$ follows Eqs.~\ref{pA_derivative_DB}-\ref{I_di}.
We then investigate the change in the frequency of each type of edge.
Assuming that a random $B$-player is selected to die, the change in $p_{AA}^{(i)}$ arises from two parts: the switching of edges connecting the focal (dead) $B$-player and its nearest neighbors, the switching of all other edges.
Under the neighborhood configuration given in Section \ref{subsubsection_pA_B}, Eq.~\ref{pAAi_AtakeB_inside} shows the change in $p_{AA}^{(i)}$ due to the former part.
The change in $p_{AA}^{(i)}$ due to the latter is
\begin{equation} \label{pAAi_AtakeB_outside_all}
\begin{split}
&\mathbb{P}\left(\Delta p_{AA}^{(i)}=\frac{\sum_{j=1}^n Nkp_{AA}^{(j)}p_{ji}^{(2)}-\sum_{j=1}^n Nkp_{AA}^{(i)}p_{ij}^{(2)}}{kN}\right) \ \\ &=\mathcal{B}\left(k_{A|B}^{(j)},k_{B|B}^{(j)}|j=1,\dots,n\right)\left[\mathbb{P}\left(A\rightarrow B\right)+\mathbb{P}\left(B\rightarrow B\right)\right].
\end{split}
\end{equation}

Suppose that a random $A$-player is selected to die.
Under the neighborhood configuration given in Section \ref{subsubsection_pA_A}, Eqs.~\ref{pAAi_AtakeA_inside} and \ref{pAAi_BtakeA_inside} capture the change in $p_{AA}^{(i)}$ due to edges between the focal $A$-player and its nearest neighbors.
The change in $p_{AA}^{(i)}$ due to the switching of other edges is
\begin{equation} \label{pAAi_AtakeA_outside_all}
\begin{split}
&\mathbb{P}\left(\Delta p_{AA}^{(i)}=\frac{\sum_{j=1}^n \left(Nkp_{AA}^{(j)}-2k_{A|A}^{(j)}\right)p_{ji}^{(2)}-\sum_{j=1}^n\left(Nkp_{AA}^{(i)}-2k_{A|A}^{(i)}\right)p_{ij}^{(2)}}{kN}\right) \ \\ &=\mathcal{A}\left(k_{A|A}^{(j)},k_{B|A}^{(j)}|j=1,\dots,n\right)\left[\mathbb{P}\left(A\rightarrow A\right)+\mathbb{P}\left(B\rightarrow A\right)\right].
\end{split}
\end{equation}

From Eqs.~\ref{pAAi_AtakeB_inside},\ref{pAAi_AtakeA_inside},\ref{pAAi_BtakeA_inside},\ref{pAAi_AtakeB_outside_all},\ref{pAAi_AtakeA_outside_all}, we obtain the time derivative of $p_{AA}^{(i)}$, given by
\begin{equation} \label{pAAi_derivative_global}
\dot{p}_{AA}^{(i)}=\sum_{j=1}^n p_{AA}^{(j)}p_{ji}^{(2)}-p_{AA}^{(i)}-\frac{2}{kN}\sum_{j=1}^n(k-1)q_{B|A}p_{ji}^{(2)}p_{AA}^{(j)}+\frac{2}{kN}\sum_{j=1}^n \left[(k-1)q_{A|B}+1\right]p_{ji}^{(1)}p_{AB}^{(j)}+O\left(\delta\right) .
\end{equation}
Analogously, we obtain the time derivatives of $p_{AB}^{(i)}$ and $p_{BB}^{(i)}$, given by
\begin{equation} \label{pABi_derivative_global}
\begin{split}
\dot{p}_{AB}^{(i)}
&=\sum_{j=1}^n p_{AB}^{(j)}p_{ji}^{(1)}-p_{AB}^{(i)} \\\
& +\frac{1}{kN}\sum_{j=1}^n(k-1)q_{B|A}p_{ji}^{(2)}p_{AA}^{(j)}\ \\
& +\frac{1}{kN}\sum_{j=1}^n\left[(k-1)\left(q_{A|A}+q_{B|B}\right)-2k\right]p_{ji}^{(1)}p_{AB}^{(j)} \ \\
& +\frac{1}{kN}\sum_{j=1}^n(k-1)q_{A|B}p_{ji}^{(0)}p_{BB}^{(j)}+O\left(\delta\right)
\end{split}
\end{equation}
and
\begin{equation} \label{pBBi_derivative_global}
\dot{p}_{BB}^{(i)}=\sum_{j=1}^n p_{BB}^{(j)}p_{ji}^{(0)}-p_{BB}^{(i)}-\frac{2}{kN}\sum_{j=1}^n(k-1)q_{A|B}p_{ji}^{(0)}p_{BB}^{(j)}+\frac{2}{kN}\sum_{j=1}^n \left[(k-1)q_{B|A}+1\right]p_{ji}^{(1)}p_{AB}^{(j)}+O\left(\delta\right) .
\end{equation}
Analogous to Eqs.~\ref{qAA_derivative_former} and \ref{qAA_derivative}, a further analysis to Eq.~\ref{pAAi_derivative_global} gives Eq.~\ref{q_AA}.
We redefine the function $\textbf{A}\left(\textbf{R}^{(s)}\right)$ to be
\begin{equation} \label{A_DB_global}
\textbf{A}\left(\textbf{R}^{(s)}\right)=
\left[
\begin{array}{ccc}
(1-2\alpha/\mu)\textbf{R}^{(2)} & 2(1+\beta)/\mu\textbf{R}^{(1)} & \textbf{0} \\
\alpha/\mu\textbf{R}^{(2)} & (1-k/\mu)\textbf{R}^{(1)} & \beta/\mu\textbf{R}^{(0)} \\
\textbf{0} & 2(1+\alpha)/\mu\textbf{R}^{(1)} & (1-2\beta/\mu)\textbf{R}^{(0)}
\end{array}
\right],
\end{equation}
where $\mu=kN$.
$\alpha$ and $\beta$ follow those defined in Eq.~\ref{A_DB}.
We can reduce the system of Eqs.~\ref{pAAi_derivative_global}-\ref{pBBi_derivative_global} to
\begin{equation} \label{system_matrix_global}
\dot{\textbf{v}}=\left[\textbf{A}\left(\bar{\textbf{P}}^{(s)}\right)-\textbf{I}\right]\textbf{v}+\textbf{A}\left(\tilde{\textbf{P}}^{(s)}\right)\textbf{b},
\end{equation}
where $\bar{\textbf{P}}^{(s)},\tilde{\textbf{P}}^{(s)},\textbf{b},\textbf{v}$ are defined in Eqs.~\ref{Ps} and \ref{system_matrix}.
Similarly, solving Eq.~\ref{system_matrix_global}, substituting the solutions into Eqs.~\ref{Iai}-\ref{Idi}, and inserting $I_{R_i}, I_{S_i}, I_{T_i}, I_{P_i}$ into Eq.~\ref{general_rule_DB}, we obtain the condition of one strategy being favored over the other under general two-strategy games.
In particular, when all games are donation games, Eq.~\ref{DB_bckxi} predicts when $A$ is favored over $B$.
$\xi_i$ can be obtained by inserting $I_{R_i}$ and $I_{T_i}$ into Eq.~\ref{xi_DB}.
The approach proposed in Section \ref{section_sensitivity} still determines whether or not the evolutionary dynamics is sensitive to the initial condition under global game transitions.

\subsection{A class of game transition patterns}
We provide an alternative approach to study a class of game transition patterns (denoted by $\Omega$): for every $s\in\{0,1,2\}$, in a Markov chain $\{M^{(s)}_t,t=0,1,2,3,\dots\}$ with state space $\{1,2,\dots,n\}$ and state transition matrix $\mathbf{P}^{(s)}$, there is only one recurrent equivalence class (and the states therein are aperiodic).
Introducing $\textbf{u}^{(s)}=\left(u_{1}^{(s)},\dots,u_{n}^{(s)}\right)$,
the solution to $\mathbf{u}^{(s)}=\mathbf{u}^{(s)}\mathbf{P}^{(s)}$ with $\sum_{j=1}^n u_{i}^{(s)}=1$ is the limiting distribution of this Markov chain.

In each time step, only one among $N$ players has the chance to modify its strategy whereas all games are likely to update.
The evolutionary rate of the game in an interaction (or in an edge) is $N/2$ times as large as that of interactants' strategies.
Therefore, for sufficiently large population size $N$, the fractions of various games reach a stationary distribution much faster than the fractions of various strategies.
For games between two $A$-players, the stationary distribution is $\mathbf{u}^{(2)}$.
Thus, in the interaction of two $A$-players, the expected payoff for each $A$-player is $\sum_{i=1}^n u_i^{(2)}R_i$.
Analogously, the stationary distribution for games between an $A$-player and a $B$-player is $\mathbf{u}^{(1)}$.
The expected payoff for $A$-player is $\sum_{i=1}^n u_i^{(1)}S_i$ and that of $B$-player is $\sum_{i=1}^n u_i^{(1)}T_i$.
For games between two $B$-players, the stationary distribution is $\textbf{u}^{(0)}$.
The expected payoff for each defector is $\sum_{i=1}^n u_i^{(0)}P_i$.
The game transition creates a situation ``as if'' all players play an ``effective'' game, with payoff structure
\begin{linenomath}
\begin{equation} \label{global_matrix}
\bordermatrix{
	& A & B \cr
	A & \sum_{i=1}^n u_i^{(2)}R_i & \sum_{i=1}^n u_i^{(1)}S_i \cr
	B & \sum_{i=1}^n u_i^{(1)}T_i & \sum_{i=1}^n u_i^{(0)}P_i \cr
}\equiv
\bordermatrix{
	& A & B \cr
	A & \bar{R} & \bar{S} \cr
	B & \bar{T} & \bar{P} \cr
},
\end{equation}
\end{linenomath}
which holds for death-birth, imitation, pairwise-comparison, and birth-death updating.

Under death-birth updating, in Eq.~\ref{general_rule_DB}, replacing $R_i$, $S_i$, $T_i$ and $P_i$ with $\bar{R}$, $\bar{S}$, $\bar{T}$ and $\bar{P}$, then inserting Eqs.~\ref{Iai}-\ref{Idi}, we reduce Eq.~\ref{general_rule_DB} to
\begin{equation} \label{global_DB}
(k+1)\sum_{i=1}^n u_i^{(2)}R_i+(k-1)\sum_{i=1}^n u_i^{(1)}S_i>(k-1)\sum_{i=1}^n u_i^{(1)}T_i+(k+1)\sum_{i=1}^n u_i^{(0)}P_i.
\end{equation}
For donation games with $R_i=b_i-c$, $S_i=-c$, $T_i=b_i$ and $P_i=0$, Eq.~\ref{global_DB} is further reduced to Eq.~\ref{DB_bckxi} with
\begin{equation}
\xi_i = -\frac{(k+1)u_i^{(2)}-(k-1)u_i^{(1)}}{2}.
\end{equation}

Similarly, under imitation updating, we can reduce Eq.~\ref{general_rule_DB} to
\begin{equation} \label{global_IM}
(k+3)\sum_{i=1}^n u_i^{(2)}R_i+(k+1)\sum_{i=1}^n u_i^{(1)}S_i>(k+1)\sum_{i=1}^n u_i^{(1)}T_i+(k+3)\sum_{i=1}^n u_i^{(0)}P_i.
\end{equation}
A further analysis of donation games leads to Eq.~\ref{IM_bckxi} with
\begin{equation} \label{global_DB_xi}
\quad \xi_i = -\frac{(k+3)u_i^{(2)}-(k+1)u_i^{(1)}}{2}.
\end{equation}

Under both pairwise-comparison and birth-death updating, $A$-players are favored over $B$-players if and only if
\begin{equation}
\sum_{i=1}^n u_i^{(2)}R_i+\sum_{i=1}^n u_i^{(1)}S_i>\sum_{i=1}^n u_i^{(1)}T_i+\sum_{i=1}^n u_i^{(0)}P_i.
\end{equation}
A simplification for donation games gives Eq.~\ref{PC_bckxi} with
\begin{equation}
\xi_i = -\frac{u_i^{(2)}-u_i^{(1)}}{2}.
\end{equation}

In this section, we solve $\mathbf{u}^{(s)}=\mathbf{u}^{(s)}\mathbf{P}^{(s)}$ and use the limiting distribution $\mathbf{u}^{(s)}$ to approximate the evolutionary process.
Eqs.~\ref{pAAi_derivative_global}-\ref{pBBi_derivative_global} actually imply this idea.
For weak selection ($\delta\ll 1$), $p_{AA}^{(i)}$, $p_{AB}^{(i)}$, $p_{BB}^{(i)}$ reach the equilibrium point much faster than $p_A$ (see Eqs.~\ref{pA_derivative_DB} and \ref{pAAi_derivative_global}-\ref{pBBi_derivative_global}).
The dynamical system thus converges quickly onto the slow manifold with $\dot{p}_{AA}^{(i)}=0$, $\dot{p}_{AB}^{(i)}=0$, $\dot{p}_{BB}^{(i)}=0$.
In the righthand of Eq.~\ref{pAAi_derivative_global}, $1/N$ occurs in the third and fourth terms.
For a sufficiently large population size $N$ ($N\gg 1$), the existence of $1/N$ may make the two terms negligible relative to the first and second terms.
This inspires the idea of using $\sum_{j=1}^n p_{AA}^{(j)}p_{ji}^{(2)}-p_{AA}^{(i)}=0$ to approximate $\dot{p}_{AA}^{(i)}=0$.
Replacing $p_{AA}^{(j)}$ with $u_{j}^{(2)}$, we have $\mathbf{u}^{(2)}\mathbf{P}^{(2)}=\mathbf{u}^{(2)}$.
The analogous analysis to Eqs.~\ref{pAAi_derivative_global} and \ref{pABi_derivative_global} gives $\mathbf{u}^{(1)}\mathbf{P}^{(1)}=\mathbf{u}^{(1)}$ and $\mathbf{u}^{(0)}\mathbf{P}^{(0)}=\mathbf{u}^{(0)}$.

\subsection{Game transitions in a fraction of interactions}
With global game transitions, games in all interactions have chances to update in each time step.
With local game transitions, games in a fraction of interactions have chances to update in each time step.
Note that with local game transitions, the interactions allowing for game transitions are spatially correlated---only games in interactions involved with the deceased individual's neighbors are likely to update.
In this section, we assume that in each time step a fraction $p$ ($0< p< 1$) of games are randomly selected to update.
In other words, in each interaction, the game has chance to update with probability $p$ and has no chance to update with probability $1-p$.
Equivalently, in each interaction, with probability $p$ the game transitions based on the probability matrix $\mathbf{P}^{(s)}$, and with probability $1-p$ the game transitions to itself.
Such a situation corresponds to game transitions based on a new probability matrix $\hat{\mathbf{P}}^{(s)}$,
\begin{align}
\hat{\mathbf{P}}^{(s)}=p\mathbf{P}^{(s)}+(1-p)\mathbf{I}. \nonumber
\end{align}
Note that the solution to $\mathbf{u}^{(s)}=\mathbf{u}^{(s)}\hat{\mathbf{P}}^{(s)}$ is the same as $\mathbf{u}^{(s)}=\mathbf{u}^{(s)}\mathbf{P}^{(s)}$.
The setting of a fraction of games being transitioned thus leads to the same results as global transition does.

\subsection{Stochastic strategies}
Up to this point, all players are assumed to be using pure strategies, namely a player cooperating (taking $A$) unconditionally or a player defecting (adopting $B$) unconditionally in each time step.
In the following, we further investigate the case where players take stochastic strategies, i.e. choosing to cooperate with a probability and to defect otherwise.
Let $s_{p}$ and $s_{q}$ denote two stochastic strategies.
Players taking $s_{p}$ choose to cooperate with probability $p$ and defect with probability $1-p$.
If taking $s_{q}$, players choose cooperation with probability $q$ and defection with probability $1-q$.
For two connected players, before one of them has a chance to update its strategy (i.e. $s_p$ or $s_q$), their actions (i.e. cooperation or defection) and games they played update many times.
Therefore, in a sufficiently large population, the fractions of various interaction scenarios (consisting of two actions and the game they play) reach a stationary distribution much faster than the fractions of various strategies.

We study the competition between $s_{p}$ and $s_{q}$.
By taking $p=1$ (pure cooperators) and $q=0$ (pure defectors), this model can recover the case of pure strategies.
In the following, we calculate the stationary distribution of interaction scenarios between players taking $s_{p}$ and players taking $s_{q}$.
Let $u_{i,r1,r2}^{(pq)}(t)$ denote the probability that in time $t$ a player with $s_p$ chooses action $r_1$, a player with $s_q$ chooses action $r_2$, and they play game $i$, where $i\in \{1,\dots,n\}$ and $r_1, r_2\in \{0,1\}$ ($0$ represents defection and $1$ means cooperation).
Then, we have
\begin{subequations}
\begin{equation}
u_{j,1,1}^{(pq)}(t+1)=\sum_{i=1}^n \left[u_{i,1,1}^{(pq)}(t)p_{ij}^{(2)}+u_{i,1,0}^{(pq)}(t)p_{ij}^{(1)}+u_{i,0,1}^{(pq)}(t)p_{ij}^{(1)}+u_{i,0,0}^{(pq)}(t)p_{ij}^{(0)}\right]pq ; \label{Eq._stochastic11}\\
\end{equation}
\begin{equation}
u_{j,1,0}^{(pq)}(t+1)=\sum_{i=1}^n \left[u_{i,1,1}^{(pq)}(t)p_{ij}^{(2)}+u_{i,1,0}^{(pq)}(t)p_{ij}^{(1)}+u_{i,0,1}^{(pq)}(t)p_{ij}^{(1)}+u_{i,0,0}^{(pq)}(t)p_{ij}^{(0)}\right]p(1-q) ; \label{Eq._stochastic10}\\
\end{equation}
\begin{equation}
u_{j,0,1}^{(pq)}(t+1)=\sum_{i=1}^n \left[u_{i,1,1}^{(pq)}(t)p_{ij}^{(2)}+u_{i,1,0}^{(pq)}(t)p_{ij}^{(1)}+u_{i,0,1}^{(pq)}(t)p_{ij}^{(1)}+u_{i,0,0}^{(pq)}(t)p_{ij}^{(0)}\right](1-p)q ; \label{Eq._stochastic01}\\
\end{equation}
\begin{equation}
u_{j,0,0}^{(pq)}(t+1)=\sum_{i=1}^n \left[u_{i,1,1}^{(pq)}(t)p_{ij}^{(2)}+u_{i,1,0}^{(pq)}(t)p_{ij}^{(1)}+u_{i,0,1}^{(pq)}(t)p_{ij}^{(1)}+u_{i,0,0}^{(pq)}(t)p_{ij}^{(0)}\right](1-p)(1-q) .
\label{Eq._stochastic00}
\end{equation}
\end{subequations}
For the game transition pattern $\Omega$, there exists a stationary distribution and we denote it by $\textbf{u}^{(pq)}=(u_{1,1,1}^{(pq)},\dots,u_{n,1,1}^{(pq)},u_{1,1,0}^{(pq)}$, $\dots,u_{n,1,0}^{(pq)},
u_{1,0,1}^{(pq)},\dots,u_{n,0,1}^{(pq)},u_{1,0,0}^{(pq)},\dots,u_{n,0,0}^{(pq)})$, where $u_{i,r_1,r_2}^{(pq)}$ indicates the stationary fraction of interactions in which two players play game $i$ and the one with strategy $s_p$ chooses action $r_1$ and the other with strategy $s_q$ chooses action $r_2$.
We rewrite Eqs.~\ref{Eq._stochastic11}-\ref{Eq._stochastic00} as
\begin{equation}
\textbf{u}^{(pq)}=\textbf{u}^{(pq)}
\left[
\begin{array}{cccc}
pq\textbf{P}^{(2)} & p(1-q)\textbf{P}^{(2)} & (1-p)q\textbf{P}^{(2)} & (1-p)(1-q)\textbf{P}^{(2)} \\
pq\textbf{P}^{(1)} & p(1-q)\textbf{P}^{(1)} & (1-p)q\textbf{P}^{(1)} & (1-p)(1-q)\textbf{P}^{(1)} \\
pq\textbf{P}^{(1)} & p(1-q)\textbf{P}^{(1)} & (1-p)q\textbf{P}^{(1)} & (1-p)(1-q)\textbf{P}^{(1)} \\
pq\textbf{P}^{(0)} & p(1-q)\textbf{P}^{(0)} & (1-p)q\textbf{P}^{(0)} & (1-p)(1-q)\textbf{P}^{(0)}
\end{array}
\right].
\end{equation}
We can get the stationary distribution $\textbf{u}^{(pq)}$ by the left eigenvector with $\sum_{r_1=0}^1\sum_{r_2=0}^1 \sum_{i=1}^n u_{i,r_1,r_2}^{pq}=1$.
Actually, letting $\bar{\textbf{u}}^{(pq)}=(u_{1}^{(pq)},\dots,u_{n}^{(pq)})$ and solving
\begin{equation}
\bar{\textbf{u}}^{(pq)} = \bar{\textbf{u}}^{(pq)}\left[pq\textbf{P}^{(2)}+(p+q-2pq)\textbf{P}^{(1)}+(1-p)(1-q)\textbf{P}^{(0)}\right]
\end{equation}
with $\sum_{i=1}^n u_{i}^{pq}=1$,
we have $u_{i,1,1}^{(pq)}=pqu_{i}^{(pq)}$, $u_{i,1,0}^{(pq)}=p(1-q)u_{i}^{(pq)}$, $u_{i,0,1}^{(pq)}=(1-p)qu_{i}^{(pq)}$, and $u_{i,0,0}^{(pq)}=(1-p)(1-q)u_{i}^{(pq)}$.
In the interaction between a player with strategy $s_{p}$ and a player with strategy $s_{q}$, the former's expected payoff is
\begin{equation}
f_{pq}= \sum_{i=1}^n u_{i}^{(pq)}\left[pqR_i+p(1-q)S_i+(1-p)qT_i+(1-p)(1-q)P_i\right].
\end{equation}
Under death-birth updating, the condition for strategy $s_p$ to be favored over $s_q$ (i.e. $\rho_{s_p}>\rho_{s_q}$) is
\begin{equation}
(k+1)f_{pp}+(k-1)f_{pq}>(k-1)f_{qp}+(k+1)f_{qq}.  \label{stochastic_strategy}
\end{equation}
We say that a stochastic strategy is more cooperative if players with such a strategy choose cooperation with a larger probability.
That is, for $p>q$, $s_p$ is more cooperative than $s_q$.
For donation games described by Eq.~\ref{donation_games}, Eq.~\ref{stochastic_strategy} can be reduced to Eq.~\ref{DB_bckxi} with
\begin{equation}
\xi_i = \frac{-(k+1)pu_{i}^{(pp)}-(k-1)qu_{i}^{(pq)}+(k-1)pu_{i}^{(qp)}+(k+1)qu_{i}^{(qq)}}{2(p-q)}.  \label{xi_stochastic_DB}
\end{equation}

Similarly, under imitation updating in donation games, natural selection favors $s_p$ over $s_q$ if Eq.~\ref{IM_bckxi} holds, where
\begin{equation}
\xi_i= \frac{-(k+3)pu_{i}^{(pp)}-(k+1)qu_{i}^{(pq)}+(k+1)pu_{i}^{(qp)}+(k+3)qu_{i}^{(qq)}}{2(p-q)}.
\end{equation}
For birth-death or pairwise-comparison updating in donation games, $s_p$ is favored over $s_q$ if Eq.~\ref{PC_bckxi} holds, where
\begin{equation}
\xi_i= \frac{-pu_{i}^{(pp)}-qu_{i}^{(pq)}+pu_{i}^{(qp)}+qu_{i}^{(qq)}}{2(p-q)}.  \label{xi_stochastic_PC}
\end{equation}

\subsection{Intuition based on ``sigma rule''}
Here we provide a few new insights into how game transitions affect the evolution of $A$-players.
In the game between $A$-players and $B$-players governed by a single payoff matrix
\begin{equation}
\bordermatrix{
& \text{\textit{A}} & \text{\textit{B}} \cr
\text{\textit{A}} & R & S \cr
\text{\textit{B}} & T & P \cr
}, 
\end{equation}
Tarnita et al have found that selection favors $A$-players over $B$-players if and only if
\begin{equation}
\sigma R+S>T+\sigma P, 
\end{equation}
which is termed as ``sigma rule'' \cite{si2009-Tarnita-p570-581}.
The coefficient $\sigma$ captures how the spatial model and its associated update rule affect evolutionary dynamics, whereas
is independent of the payoffs.
For an infinite random regular graph under death-birth updating, $\sigma =\left(k+1\right) /\left(k-1\right)$. 

When all interactions are governed by a fixed donation game with a donation cost $c$ and benefit $b_1$, substituting $R=b_1-c$, $S=-c$, $T=b_1$ and $P=0$ into the sigma rule gives the condition of $A$-players being favored over $B$-players.
Intriguingly, Eq.~\ref{DB_bckxi} can be phrased in the form of a sigma rule with $R=b_1-c+\frac{2}{k+1}\sum_{i=2}^n \xi_i\Delta b_{1i}$, $S=-c$, $T=b_1$ and $P=0$.
With game transitions, evolution proceeds ``as if'' all interactions are governed by an effective game with payoff matrix
\begin{equation}
\bordermatrix{
& \text{\textit{A}} & \text{\textit{B}} \cr
\text{\textit{A}} & b_1-c+\frac{2}{k+1}\sum_{i=2}^n \xi_i\Delta b_{1i} & -c \cr
\text{\textit{B}} & b_1 & 0 \cr
}.
\end{equation}
Compared with the fixed donation game, mutual cooperation brings each player an extra payoff $\frac{2}{k+1}\sum_{i=2}^n \xi_i\Delta b_{1i}$ in this effective game.
This payoff depends on two factors: game transition patterns (described by $\xi_i$), variations in different games (described by $\Delta b_{1i}$).

For an infinite random regular graph under pairwise-comparison updating, $\sigma=1$.
Analogously, Eq.~\ref{PC_bckxi} can be phrased in the form of a sigma rule with $R=b_1-c+2\sum_{i=2}^n \xi_i\Delta b_{1i}$, $S=-c$, $T=b_1$ and $P=0$.
With game transitions, evolution proceeds ``as if" all interactions are governed by an effective game with payoff structure
\begin{equation}
\bordermatrix{
& \text{\textit{A}} & \text{\textit{B}} \cr
\text{\textit{A}} & b_1-c+2\sum_{i=2}^n \xi_i\Delta b_{1i} & -c \cr
\text{\textit{B}} & b_1 & 0 \cr
}.
\end{equation}
\section{Representative examples} \label{section_examples}
In Section \ref{section_local} and \ref{section_global}, we derive the general condition of one strategy to be favored over the other strategy, which requires to solve a set of equations.
In this section, we study four representative interaction scenarios and provide explicit expressions.
\subsection{Evolutionary dynamics with state-independent game transitions}
If the game to be played in the next time step is independent of the game played in the past, the game transition is state-independent.
That is, $p_{im}^{(s)}=p_{jm}^{(s)}$ for all $i$ and $j$.
The number of $A$-players determines the game to be played.
Let $p_{im}^{(s)}=p_{m}^{(s)}$.
For local game transitions, focusing on game transition patterns under which the evolutionary outcome is independent of the initial condition, we have
\begin{equation}
\begin{split}
&\textbf{DB}: \xi_i = \frac{\left(- 6k^4 - 2k^3 - 3k^2 - 6k - 2\right)p_{i}^{(2)}+\left(6k^4 - 11k^3 + 3k^2 + 6k + 4\right)p_{i}^{(1)}+\left(k^3 - 2\right) p_{i}^{(0)}}{12k^3}; \ \\
&\textbf{IM}: \xi_i = \frac{\left(- 6k^4 - 20k^3 - 13k^2 - 14k - 6\right)p_{i}^{(2)}+\left(6k^4 + 7k^3 - k^2 + 10k + 12\right) p_{i}^{(1)}+\left(k^3 + 2k^2 + 4k - 6\right) p_{i}^{(0)}}{12k^2(k+ 1)}; \ \\
&\textbf{PC}: \xi_i = \frac{\left(- 20k^2 + 8k - 2\right) p_{i}^{(2)}+\left(19k^2 - 10k + 4\right) p_{i}^{(1)}+\left(k^2 + 2k - 2\right)p_{i}^{(0)}}{24k(2k - 1)}. \
\end{split}
\end{equation}
For global game transitions, focusing on the game transition pattern $\Omega$ introduced in Section \ref{section_global}, we have
\begin{equation}
\begin{split}
&\textbf{DB}: \quad \xi_i = -\frac{(k+1)p_i^{(2)}-(k-1)p_i^{(1)}}{2}; \ \\
&\textbf{IM}: \quad \xi_i = -\frac{(k+3)p_i^{(2)}-(k+1)p_i^{(1)}}{2}; \ \\
&\textbf{PC/DB}: \quad \xi_i = -\frac{p_i^{(2)}-p_i^{(1)}}{2}. \
\end{split}
\end{equation}
In particular, if $p_{m}^{(s)}=1/n$ for all $m$ and $s$, the game transition is fully stochastic.
In the next time step, any game occurs with the equal probability.
For both local and global game transitions, weak selection favors $A$ over $B$ if
\begin{equation}
\begin{split}
&\textbf{DB}: \quad (k+1)\sum_{i=1}^n R_i+(k-1)\sum_{i=1}^n S_i-(k-1)\sum_{i=1}^n T_i- (k+1)\sum_{i=1}^n P_i>0; \ \\
&\textbf{IM}: \quad (k+3)\sum_{i=1}^n R_i+(k+1)\sum_{i=1}^n S_i-(k+1)\sum_{i=1}^n T_i- (k+3)\sum_{i=1}^n P_i>0; \ \\
&\textbf{PC}: \quad \sum_{i=1}^n R_i+\sum_{i=1}^n S_i-\sum_{i=1}^n T_i-\sum_{i=1}^n P_i>0. \
\end{split}
\end{equation}
Let $\bar{R}=\left(\sum_{i=1}^n R_i\right)/n$, $\bar{S}=\left(\sum_{i=1}^n S_i\right)/n$, $\bar{T}=\left(\sum_{i=1}^n T_i\right)/n$, $\bar{P}=\left(\sum_{i=1}^n P_i\right)/n$.
We find that the evolutionary process with stochastic and diverse games can be approximated by that of a static and unified game.

\subsection{Evolutionary dynamics with strategy-independent game transitions}
If the game to be played in the next time step is independent of players' strategic actions in the past, the game transition is strategy-independent.
That is, $\textbf{P}^{(2)}=\textbf{P}^{(1)}=\textbf{P}^{(0)}$.
Let $\textbf{P}^{(2)}=\textbf{P}^{(1)}=\textbf{P}^{(0)}=\textbf{P}$.
Here, we consider the game transition pattern $\Omega$ introduced in Section \ref{section_global}.

For global game transitions, we introduce a vector $\textbf{u}=\left(u_1,u_2,\dots,u_n\right)$, which satisfies $\textbf{u}=\textbf{uP}$.
Then, we have
\begin{equation}
\begin{split}
&\textbf{DB}: \xi_i = -u_i ; \\
&\textbf{IM}: \xi_i = -u_i ; \\
&\textbf{PC/BD}: \quad \xi_i = 0 .
\end{split}
\end{equation}
Note that in pairwise-comparison or birth-death updating, $\xi_i=0$ means that cooperation can never evolve regardless of the benefit provided by a cooperative behavior in the donation game.
In other words, if the game is independent of strategic actions, game transitions cannot promote cooperation.

\subsection{Evolutionary dynamics with game transitions between two states ($n=2$)} \label{section_example2}
Given the theoretical significance of two states, we provide a systematic investigation of game transitions between two donation games.
According to Eq.~\ref{DB_bckxi}, under death-birth updating, the general rule for cooperation to be favored over defection is
\begin{equation} \label{fig2_DB}
\frac{b_1}{c}>k-\xi \frac{\Delta b}{c}.
\end{equation}

For local game transitions, focusing on game transition patterns under which the evolutionary outcome is insensitive to the initial condition,
we have
\begin{equation} \label{local2state}
\xi = \int_{0}^{1}\frac{\beta_3 y^3+\beta_2 y^2+\beta_1 y+\beta_0}{\alpha_2 y^2+\alpha_1 y+\alpha_0}dy,
\end{equation}
where
\begin{equation}
\begin{split}
\alpha_2 &= -k(P_0-1)\left[2(P_1-1)(P_2-1)k^4 + (P_1 + 2P_2 - 3P_1P_2)k^3 + (5P_1P_2 - 4P_2)k^2 - 3P_1P_2k + 2P_1P_2\right] ; \\
\alpha_1 &= k(k - 2)\left[2(P_2-P_0)(P_1-1)k^2 + 2(P_0 - P_1)P_2 k + P_0P_1 - 4P_0P_2 + 3P_1P_2\right] ; \\
\alpha_0 &= k(k - 2)^2(P_0P_1 - 2P_0P_2 + P_1P_2) ; \\
\beta_3 &= (k + 1)(k - 2)^3\left[p_{21}^{(2)}(P_1 - P_0)+p_{21}^{(1)}(P_0 - P_2)+p_{21}^{(0)}(P_2 - P_1)\right] ; \\
\beta_2 &= -p_{21}^{(2)}(k - 2)^2\left[(P_0 - 2P_1 + 1)k^3 + (1 - P_0)k^2 + (3P_0 - 2P_1)k + 4P_0 - 4P_1\right] ; \\
&-p_{21}^{(1)}(k - 2)^2\left[(2P_2 - 2)k^3 + (2P_0 - 2)k^2 + (P_2 - 3P_0)k - 4P_0 + 4P_2\right] \\
&-p_{21}^{(0)}(k - 2)^2\left[(1 - P_2)k^3 + (P_2 - 2P_1 + 1)k^2 + (2P_1 - P_2)k + 4P_1 - 4P_2\right]+\alpha_0 ; \\
\beta_1 &= p_{21}^{(2)}(k - 2)\big[- 2(P_0 - 1)(P_1 - 1)k^5 + P_1(P_0 - 1)k^4 + (P_0 + 4P_1 - 2P_0P_1 - 3)k^3 \\
& \hspace{2.05cm} -2(P_0 - 1)(P_1 - 2)k^2 + P_0(P_1 - 2)k - 3P_0 + 5P_1 - 2P_0P_1\big] \\
& + p_{21}^{(1)}(k - 2)\big[(2P_0 - 2)(P_2 - 1)k^5 - (P_0 - 1)(P_2 - 1)k^4 + (2P_2 - 3)(P_0 - 1)k^3 \\
& \hspace{2.4cm} + (2P_0P_2 - 4P_2 - 6P_0 + 6)k^2 + (2P_0 + 3P_2 - P_0P_2)k + 3P_0 - 5P_2 + 2P_0P_2\big] \\
& + p_{21}^{(0)}(k - 2)\left[(1 - P_2)k^4 + (P_2 - P_1)k^3 + (4P_1 - 2)k^2 - 3P_2k - 5P_1 + 5P_2\right]+ \alpha_1 ; \\
\beta_0 &= -p_{21}^{(2)}(1 - P_0)\left[(2 - 2P_{1})k^5 + P_{1}k^4 + (2 - 2P_{1})k^3 + (3 - 2P_{1})k^2 + P_{1}k - 2P_{1}\right] \\
&+p_{21}^{(1)}(1-P_0)\left[- 2P_{2}k^4 + (3P_{2} + 1)k^3 + (4 - 5P_{2})k^2 + 3P_{2}k - 2P_{2}\right] \\
&-p_{21}^{(0)}\left[(P_1 + P_2 - 1)k^3 + (1 - 2P_2 - 2P_1)k^2 + (3P_2 - P_1)k + 2P_1 - 2P_2\right]+\alpha_2 .
\end{split}
\end{equation}
In the above equations, $P_2=p_{11}^{(2)}-p_{21}^{(2)}$, $P_1=p_{11}^{(1)}-p_{21}^{(1)}$, and $P_0=p_{11}^{(0)}-p_{21}^{(0)}$.

For global game transitions, focusing on the game transition pattern $\Omega$ introduced in Section \ref{section_global}, we have
\begin{equation} \label{global2state}
\xi = \frac{(k-1)p_{12}^{(1)}p_{21}^{(2)}-(k+1)p_{21}^{(1)}p_{12}^{(2)}-2p_{12}^{(1)}p_{12}^{(2)}}{2\left(p_{12}^{(1)}+p_{21}^{(1)}\right)\left(p_{12}^{(2)}+p_{21}^{(2)}\right)}.
\end{equation}

We proceed with a specific game transition pattern involving two states: mutual cooperation leads to game 1 and all other action profiles lead to game 2.
Replacing $A$ with cooperation ($C$) and $B$ with defection ($D$), the game transition matrix is Eq.~\ref{fig2_matrix}.
In Section 2,
we have shown that the evolutionary process is insensitive to the initial condition.
Inserting Eq.~\ref{fig2_matrix} into \ref{local2state}, we have Eq.~\ref{fig2_DB} with $\xi =\left(6k^{4}-10k^{3}+3k^{2}+6k+2\right)/\left(12k^{3}\right)$ for local transitions.
Such a transition pattern is a case of $\Omega$.
Inserting Eq.~\ref{fig2_matrix} into \ref{global2state}, we have Eq.~\ref{fig2_DB} with $\xi =\left(k-1\right) /2$ for global transitions.
Eq.~\ref{fig2_DB} corresponds to Eq. 1 in the main text.
For pairwise-comparison updating, analogously, we have
\begin{equation}\label{fig2_PC}
\rho_{C}>\rho_{D} \iff \xi\frac{\Delta b}{c} > 1 ,
\end{equation}
where $\xi =\left(10k^{2}-4k+1\right) /\left(24k^{2}-12k\right)$ for local game transitions and $\xi=1/2$ for global game transitions.
Eq.~\ref{fig2_PC} corresponds to Eq. 2 in the main text.

With the same game transition pattern, in the competition between two stochastic strategies, i.e. $s_p$ and $s_q$ ($p>q$), from Eqs.~\ref{xi_stochastic_DB} and \ref{xi_stochastic_PC}, we have
\begin{equation} \label{stochastic2global}
\rho_{s_p}>\rho_{s_q} \iff \frac{b_{1}}{c} > k - \frac{(k+1)\left(p^2+q^2\right)+2pq-2}{2}\frac{\Delta b}{c}
\end{equation}
under death-birth updating and
\begin{equation} \label{stochastic2local}
\rho_{s_p}>\rho_{s_q} \iff \frac{p^2+q^2}{2}\frac{\Delta b}{c} > 1
\end{equation}
under birth-death or pairwise-comparison updating.
Eq.~\ref{stochastic2global} shows that for a large value of $p$, game transitions lower the threshold for $\rho_{s_p}>\rho_{s_q}$ relative to playing a fixed game.
Eq.~\ref{stochastic2local} shows that game transitions make it possible for cooperative stochastic strategies to be favored over less cooperative stochastic strategies under birth-death or pairwise-comparison updating, which has not been observed when players play a fixed game.

We next consider a game transition pattern under which the evolutionary outcome relies on the initial condition.
The game transition matrices are shown in Eq.~\ref{pattern_state2}.
$\textbf{M}$ in Eq.~\ref{M_proposition} is given by
\begin{equation}
\textbf{M} =
\bordermatrix{
& E_{AA}^{(1)} & E_{AA}^{(2)} & E_{AB}^{(1)} & E_{AB}^{(2)} & E_{BB}^{(1)} & E_{BB}^{(2)} \cr
E_{AA}^{(1)} & 1/2 & 0 & 1/2 & 0 & 0 & 0 \cr
E_{AA}^{(2)} & 0 & 1/2 & 0 & 1/2 & 0 & 0 \cr
E_{AB}^{(1)} & 1/3 & 0 & 1/3 & 0 & 1/3 & 0 \cr
E_{AB}^{(2)} & 0 & 1/3 & 0 & 1/3 & 0 & 1/3 \cr
E_{BB}^{(1)} & 0 & 0 & 1/2 & 0 & 1/2 & 0 \cr
E_{BB}^{(2)} & 0 & 0 & 0 & 1/2 & 0 & 1/2
}.
\end{equation}
Switching a few row entries and column entries, we have
\begin{equation}
\tilde{\textbf{M}} =
\bordermatrix{
& E_{AA}^{(1)} & E_{AB}^{(1)} & E_{BB}^{(1)} & E_{AA}^{(2)} & E_{AB}^{(2)} & E_{BB}^{(2)} \cr
E_{AA}^{(1)} & 1/2 & 1/2 & 0 & 0 & 0 & 0 \cr
E_{AB}^{(1)} & 1/3 & 1/3 & 1/3 & 0 & 0 & 0 \cr
E_{BB}^{(1)} & 0 & 1/2 & 1/2 & 0 & 0 & 0 \cr
E_{AA}^{(2)} & 0 & 0 & 0 & 1/2 & 1/2 & 0 \cr
E_{AB}^{(2)} & 0 & 0 & 0 & 1/3 & 1/3 & 1/3 \cr
E_{BB}^{(2)} & 0 & 0 & 0 & 0 & 1/2 & 1/2
}.
\end{equation}
By analyzing $\tilde{\textbf{M}}$, we get $E_{AA}^{(1)},E_{AB}^{(1)},E_{BB}^{(1)}$ belonging to one closed communicating class, and $E_{AA}^{(2)},E_{AB}^{(2)},E_{BB}^{(2)}$ belonging to the other class.
Thus, the original system can be reduced into two subsystems, one consisting of $E_{AA}^{(1)},E_{AB}^{(1)},E_{BB}^{(1)}$ and the other consisting of $E_{AA}^{(2)},E_{AB}^{(2)},E_{BB}^{(2)}$.
Here, we deal with the subsystem consisting of $E_{AA}^{(1)},E_{AB}^{(1)},E_{BB}^{(1)}$.
Substituting $\textbf{P}^{(2)}$, $\textbf{P}^{(1)}$, and $\textbf{P}^{(0)}$ into Eq.~\ref{system_matrix}, we have $\dot{\textbf{v}}=\bar{\textbf{A}}\textbf{v}$.
$\bar{\textbf{A}}$'s eigenvalues are $\lambda_1 = 0,\lambda_2 = -(k-2)/(kN),\lambda_3 = -(2k-2)/(kN)$, where $\lambda_2$ and $\lambda_3$ are negative for $k>2$.
Decomposing $\bar{\textbf{A}}$ into $\bar{\textbf{A}}=\textbf{VDV}^{-1}$, where
\begin{equation}
\textbf{V}=\left[
\begin{array}{ccc}
-\frac{p_A(kp_A - 2p_A + 1)}{(p_A - 1)(k + 2p_A - kp_A - 1)} & -\frac{kp_A - 2p_A + 1}{k + 2p_A - kp_A - 1} & 1\\
\frac{p_A(k - 2)}{k + 2p_A - kp_A - 1}                       & -\frac{k + 4p_A - 2kp_A - 2}{2(k + 2p_A - kp_A - 1)} & -1\\
1 & 1 & 1
\end{array}
\right],
\quad
\textbf{D}=\left[
\begin{array}{ccc}
0 & 0 & 0\\
0 & -\frac{k-2}{kN} & 0\\
0 & 0 & -\frac{2k-2}{kN}
\end{array}
\right] ,
\end{equation}
we can get $p_{AA}^{(1)}(t)$, $p_{AB}^{(1)}(t)$ and $p_{BB}^{(1)}(t)$ as a function of time $t$, given by
\begin{equation}
\left[
\begin{array}{c}
p_{AA}^{(1)}(t) \\
p_{AB}^{(1)}(t) \\
p_{BB}^{(1)}(t)
\end{array}
\right]
=\textbf{V}e^{\textbf{D}t}\textbf{V}^{-1}
\left[
\begin{array}{c}
p_{AA}^{(1)}(t_0) \\
p_{AB}^{(1)}(t_0) \\
p_{BB}^{(1)}(t_0)
\end{array}
\right],
\end{equation}
where $p_{AA}^{(1)}(t_0),p_{AB}^{(1)}(t_0),p_{BB}^{(1)}(t_0)$ are the initial values of $p_{AA}^{(1)},p_{AB}^{(1)},p_{BB}^{(1)}$.
When $t$ approaches infinity, we have
\begin{subequations}
\begin{equation}
p_{AA}^{(1)}(\infty)=\frac{(k-2)p_A^2+p_A}{k-1}\left(p_{AA}^{(1)}(t_0)+2p_{AB}^{(1)}(t_0)+p_{BB}^{(1)}(t_0)\right) ; \label{Twogame2pAA}\\
\end{equation}
\begin{equation}
p_{AB}^{(1)}(\infty)=-\frac{(k-2)p_A^2-(k-2)p_A}{k-1}\left(p_{AA}^{(1)}(t_0)+2p_{AB}^{(1)}(t_0)+p_{BB}^{(1)}(t_0)\right) ; \label{Twogame2pAB}\\
\end{equation}
\begin{equation}
p_{BB}^{(1)}(\infty)=\frac{(k-2)p_A^2+(3-2k)p_A+k-1}{k-1}\left(p_{AA}^{(1)}(t_0)+2p_{AB}^{(1)}(t_0)+p_{BB}^{(1)}(t_0)\right) . \label{Twogame2pBB}
\end{equation}
\end{subequations}
Note that $p_{AA}^{(1)}(t_0)+2p_{AB}^{(1)}(t_0)+p_{BB}^{(1)}(t_0)$ is the initial frequency of game 1, denoted by $p^{(1)}$.
Then, we have $p_{AA}^{(1)}$, $p_{AB}^{(1)}$, and $p_{BB}^{(1)}$ as functions of $p_A$ and $p^{(1)}$.
Analogously, $p_{AA}^{(2)}$, $p_{AB}^{(2)}$, and $p_{BB}^{(2)}$ are functions of $p_A$ and $p^{(2)}$.
Substituting $p_{AA}^{(1)}$, $p_{AB}^{(1)}$, $p_{BB}^{(1)}$ and $p_{AA}^{(2)}$, $p_{AB}^{(2)}$, $p_{BB}^{(2)}$ into Eqs.~\ref{Iai}-\ref{Idi}, we can reduce Eq.~\ref{general_rule_DB} to
\begin{equation} \label{reduced_2games}
(k+1)\sum_{i=1}^2 p^{(i)}R_i+(k-1)\sum_{i=1}^2 p^{(i)}S_i-(k-1)\sum_{i=1}^2 p^{(i)}T_i-(k+1)\sum_{i=1}^2 p^{(i)}P_i>0.
\end{equation}
For donation games, we have
\begin{equation}
\frac{b_1}{c}>k+p^{(2)}\frac{\Delta b_{12}}{c} .
\end{equation}
$\xi=-p^{(2)}$ shows that the evolutionary outcome is sensitive to the initial fractions of various games.

Based on Eqs.~\ref{A_DB_global} and \ref{system_matrix_global}, we can perform the analogous study under global game transitions. With the game transition pattern given in Eq.~\ref{pattern_state2}, we have $\xi=-p^{(2)}$.

\subsection{Evolutionary dynamics with probabilistic game transitions among three states ($n=3$)}
All examples that we examine in the main text exhibit a deterministic game transition.
That is, the game to be played in the next time step is not probabilistic.
In this section, we present an example with stochastic game transitions among three states.
Game 1 is the most valuable and game 3 is the least valuable, i.e., $b_1>b_2>b_3$.
The game transition matrices are given by
\begin{equation}
\textbf{P}^{(2)}
=
\left[
\begin{matrix}
1 & 0 & 0 \\
p & 1-p & 0 \\
p & 0 & 1-p
\end{matrix}
\right],
\quad
\textbf{P}^{(1)}
=
\left[
\begin{matrix}
1-p & p & 0 \\
0 & 1 & 0 \\
0 & p & 1-p
\end{matrix}
\right],
\quad
\textbf{P}^{(0)}
=
\left[
\begin{matrix}
1-p & 0 & p \\
0 & 1-p & p \\
0 & 0 & 1
\end{matrix}
\right].
\end{equation}
Mutual cooperation (mutual defection) is prone to yield game $1$ (game $3$) and unilateral cooperation/defection a moderately-valuable game 2.
The game transition occurs with a probability $p$ and players play the old game with a probability $1-p$.
For $p=0$, by virtue of the approach in Section \ref{section_sensitivity}, we know that the evolutionary outcome relies on the initial condition.
We can refer to the example in Section 4\ref{section_example2} to derive the evolutionary outcome.
For $p>0$, the evolutionary outcome is independent of the initial condition.
Note that $p = 1$ corresponds to the deterministic case.
Under death-birth updating, the general rule for cooperation to be favored over defection is
\begin{equation}
\rho_C>\rho_D \Longleftrightarrow \frac{b_1}{c}>k-\xi_2\frac{\Delta b_{12}}{c}-\xi_3\frac{\Delta b_{13}}{c}.
\end{equation}
For local game transitions, we have
\begin{equation}
\xi_2=\frac{\left(6k^5 - 15k^4 + 24k^3 - 24k^2 + 15k - 6\right) p^2 + \left(4k^4 - 21k^3 + 32k^2 - 19k + 14\right)p - 2k^2 + 8k - 8}{\left(12k^4 - 18k^3 + 30k^2 - 18k + 12\right) p^2 + \left(18k^3 - 36k^2 + 36k - 24\right) p + 6k^2 - 18k + 12}
\end{equation}
and
\begin{equation}
\xi_3=\frac{\left(k^4 + 2k^2 - 7k + 2\right) p - 2k^2 + 5k - 2}{\left(12k^4 - 18k^3 + 30k^2 - 18k + 12\right) p^2 + \left(18k^3 - 36k^2 + 36k - 24\right) p + 6k^2 - 18k + 12}.
\end{equation}
Here, we analyze how the probabilistic transition measured by $p$ affects the critical benefit-to-cost ratio $\left(b_1/c\right)^{*}$.
Analyzing $\left(b_1/c\right)^{*}=k-\xi_2\Delta b_{12}/c-\xi_3\Delta b_{13}/c$, we get a threshold
\begin{equation}
\left(\frac{\Delta b_{13}}{\Delta b_{12}}\right)^*=5 - \frac{3\left(k + 4\right)\left(k^2 + 2k - 2\right)}{2k^5 - 3k^4 + 6k^2 + 2k - 8}.
\end{equation}
If $\Delta b_{12}/\Delta b_{13}$ is lower than $\left(\Delta b_{12}/\Delta b_{13}\right)^*$, $\left(b_1/c\right)^{*}$ decreases monotonically in $p$.
If $\Delta b_{12}/\Delta b_{13}$ is larger than $\left(\Delta b_{12}/\Delta b_{13}\right)^*$, $\left(b_1/c\right)^{*}$ is a non-monotonic function of $p$.
Specifically, as $p$ increases starting from a value slightly larger than $0$, $\left(b_1/c\right)^{*}$ decreases first and then increases.
The optimal transition probability $p$ for collective cooperation is
\begin{equation}
p^*=\frac{\phi_2r+\phi_3+\sqrt{\phi_4r^2+\phi_5r+\phi_6}}{\phi_0r+\phi_1},
\end{equation}
where
\begin{equation}
\begin{split}
\phi_0 &= k^4 + 2k^2 - 7k + 2 ; \\
\phi_1 &= - 5k^4 + 6k^3 - 4k^2 + 11k + 2 ; \\
\phi_2 &= \left(2k - 1\right)\left(k - 2\right) ; \\
\phi_3 &= \left(k - 2\right)\left(3k^2 - 4k - 1\right) ; \\
\phi_4 &= \left(- k^3 - 4k^2 + 2k + 2\right)\phi_7 ; \\
\phi_5 &= \left(k^3 + 13k^2 - 8k - 8\right)\phi_7 ; \\
\phi_6 &= \left(2k^3 - 10k^2 + 8k + 8\right)\phi_7 ; \\
\phi_7 &= \frac{-k^2\left(k - 2\right)\left( k^2 - k - 1\right)\left(k^2 + 2k - 2\right)}{\left(2k^2 - k + 2\right)\left(k^2 - k + 1\right)} .
\end{split}
\end{equation}

Depending on the variations in different games, probabilistic game transitions can strengthen the promotive effects of game transitions on the evolution of cooperation in a few cases, whereas weaken them in other cases.
The conclusion holds under other updating rules like imitation and pairwise-comparison updating.

For global game transitions, the related parameters are $\xi_2 = (k-1)/2$ and $\xi_3 = 0$.
In this case, probabilistic transitions do not alter the effects of game transitions on the evolution of cooperation.

\makeatletter
\@fpsep\textheight
\makeatother

\setcounter{figure}{0}
\renewcommand{\thefigure}{S\arabic{figure}}

\clearpage
\newpage

\begin{figure}
\centering
\includegraphics[width=1\textwidth]{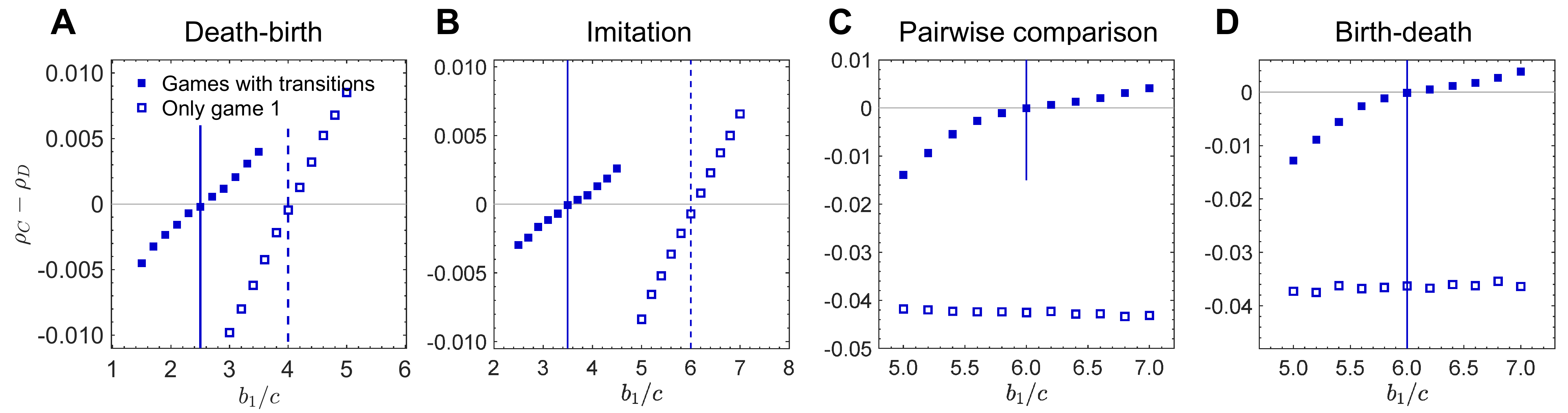}
\caption{\label{fig:6} \textbf{Game transitions can promote cooperation}.
	We study the transition between two donation games: a cooperator pays a cost $c$ to bring its opponent a benefit $b_1$ in game 1 or $b_2$ in game 2;
	defectors forgo this donation.
	$b_1$ is larger than $b_2$.
	Mutual cooperation leads to game 1 and other action profiles lead to game 2.
	We examine death-birth (\textbf{A}), imitation (\textbf{B}), pairwise-comparison (\textbf{C}), and birth-death (\textbf{D}) updating on random regular graphs.
	The cross points of the dots and the horizontal lines mark the critical benefit-to-cost ratios for cooperation to be favored over defection, i.e. $\rho_C>\rho_D$, by numerical simulations.
	The vertical lines give the analytical critical benefit-to-cost ratios.
	Under death-birth and imitation updating, game transitions reduce the critical benefit-to-cost for $\rho_C>\rho_D$.
	Under pairwise-comparison and birth-death updating, game transitions make it possible for $\rho_C>\rho_D$.
	We take $N=500$, $k=4$, $\delta=0.01$, $c = 1$.
	Other parameters: $b_2 = b_1-1$ for death-birth and imitation updating, $b_2 = 4$ for pairwise-comparison and birth-death updating.
	Each simulation runs until the population reaches fixation and each point is averaged over $10^6$ runs.
}
\end{figure}

\clearpage
\newpage

\begin{figure}
\centering
\includegraphics[width=1\textwidth]{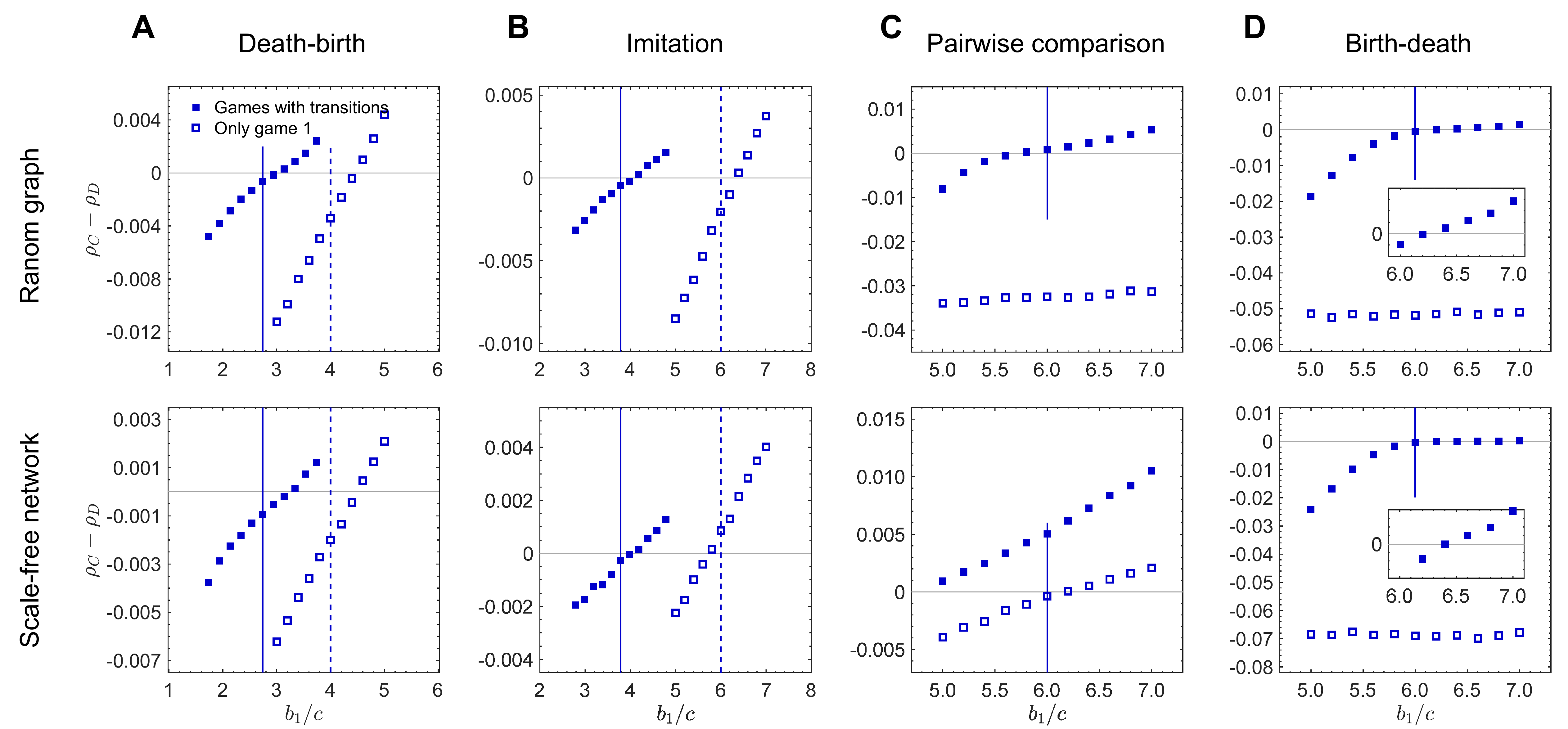}
\caption{\label{fig:7} \textbf{Game transitions can promote cooperation on social networks}.
	We study the transition between two donation games: a cooperator pays a cost $c$ to bring its opponent a benefit $b_1$ in game 1 or $b_2$ in game 2;
	defectors pay no costs and provide no benefits.
	$b_1$ is larger than $b_2$.
	Mutual cooperation allows for game 1 and other action profiles lead to game 2.
	We examine death-birth (\textbf{A}), imitation (\textbf{B}), pairwise-comparison (\textbf{C}), and birth-death (\textbf{D}) updating on random graphs \cite{si1960-Erdoes-p17-61_s} and scale-free networks \cite{si1999-Barabasi-p509-512_s,si2002-Albert-p47-97_s}.
	The cross points of the dots and the horizontal lines mark the critical benefit-to-cost ratios for cooperation to be favored over defection by numerical simulations.
	The vertical lines give the analytical critical benefit-to-cost ratios based on random regular graphs.
	The average degree of the random regular graph and the scale-free networks is $4$.
	Other parameters are the same as those in Fig. \ref{fig:6}.
	Game transitions reduce the critical benefit-to-cost for the success of cooperators ($\rho_C>\rho_D$).
}
\end{figure}

\clearpage
\newpage

\begin{figure}
\centering
\includegraphics[width=0.9\textwidth]{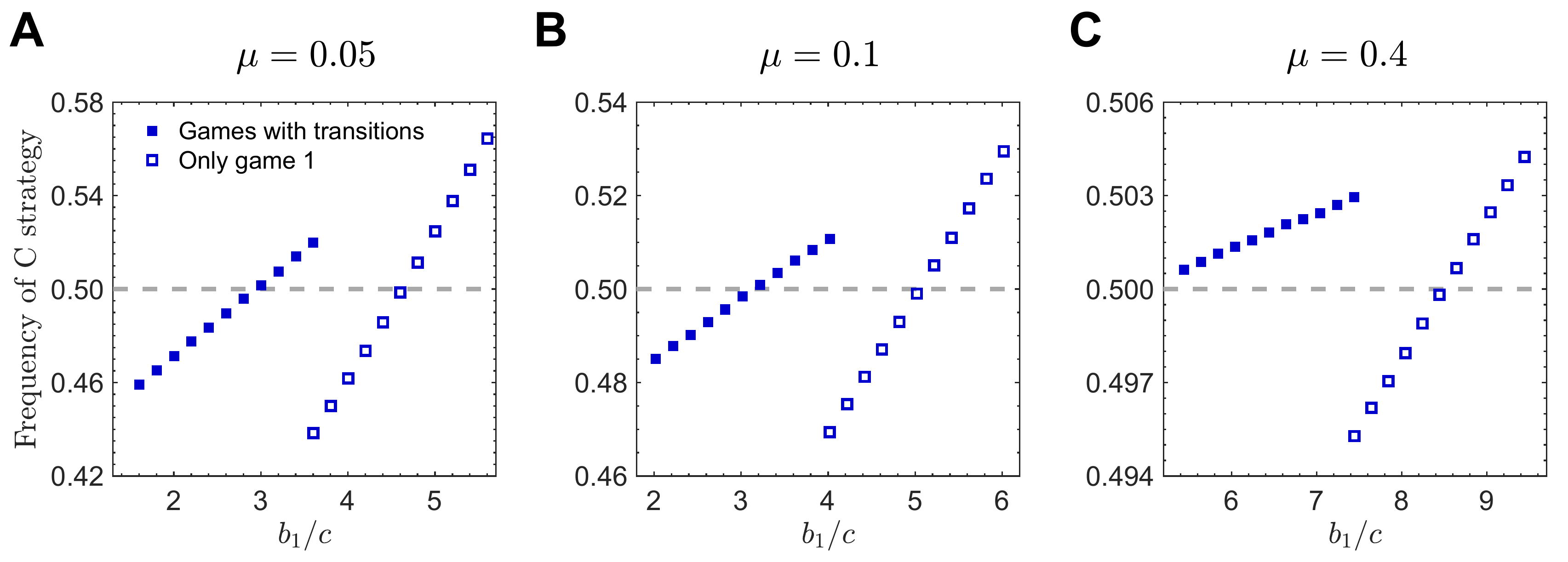}
\caption{\label{fig:8} \textbf{Game transitions can promote cooperation in the presence of mutation or random strategy exploration}.
	We study the transition between two donation games: a cooperator pays a cost $c$ to bring its opponent a benefit $b_1$ in game 1 or $b_2$ in game 2;
	defectors pay no costs and provide no benefits.
	$b_1$ is larger than $b_2$.
	Mutual cooperation allows for game 1 and other action profiles lead to game 2.
	We investigate death-birth updating on random regular graphs.
	With probability $1-\mu$, the empty site is occupied by the neighbor's offspring.
	With probability $\mu$, the empty is occupied by a cooperator or a defector with equal probability.
	Here, the frequency of cooperative strategies $\langle f_C\rangle$ is used to measure the success of cooperators.
	Cooperation is favored over defection if $\langle f_C\rangle > 1/2$.
	We obtain each data point by averaging $\langle f_C\rangle$ in 100 independent runs.
	For each run, $\langle f_C\rangle$ is obtained by averaging the frequency of cooperative strategies in the last $2\times 10^7$ time steps.
	We take $N=500$, $k=4$, $b_2 = b_1-1$, $\delta=0.01$, and $c = 1$.
	Other parameters: $\mu = 0.05$ (\textbf{A}), $\mu = 0.1$ (\textbf{B}), and $\mu = 0.4$ (\textbf{C}).
	The cross points of the dots and the horizontal lines mark the critical benefit-to-cost ratios for cooperation being favored over defection ($\langle f_C\rangle > 1/2$).
	Game transitions reduce the critical benefit-to-cost for the success of cooperators.
}
\end{figure}

\clearpage
\newpage

\begin{figure}
\centering
\includegraphics[width=0.9\textwidth]{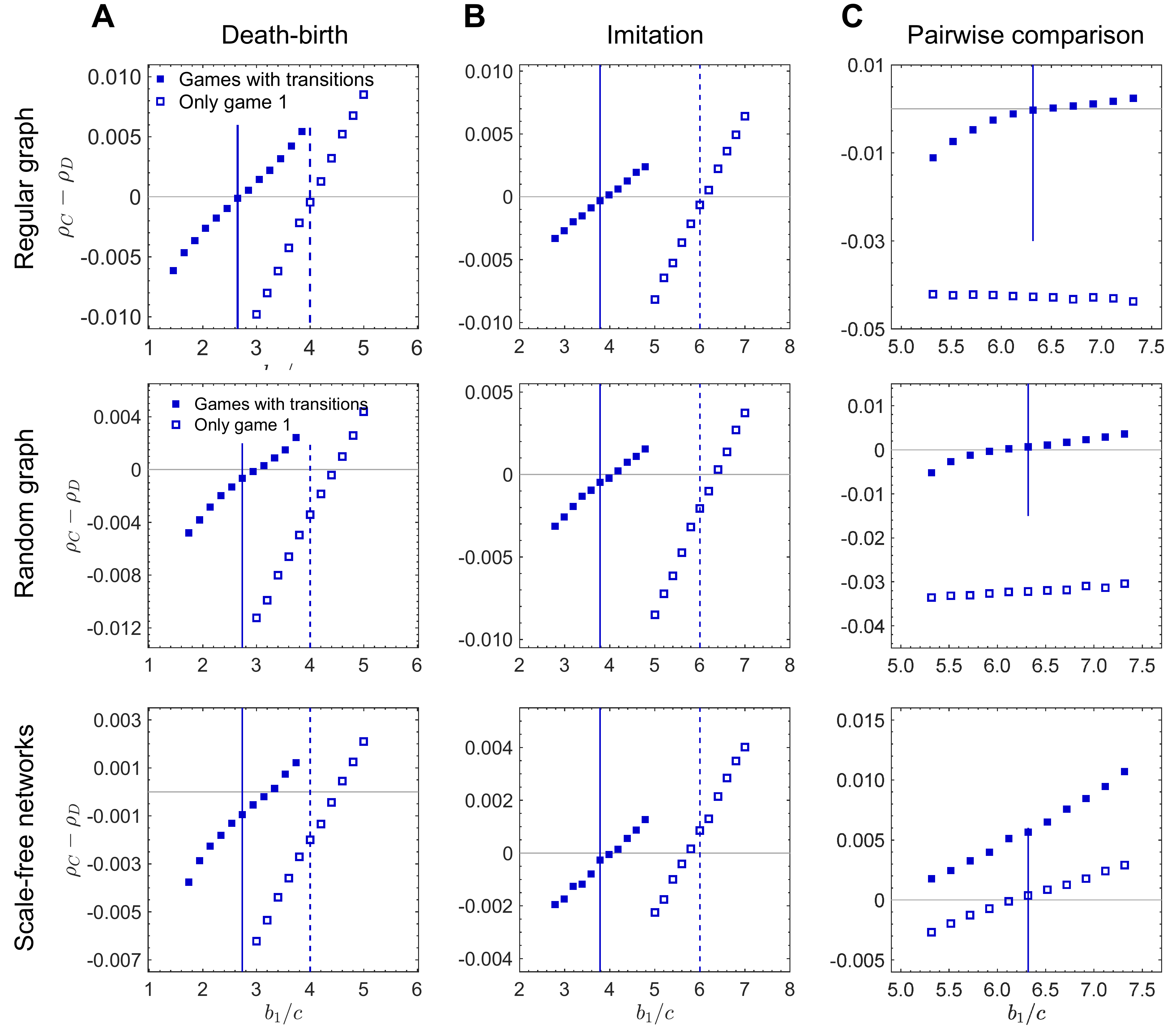}
\caption{\label{fig:10}\textbf{Local game transitions can promote cooperation}.
	We study the transition between two donation games: a cooperator pays a cost $c$ to bring its opponent a benefit $b_1$ in game 1 or $b_2$ in game 2;
	defectors pay no costs and provide no benefits.
	$b_1$ is larger than $b_2$.
	Mutual cooperation allows for game 1 and other action profiles lead to game 2.
	We examine death-birth (\textbf{A}), imitation (\textbf{B}), and pairwise-comparison (\textbf{C}) on random regular graphs, random graphs, and scale-free networks.
	The cross points of the dots and the horizontal lines mark the critical benefit-to-cost ratios for cooperation to be favored over defection by numerical simulations.
	The vertical lines give the analytical critical benefit-to-cost ratios.
	Game transitions reduce the critical benefit-to-cost for the success of cooperators ($\rho_C>\rho_D$).
	All parameters are the same as those in Fig. \ref{fig:6}.
}
\end{figure}

\clearpage
\newpage

\begin{figure}
\centering
\includegraphics[width=0.9\textwidth]{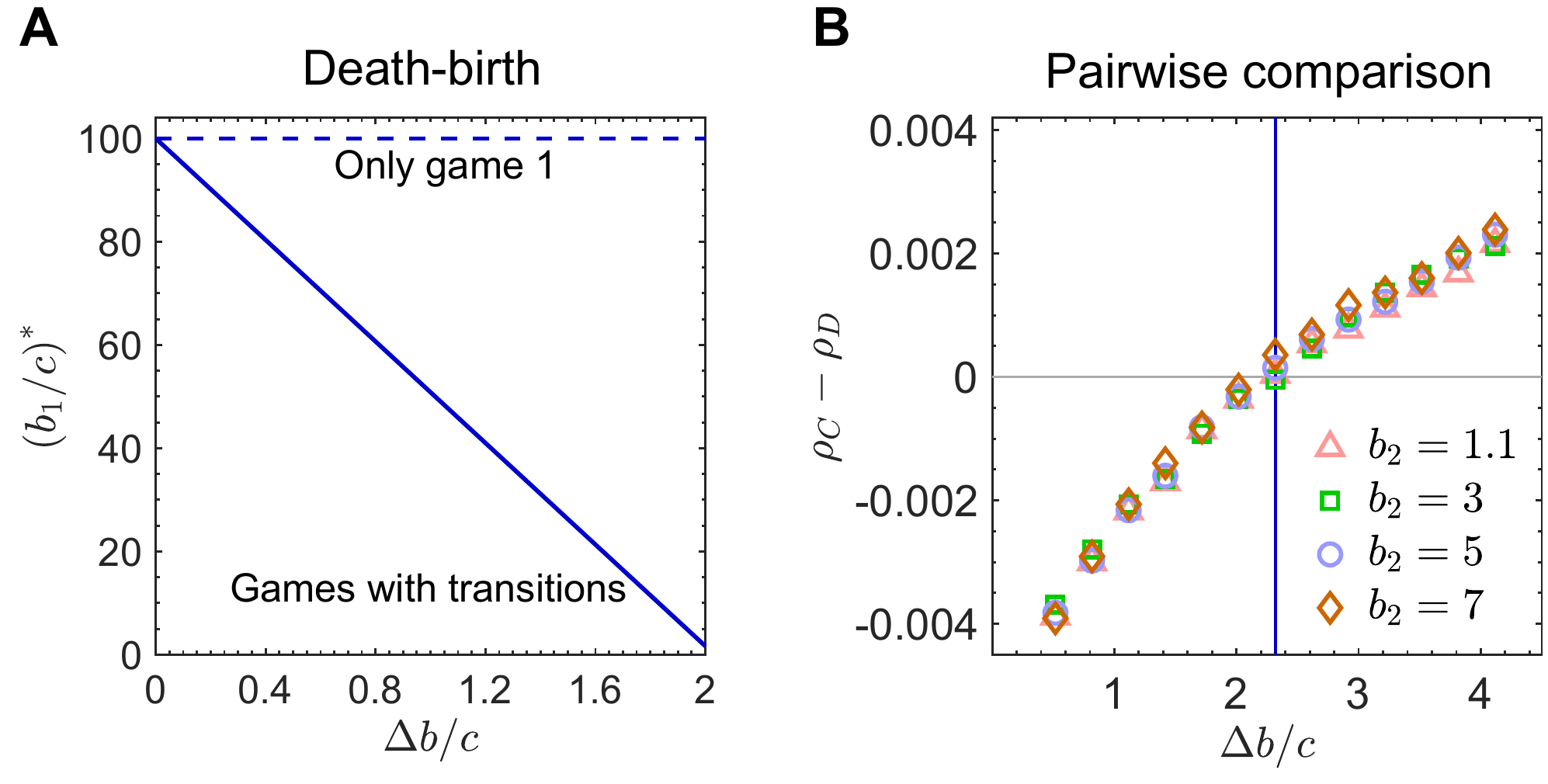}
\caption{\label{fig:11} \textbf{With local game transitions a small variation in different games can promote cooperation markedly}.
	We study the transition between two donation games: a cooperator pays a cost $c$ to bring its opponent a benefit $b_1$ in game 1 or $b_2$ in game 2;
	defectors forgo the helping behavior.
	$b_1$ is larger than $b_2$.
	Mutual cooperation allows for game 1 and other action profiles lead to game 2.
	We examine death-birth (\textbf{A}) and pairwise-comparison (\textbf{B}) updating on random regular graphs.
	Under death-birth updating, a small difference between $b_1$ and $b_2$ ($\Delta b = b_1-b_2$) greatly reduces the critical benefit-to-cost ratio (\textbf{A}).
	Under pairwise-comparison updating, the difference between games, $b_1-b_2$, rather than the individual value of $b_1$ and $b_2$, determines the success of cooperators (\textbf{B}).
	Apart from $b_1$ and $b_2$, all other parameters follow Fig. \ref{fig:6}.
}
\end{figure}


\begin{thebibliography}{50}
	
	\bibitem{sigmund:PUP:2010}
	Sigmund K (2010) {\em The calculus of selfishness}.
	\newblock (Princeton University Press).
	
	\bibitem{2006-Nowak-p1560-1563}
	Nowak MA (2006) Five rules for the evolution of cooperation.
	\newblock {\em Science} 314(5805):1560--1563.
	
	\bibitem{1964-Hamilton-p1-16}
	Hamilton WD (1964) {The genetical evolution of social behaviour. I}.
	\newblock {\em J. Theor. Biol.} 7(1):1--16.
	
	\bibitem{hamilton:JTB:1964b}
	Hamilton WD (1964) The genetical evolution of social behaviour. {II}.
	\newblock {\em J. Theor. Biol.} 7(1):17--52.
	
	\bibitem{1992-Nowak-p826-829}
	Nowak MA, May RM (1992) Evolutionary games and spatial chaos.
	\newblock {\em Nature} 359(6398):826--829.
	
	\bibitem{2006-Ohtsuki-p502-505}
	Ohtsuki H, Hauert C, Lieberman E, Nowak MA (2006) A simple rule for the
	evolution of cooperation on graphs and social networks.
	\newblock {\em Nature} 441(7092):502--505.
	
	\bibitem{2007-Taylor-p469-469}
	Taylor PD, Day T, Wild G (2007) Evolution of cooperation in a finite
	homogeneous graph.
	\newblock {\em Nature} 447(7143):469--472.
	
	\bibitem{2017-Allen-p227-230}
	Allen B, et~al. (2017) Evolutionary dynamics on any population structure.
	\newblock {\em Nature} 544(7649):227--230.
	
	\bibitem{2019-Qi-p20190041-20190041}
	Su Q, Li A, Wang L, Stanley HE (2019) Spatial reciprocity in the evolution of
	cooperation.
	\newblock {\em Proc. R. Soc. B-Biol. Sci.} 286(1900):20190041.
	
	\bibitem{wilson:EE:1992}
	Wilson DS, Pollock GB, Dugatkin LA (1992) Can altruism evolve in purely viscous
	populations?
	\newblock {\em Evol. Ecol.} 6(4):331--341.
	
	\bibitem{taylor:EE:1992}
	Taylor PD (1992) Altruism in viscous populations {\textemdash} an inclusive
	fitness model.
	\newblock {\em Evol. Ecol.} 6(4):352--356.
	
	\bibitem{mitteldorf:JTB:2000}
	Mitteldorf J, Wilson DS (2000) Population viscosity and the evolution of
	altruism.
	\newblock {\em J. Theor. Biol.} 204(4):481--496.
	
	\bibitem{2014-Rand-p17093-17098}
	Rand DG, Nowak MA, Fowler JH, Christakis NA (2014) Static network structure can
	stabilize human cooperation.
	\newblock {\em Proc. Natl. Acad. Sci. U.S.A.} 111(48):17093.
	
	\bibitem{2015-Mastrandrea-p136497-136497}
	Mastrandrea R, Fournet J, Barrat A (2015) Contact patterns in a high school: A
	comparison between data collected using wearable sensors, contact diaries and
	friendship surveys.
	\newblock {\em PLoS ONE} 10(9):e0136497.
	
	\bibitem{traud:SIAMR:2011}
	Traud AL, Kelsic ED, Mucha PJ, Porter MA (2011) Comparing community structure
	to characteristics in online collegiate social networks.
	\newblock {\em SIAM Review} 53(3):526--543.
	
	\bibitem{traud:PA:2012}
	Traud AL, Mucha PJ, Porter MA (2012) {Social structure of {F}acebook networks}.
	\newblock {\em Phys. A} 391(16):4165--4180.
	
	\bibitem{nr}
	Rossi RA, Ahmed NK (2015) The network data repository with interactive graph
	analytics and visualization in {\em Proceedings of the Twenty-Ninth AAAI
		Conference on Artificial Intelligence}.
	\newblock (AAAI, Menlo Park, CA), p. 4292–4293.
	
	\bibitem{2018-Hilbe-p246-249}
	Hilbe C, \v{S}imsa v, Chatterjee K, Nowak MA (2018) Evolution of cooperation in
	stochastic games.
	\newblock {\em Nature} 559(7713):246--249.
	
	\bibitem{2014-Levin-p10838-10845}
	Levin SA (2014) Public goods in relation to competition, cooperation, and
	spite.
	\newblock {\em Proc. Natl. Acad. Sci. U.S.A.} 111(Suppl 3):10838--10845.
	
	\bibitem{2013-Franzenburg-p781-781}
	Franzenburg S, et~al. (2013) Bacterial colonization of hydra hatchlings follows
	a robust temporal pattern.
	\newblock {\em ISME J.} 7(4):781.
	
	\bibitem{2013-McFall-Ngai-p3229-3236}
	McFall-Ngai M, et~al. (2013) Animals in a bacterial world, a new imperative for
	the life sciences.
	\newblock {\em Proc. Natl. Acad. Sci. U.S.A.} 110(9):3229--3236.
	
	\bibitem{2008-Acar-p471-471}
	Acar M, Mettetal JT, van Oudenaarden A (2008) Stochastic switching as a
	survival strategy in fluctuating environments.
	\newblock {\em Nat. Genet.} 40(4):471.
	
	\bibitem{2007-Rankin-p643-651}
	Rankin DJ, Bargum K, Kokko H (2007) The tragedy of the commons in evolutionary
	biology.
	\newblock {\em Trends Ecol. Evol} 22(12):643 -- 651.
	
	\bibitem{2004-Nowak-p646-650}
	Nowak MA, Sasaki A, Taylor C, Fudenberg D (2004) Emergence of cooperation and
	evolutionary stability in finite populations.
	\newblock {\em Nature} 428(6983):646--650.
	
	\bibitem{2010-Wu-p46106-46106}
	Wu B, Altrock PM, Wang L, Traulsen A (2010) Universality of weak selection.
	\newblock {\em Phys. Rev. E} 82(4):046106.
	
	\bibitem{2013-Wu-p1003381-1003381}
	Wu B, Garc\'{i}a J, Hauert C, Traulsen A (2013) Extrapolating weak selection in
	evolutionary games.
	\newblock {\em PLoS Comput. Biol.} 9(12):e1003381.
	
	\bibitem{1998-Watts-p440-442}
	Watts DJ, Strogatz SH (1998) Collective dynamics of ‘small-world’ networks.
	\newblock {\em Nature} 393(6684):440--442.
	
	\bibitem{1958-Moran-p60-60}
	Moran PAP (1958) Random processes in genetics.
	\newblock {\em Math. Proc. Cambridge Philos. Soc.} 54(01):60.
	
	\bibitem{1998-Szabo-p69-73}
	Szab{\'{o}} G, T{\H{o}}ke C (1998) Evolutionary prisoner's dilemma game on a
	square lattice.
	\newblock {\em Phys. Rev. E} 58(1):69--73.
	
	\bibitem{2007-Traulsen-p522-529}
	Traulsen A, Pacheco JM, Nowak MA (2007) Pairwise comparison and selection
	temperature in evolutionary game dynamics.
	\newblock {\em J. Theor. Biol.} 246(3):522--529.
	
	\bibitem{1960-Erdoes-p17-61}
	Erd\"{o}s P, R\'{e}nyi A (1960) On the evolution of random graphs.
	\newblock {\em Publ. Math. Inst. Hung. Acad. Sci.} 5:17--61.
	
	\bibitem{1999-Barabasi-p509-512}
	Barab\'{a}si AL, Albert R (1999) Emergence of scaling in random networks.
	\newblock {\em Science} 286(5439):509--512.
	
	\bibitem{2009-Tarnita-p570-581}
	Tarnita CE, Ohtsuki H, Antal T, Fu F, Nowak MA (2009) Strategy selection in
	structured populations.
	\newblock {\em J. Theor. Biol.} 259(3):570--581.
	
	\bibitem{2014-Allen-p113-151}
	Allen B, Nowak MA (2014) Games on graphs.
	\newblock {\em EMS Surv. Math. Sci.} 1(1):113--151.
	
	\bibitem{1991-Bull-p63-74}
	Bull JJ, Rice WR (1991) Distinguishing mechanisms for the evolution of
	co-operation.
	\newblock {\em J. Theor. Biol.} 149(1):63--74.
	
	\bibitem{2004-Sachs-p135-160}
	Sachs J, Mueller U, Wilcox T, Bull J (2004) The evolution of cooperation.
	\newblock {\em Q. Rev. Biol.} 79(2):135--160.
	
	\bibitem{schardl:PRPB:1997}
	Schardl CL, Clay K (1997) {Evolution of Mutualistic Endophytes from Plant
		Pathogens} in {\em Plant Relationships Part B}.
	\newblock (Springer Berlin Heidelberg), pp. 221--238.
	
	\bibitem{cheplick:OUP:2009}
	Cheplick GP, Faeth S (2009) {\em {Ecology and Evolution of the Grass-Endophyte
			Symbiosis}}.
	\newblock (Oxford University Press).
	
	\bibitem{2000-Amaral-p11149-11152}
	Amaral LAN, Scala A, Barth\'{e}l\'{e}my M, Stanley HE (2000) Classes of
	small-world networks.
	\newblock {\em Proc. Natl. Acad. Sci. U.S.A.} 97(21):11149--11152.
	
	\bibitem{2012-Allen-p97-105}
	Allen B, Traulsen A, Tarnita CE, Nowak MA (2012) How mutation affects
	evolutionary games on graphs.
	\newblock {\em J. Theor. Biol.} 299:97--105.
	
	\bibitem{2019-Peter-p20140663-20140663}
	Ashcroft P, Altrock PM, Galla T (2019) Fixation in finite populations evolving
	in fluctuating environments.
	\newblock {\em J. R. Soc. Interface} 11(100):20140663.
	
	\bibitem{2013-Assaf-p238101-238101}
	Assaf M, Mobilia M, Roberts E (2013) Cooperation dilemma in finite populations
	under fluctuating environments.
	\newblock {\em Phys. Rev. Lett.} 111(23):238101.
	
	\bibitem{2015-McAvoy-p1004349-1004349}
	McAvoy A, Hauert C (2015) Asymmetric evolutionary games.
	\newblock {\em PLoS Comput. Biol.} 11(8):e1004349.
	
	\bibitem{2016-Weitz-p7518-7525}
	Weitz JS, Eksin C, Paarporn K, Brown SP, Ratcliff WC (2016) An oscillating
	tragedy of the commons in replicator dynamics with game-environment feedback.
	\newblock {\em Proc. Natl. Acad. Sci. U.S.A.} 113(47):E7518--E7525.
	
	\bibitem{2016-Gokhale-p28-42}
	Gokhale CS, Hauert C (2016) Eco-evolutionary dynamics of social dilemmas.
	\newblock {\em Theor. Popul. Biol.} 111:28--42.
	
	\bibitem{2006-Hauert-p2565-2571}
	Hauert C, Holmes M, Doebeli M (2006) Evolutionary games and population
	dynamics: maintenance of cooperation in public goods games.
	\newblock {\em Proc. R. Soc. B-Biol. Sci.} 273(1600):2565--2571.
	
	\bibitem{2014-Stewart-p17558-17563}
	Stewart AJ, Plotkin JB (2014) Collapse of cooperation in evolving games.
	\newblock {\em Proc. Natl. Acad. Sci. U.S.A.} 111(49):17558--17563.
	
	\bibitem{2019-Tilman-p-}
	Tilman AR, Plotkin JB, Ak\c{c}ay E (2019) Evolutionary games with environmental
	feedbacks.
	\newblock bioRxiv: 10.1101/493023.
	
	\bibitem{2019-Hauert-p347-360}
	Hauert C, Saade C, McAvoy A (2019) Asymmetric evolutionary games with
	environmental feedback.
	\newblock {\em J. Theor. Biol.} 462:347--360.
	
	\bibitem{2019-Su-p1006947-1006947}
	Su Q, Zhou L, Wang L (2019) Evolutionary multiplayer games on graphs with edge
	diversity.
	\newblock {\em PLoS Comput. Biol.} 15(4):e1006947.
	
\end{thebibliography}

\begin{thebibliography}{13}

\bibitem{si2004-Nowak-p646-650}
Nowak MA, Sasaki A, Taylor C, Fudenberg D (2004) Emergence of cooperation and
evolutionary stability in finite populations.
\newblock {\em Nature} 428(6983):646--650.

\bibitem{si2004-Ewens}
Ewens WJ (2004) {\em Mathematical Population Genetics. I. Theoretical
	Introduction}.
\newblock (New York: Springer).

\bibitem{si2010-Wu-p46106-46106}
Wu B, Altrock PM, Wang L, Traulsen A (2010) Universality of weak selection.
\newblock {\em Phys. Rev. E} 82(4):046106.

\bibitem{si2013-Wu-p1003381-1003381}
Wu B, Garc\'{i}a J, Hauert C, Traulsen A (2013) Extrapolating weak selection in
evolutionary games.
\newblock {\em PLoS Comput. Biol.} 9(12):e1003381.

\bibitem{si2006-Ohtsuki-p502-505}
Ohtsuki H, Hauert C, Lieberman E, Nowak MA (2006) A simple rule for the
evolution of cooperation on graphs and social networks.
\newblock {\em Nature} 441(7092):502--505.

\bibitem{si2001-Khalil}
Khalil HK (2001) {\em Nonlinear Systems}.
\newblock (Prentice Hall).

\bibitem{si2004-Gardiner}
Gardiner CW (2004) {\em Handbook of Stochastic Methods}.
\newblock (Springer).

\bibitem{si2015-McAvoy-p1004349-1004349}
McAvoy A, Hauert C (2015) Asymmetric evolutionary games.
\newblock {\em PLoS Comput. Biol.} 11(8):e1004349.

\bibitem{si2019-Su-p1006947-1006947}
Su Q, Zhou L, Wang L (2019) Evolutionary multiplayer games on graphs with edge
diversity.
\newblock {\em PLoS Comput. Biol.} 15(4):e1006947.

\bibitem{si2009-Tarnita-p570-581}
Tarnita CE, Ohtsuki H, Antal T, Fu F, Nowak MA (2009) Strategy selection in
structured populations.
\newblock {\em J. Theor. Biol.} 259(3):570--581.

\bibitem{si1960-Erdoes-p17-61_s}
Erd\"{o}s P, R\'{e}nyi A (1960) On the evolution of random graphs.
\newblock {\em Publ. Math. Inst. Hung. Acad. Sci.} 5:17--61.

\bibitem{si1999-Barabasi-p509-512_s}
Barab\'{a}si AL, Albert R (1999) Emergence of scaling in random networks.
\newblock {\em Science} 286(5439):509--512.

\bibitem{si2002-Albert-p47-97_s}
Albert R, Barab\'asi AL (2002) Statistical mechanics of complex networks.
\newblock {\em Rev. Mod. Phys.} 74(1):47--97.

\end{thebibliography}
\end{document}